# Signatures of Heavy $Z'$ in the Extra U(1) Superstring Inspired Model: RGEs Analysis

Pranav Saxena[1,♣], Prachi Parashar[2,♦], N.K. Sharma[1,♥], Ashok K. Nagawat[1,♠], and Sardar Singh[1,*]

[1]*Theoretical High Energy Physics Group, Department of Physics, University of Rajasthan, Jaipur – 302004 (India)*
[2]*Dept. of Physics & Astronomy, Oklahoma University, Oklahoma, Norman 73019 (USA)*

## Abstract

In the extra U(1) superstring inspired model, we examine the electroweak and $U(1)'$ symmetry breaking with the singlet and exotic quark $D$, $D^c$ along with the study of heavy $Z'$ boson in accordance with the top quark mass region. For this, we have done the analysis of complete renormalization group equations (RGEs) pertaining to the anomaly free $E_6$ - η model of rank 5. The $Z'$ is found of the order of *TeV* or above with allowed small $Z-Z'$ mixing angle, for which the large singlet VEV is required. This is done by considering the non-universality of Yukawa couplings at GUT scale since these do not obey the $E_6$ relationship and also satisfies the unitarity constraints both at GUT and weak scale, and rest of the parameters, i.e., gaugino masses, tri-linear couplings, and soft supersymmetric breaking masses are kept universal at GUT scale with the gauge couplings unification. The large value of Yukawa couplings (order of 1) triggered the symmetry breaking radiatively and induces the effective μ parameter at the electroweak scale and lead to a viable low energy spectrum at weak scale.

♣ pranav@uniraj.erent.in
♦ prachi@nhn.ou.edu
♥ nksharma@uniraj.ernet.in
♠ nagawat@uniraj.ernet.in
* ssingh@uniraj.ernet.in

## 1. INTRODUCTION

For doing the physics beyond SM at low energy, we have made the efforts to obtain the phenomenological acceptable low energy spectrum through the analysis of Renormalization Group Equations (RGEs) in the considered simplest $E_6 - \eta$, namely, $SU(3)_C \times SU(2)_L \times U(1)_Y \times U(1)_\eta$ the rank –5 model [1-5], with all elements that are necessary for a realistic model. The model contains two Higgs doublet $(H, \overline{H})$ and couple to a singlet (SM like) N, which carries a non-trivial $U(1)'$ quantum charge. The particle contents and their quantum charges are specified in Table 1. In addition, this model satisfies the required phenomenological constraints [6-9]: (i) There is non-anomalous $U(1)'$ group such that the masses come from symmetry breaking in the observable sector, (ii) The soft SUSY breaking is such that all soft scalar masses are positive and have same order of magnitude at $M_G$, where symmetry breaks in a hidden sector, (iii) Model is anomaly free i.e. $U(1)'$ quantum numbers satisfy $Y_1' + Y_2' + Y_N' = 0$ and $Y_D' + Y_{D^c}' + Y_N' = 0$ which allows the coupling $NH\overline{H}$ and $ND^cD$ in superpotential. The advantages with the model containing two Higgs doublet and one singlet is more realistic than that the model contain one Higgs doublet and one singlet discussed in Ref. [7] because of, (i) there is automatically anomaly cancellation in the off-diagonal mass matrices of $Z - Z'$ due to $Y_1'Y_2' > 0$, and (ii) presence of tri-linear couplings in the superpotential (forbidden by SU(2) in the model of [7]) qualitatively changes the Higgs potential and thus gives a rich pattern of symmetry breaking for which $Y_1'Y_N' > 0$ is no longer required as for the model of [7].

Therefore, the Extra U(1) superstring inspired model provides an interacting phenomenology in acceptable low energy parametric space e.g. $U(1)'$ breaking at the electroweak scale in supersymmetric extension of SM could solve the $\mu$-problem [7, 9-13], inducing an effective $\mu$- term ($NH\overline{H}$) at the electroweak scale by $U(1)'$ breaking. Due to the term $\lambda_H NH\overline{H}$ in superpotential, a numerous possibility of symmetry breaking scenario are found to have a phenomenologically acceptable spectrum leading to small mixing angle ($\approx 10^{-3}$) and $M_{Z'}$ to the order of 1 *TeV*. There is no strong motivation to think that an extra $Z'$ would actually to be light enough to be observed at future colliders. Even within ordinary GUT's, there is no robust predictions for the mass of a $Z'$. In particular, the occurrence of light $Z'$ (leptophobic) in $\eta$ model is ruled out [6] because of the effect of kinetic mixing is too small to make $Z'$ leptophobic. Also light $Z'\left(M_{Z'} < M_Z\right)$ has been ruled out in direct searches in $e^+e^-$ collider since the leptophobic couplings are too weaker to experimental sensitivities. Some strong and concrete predictions have been made by some authors [14,15] for the possible excess of $Z \to b\overline{b}$ events at CERN $e^+e^-$ collider LEP for the relative coupling of the ordinary and exotic particles to $Z'$ in the GUT framework, but the recent LEP measurements, especially from ALEPH [16], have found this possibility is more weakened.

In superstring motivated models, it is often the case that electroweak and $U(1)'$ breaking are both driven by soft supersymmetry breaking parameters so one typically expects that $Z'$ mass to be the same order as the electroweak scale, i.e. less than a TeV or so [7], so that such particles, if they exist, should be easily observed and their couplings studied at future colliders or at Fermi lab Tevatron [17]. Bounds from the direct searches at the Fermi lab Tevatron ($p\overline{p} \to Z' \to l^+l^-$) [18] and precision electroweak measurements [19] on the $Z'$ mass and mixing angle are stringent. The constraints depend on the particular $Z'$ couplings, but in typical models one requires $M_{Z'} > (500-800)$ *GeV* and the mixing angle to be smaller than a few $\times 10^{-3}$. There are actually hints of deviations from the standard model in the NuTeV experiments [20] and in atomic parity violations [21], which could possibly be early sign of $Z'$ [22]. It has shown in a number of examples [6, 8] that there are small but not overly tuned corners of parameters space which can yield acceptable $Z'$ parameters. In future, it may be possible to test $Z'$ at the LHC for masses to the order of 10 *TeV* and its couplings at the LHC or NLC should be possible to the order of few *TeV*. Thus the mass of additional gauge boson can range from electroweak scale (*O TeV*) to Planck scale [23].

With these motivations we have done the RGEs analysis to study $U(1)'$ breaking where the minima of the Higgs and Singlet fields occur at weak scale such that $M_{Z_2}$ to be the order of few TeV with small mixing angle ($\leq 10^{-3}$) and exotic particles. We have also discussed the phenomenological features of the obtained low energy spectrum from large Singlet VEV scenario and finally summarize the work done in conclusion.

**Table 1**

Transformation properties under $SU(3)_c \times SU(2)_L \times U(1)_Y \times U(1)_E$ of the chiral matter superfields contained in the 27 of $E_6$. In terms of $Y_L$, $Y_R$ associated with the maximal subgroup of $E_6$, the conventional weak hypercharge $Y$ and new hypercharge $Y'(Y_E)$ are given by

$$Y = \frac{Y_L + Y_R}{2}, \qquad Y'(Y_E) = \frac{Y_L - Y_R}{2}.$$

Properly normalized $Y$ and $Y'(Y_E)$ are given by

$$\widehat{Y} = \sqrt{\frac{3}{5}} Y, \qquad \widehat{Y}' = \sqrt{\frac{3}{5}} Y'.$$

| **Chiral Multiplets** | **Quantum Numbers** | | | | |
|---|---|---|---|---|---|
| | $SU(3)_C$ | $SU(2)_L$ | $U(1)_Y$ | $U(1)_E$ | Normalized $(Y')$ |
| $Q \cong \begin{pmatrix} u \\ d \end{pmatrix}$ | 3 | 2 | 1/6 | 1/3 | $Y'_Q = \sqrt{\frac{1}{15}}$ |
| $u^c$ | 3 | 1 | -2/3 | 1/3 | $Y'_u = \sqrt{\frac{1}{15}}$ |
| $d^c$ | 3 | 1 | 1/3 | -1/6 | $Y'_d = \frac{-1}{2\sqrt{15}}$ |
| $L \cong \begin{pmatrix} \nu \\ e \end{pmatrix}$ | 1 | 2 | -1/2 | -1/6 | $Y'_L = \frac{-1}{2\sqrt{15}}$ |
| $e^c$ | 1 | 1 | 1 | 1/3 | $Y'_e = \sqrt{\frac{1}{15}}$ |
| $\nu^c$ | 1 | 1 | 0 | 5/6 | $Y'_{\nu^c} = \frac{\sqrt{5}}{2\sqrt{3}}$ |
| $H \cong \begin{pmatrix} H^+ \\ H^0 \end{pmatrix}$ | 1 | 2 | ½ | -2/3 | $Y'_1 = \frac{-2}{\sqrt{15}}$ |
| $\overline{H} \cong \begin{pmatrix} \overline{H}^0 \\ \overline{H}^- \end{pmatrix}$ | 1 | 2 | -1/2 | -1/6 | $Y'_2 = \frac{-1}{2\sqrt{15}}$ |
| $N$ | 1 | 1 | 0 | 5/6 | $Y'_N = \frac{\sqrt{5}}{2\sqrt{3}}$ |
| $D$ | $\overline{3}$ | 1 | -1/3 | -2/3 | $Y'_D = \frac{-2}{\sqrt{15}}$ |
| $D^c$ | $\overline{3}$ | 1 | 1/3 | -1/6 | $Y'_{D^c} = \frac{-1}{2\sqrt{15}}$ |

## 2. $SU(3)_C \times SU(2)_L \times U(1)_Y \times U(1)_E$ MODEL: U(1)' BREAKING

It is generally recognized in supersymmetric theories that the Yukawa couplings of the heavy fields with the masses of $O(M_Z)$ play an essential role in triggering the gauge symmetry breakings, namely, the soft supersymmetry breaking mass squared terms of the relevant Higgs fields can be rendered negative by the radiative corrections due to the Yukawa couplings [24,25]. In particular, the top quark will give a dominant contribution to realize the phenomenologically desired gauge symmetry breakings.

In this thesis we consider a minimal case that the low energy gauge symmetry is described by a rank 5 subgroup G of $E_6$ via Wilson loop breaking mechanism [26-29]:

$$G = SU(3)_C \times SU(2)_W \times U(1)_Y \times U(1)_E \qquad (1)$$

The most general superpotential [30-35] of the model is

$$W = \lambda_t t^c QH + \lambda_b b^c Q\overline{H} + \lambda_\tau \tau^c l\overline{H} + \lambda_H N\overline{H}H + \lambda_D NDD^c \qquad (2)$$

Some residual freedom is left in the choice of non-zero Yukawa couplings in the superpotential. Therefore, we have considered only those couplings, which triggered the gauge symmetry breaking and play an important role in determining the VEVs of the scalar fields. The superpotential relevant for our analysis is

$$W = \lambda_t t^c QH + \lambda_D ND^c D + \lambda_H N\overline{H}H\ , \qquad (3)$$

where only the third generation of fermions are supposed to acquire the masses through the symmetry breaking. Some comments on the superpotential (3) are in order:

(i) The coupling $\lambda_H N\overline{H}H$ plays an important role in generating $m_{1/2} \neq 0$. Moreover, the coupling $\lambda_H$ can by itself driven $m_N^2 < 0$ as required for the gauge symmetry breaking. Such a coupling is also needed to avoid the appearances the massless Higgs fields in the particle spectrum. When $m_N^2 < 0$, one gets also $< N > \neq 0$, and in this sense this serves an effective $\mu \overline{H} H$ coupling with the additional bonus that now phenomenologically desirable scale of $\mu = \lambda_H < N >$ to the order of TeV has a natural explanation.

(ii) The low energy gauge group symmetry does not prevent the mixing between some chiral fields with the same quantum numbers, e.g. $D^c$ and $d^c$. Such mixing must be negligible from a phenomenological point of view to explain the tiny flavor changing neutral current (FCNC) processes and lepton number violations.

(iii) The unconventional exotic quarks ($D$, $D^c$) have the same $U(1)'$ charge as $H, \overline{H}$ and thus coupling of the form $\lambda_D N D D^c$ is essential for giving masses to the unconventional D-quark. The coupling $\lambda_D$ has also an important role in helping to drive $m_N^2 < 0$.

(iv) We have retained only specific terms for the third generation among various possible terms. The supersymmetric coupling $\lambda_t$ and $\lambda_D$ together with the associated A-terms have important renormalization effect on the effective Higgs potential which is relevant for the low energy gauge symmetry breaking.

(v) The effects of other possible terms such as $\lambda_b b^c Q \overline{H}$ and $\lambda_\tau \tau^c l \overline{H}$ etc. is neglected safely as suggested by the actual mass spectra and the mixingof quarks and leptons. Thus in general, we neglect everywhere intergenerational mixings.

(vi) We consider only third generation of Higgs doublets, which are relevant for the gauge symmetry breaking to develop non-vanishing VEVs.

In addition to the Supersymmetric terms described by superpotential and the gauge symmetry, the effective low energy lagrangian will include soft SUSY breaking terms i.e. mass squared terms for the scalar fields, the so called associated A parameters and mass term for gauginos $\chi_a$ associated with the gauge group. Thus

$$\mathcal{L}_{soft} = \sum_i m_i^2 |\phi_i|^2 + (\lambda_t A_t t^c QH + \lambda_D A_D N\tilde{D}^c \tilde{D} + \lambda_H A_H NH\overline{H}) + \sum_a \left(\frac{M_a}{2}\right)(\chi_a \chi_a + \ldots) + h.c. \tag{4}$$

The effective scalar potential for the Higgs fields $H, \overline{H}, N$ is derived from the low energy lagrangian including the soft SUSY breaking terms and the gauge interaction [30]. Thus the Higgs potential has contributions from F- terms, D- terms and $L_{soft}$ :

$$V_{Higgs} = V_F + V_D + V_{soft}, \tag{5a}$$

where

$$V_F = \lambda_H^2 \left(|H|^2 |N|^2 + |\overline{H}|^2 |N|^2 + |H|^2 |\overline{H}|^2\right), \tag{5b}$$

$$V_D = \frac{g_2^2}{2}\left(\left|H^+ H\right|^2\right) + \frac{\frac{3}{5} g_1^2}{2}\left(\frac{1}{2}|H|^2 - \frac{1}{2}|\overline{H}|^2\right)^2 + \frac{\frac{3}{5} g_E^2}{2}\left(Y_1' |H|^2 + Y_2' |\overline{H}|^2 + Y_N' |N|^2\right)^2 \tag{5c}$$

$$V_{soft} = m_H^2 |H|^2 + m_{\overline{H}}^2 |\overline{H}|^2 + m_N^2 |N|^2 + \left(\lambda_H A_H H\overline{H}N + h.c.\right) + \tag{5d}$$

The VEVs of the Higgs doublets and Singletare defined [1, 2, 30, 31] as

$$< H >= \begin{pmatrix} 0 \\ \upsilon \end{pmatrix}, < \overline{H} >= \begin{pmatrix} \overline{\upsilon} \\ 0 \end{pmatrix}, < N >= x. \tag{6}$$

The breaking is realized when the Higgs multiplets have non-vanishing VEVs and $\upsilon, \overline{\upsilon}$ and $x$ are real and positive i.e.

$$\upsilon \geq 0\,;\ \overline{\upsilon} \geq 0\,;\ x \geq 0\,. \qquad (7)$$

In simplifying the Higgs potential (2.5a), we assume [30, 31] that all other scalar fields different from $H, \overline{H}, N$ have vanishing VEVs. For such a case, it has been claimed [36, 37] that no other VEV arise if all soft masses $m_i^2$ (apart from $m_H^2$, $m_{\overline{H}}^2$, $m_N^2$) are positive at weak scale i.e.

$$\widetilde{m}_Q^2,\ \widetilde{m}_{t^c}^2,\ \widetilde{m}_D^2,\ \widetilde{m}_{D^C}^2,\ \widetilde{m}_{b^c}^2,\ \widetilde{m}_L^2,\ \widetilde{m}_{e^c}^2,\ \widetilde{m}_{\nu^c}^2 \geq 0\,, \qquad (8)$$

and the following conditions are satisfied [30, 31]

$$A_t^2 \leq 3\left(\widetilde{m}_Q^2 + \widetilde{m}_{t^c}^2 + m_H^2\right), \qquad (9a)$$

$$A_D^2 \leq 3\left(\widetilde{m}_D^2 + \widetilde{m}_{D^c}^2 + m_N^2\right), \qquad (9b)$$

These will ensure to avoid the charge and color breaking. Then we obtain with real and positive $\upsilon, \overline{\upsilon}, x$

$$\begin{aligned} V_{Higgs} = & m_H^2 \upsilon^2 + m_{\overline{H}}^2 \overline{\upsilon}^2 + m_N^2 x^2 + \lambda_H A_H x \upsilon(\upsilon^2 + \overline{\upsilon}^2) + \\ & \lambda_H^2 \left(\upsilon^2 x^2 + \overline{\upsilon}^2 x^2 + \upsilon^2 \overline{\upsilon}^2\right) + \tfrac{1}{8}\left(g_2^2 + \tfrac{3}{5} g_1^2\right)\left(|\upsilon|^2 - |\overline{\upsilon}|^2\right) + \\ & \tfrac{1}{120} g_E^2 \left(5x^2 - 4\upsilon^2 - \overline{\upsilon}^2\right) + \ldots\ldots\ldots\ldots \end{aligned} \qquad (10)$$

The gauge invariance of W demands $Y_1' + Y_2' + Y_N' = 0$ under U(1)' transformation. The effective $\mu$- parameter generated by the VEV of Singlet $N$, i.e., $< N >= x$ provides $\mu = \lambda_H . x$ .

Making use of minimization condition for the scalar potential, when all VEVs are non-zero, we get

$$-m_H^2 = m_3^2 \tan\beta + \lambda_H^2 (x^2 + \overline{\upsilon}^2) + \frac{G^2}{4} \upsilon_{\max.}^2 Cos2\beta + g_E^2 Y_1' \left[Y_1' \upsilon^2 + Y_2' \overline{\upsilon}^2 + Y_N' x^2\right], \quad (11a)$$

$$-m_{\overline{H}}^2 = m_3^2 Cot\beta + \lambda_H^2 (x^2 + \upsilon^2) + \frac{G^2}{4} \upsilon_{\max.}^2 Cos2\beta + g_E^2 Y_2' \left[Y_1' \upsilon^2 + Y_2' \overline{\upsilon}^2 + Y_N' x^2\right], \quad (11b)$$

$$-m_N^2 = m_3^2 \frac{\upsilon_{\max.}^2}{x^2} \sin\beta \cos\beta + \lambda_H^2 \upsilon_{\max.}^2 + g_E^2 Y_N' \left[Y_1' \upsilon^2 + Y_2' \overline{\upsilon}^2 + Y_N' x^2\right]. \qquad (11c)$$

where

$$\tan\beta = \frac{\overline{\upsilon}}{\upsilon}, \qquad \upsilon^2_{\max.} = \left(\upsilon^2 + \overline{\upsilon}^2\right);$$

$$m_3^2 = \lambda_H A_H x = A_H \mu, \; G = g_2^2 + \tfrac{3}{5} g_1^2 .$$

The spectrum of physical Higgses after symmetry breaking consists of three neutral scalars, one pseudoscalar and a pair of charged Higgses. The mass matrices for the charged and neutral Higgs fields [30-33] are given by

$$\begin{array}{c} \\ H^{+*} \\ \overline{H}^{-} \end{array} \begin{array}{c} \begin{array}{cc} \qquad H^{+} \qquad & \qquad \overline{H}^{-*} \end{array} \\ \left[\begin{array}{cc} -\lambda_H A_H \dfrac{x\overline{\upsilon}}{\upsilon} + \left(g_2^2/2 - \lambda_H^2\right)\overline{\upsilon}^2 & -\lambda_H A_H x + \left(g_2^2/2 - \lambda_H^2\right)\overline{\upsilon}\upsilon \\ SYMMETRIC & -\lambda_H A_H \dfrac{x\overline{\upsilon}}{\upsilon} + \left(g_2^2/2 - \lambda_H^2\right)\overline{\upsilon}^2 \end{array}\right] \end{array}, \quad (12)$$

and

$$\begin{array}{c|cccccc} & H_R^0 & H_I^0 & \overline{H}_R^0 & \overline{H}_I^0 & N_R & N_I \\ \hline H_R^0 & \left(g_2^2 + \frac{3}{5}g_1^2 + \frac{16}{15}g_E^2\right)\upsilon^2 - 2\lambda_H A_H \overline{\upsilon}x/\upsilon & 0 & \left(4\lambda_H^2 - g_2^2 - \frac{3}{5}g_1^2 + \frac{4}{15}g_E^2\right)\upsilon\overline{\upsilon} + 2\lambda_H A_H x & 0 & \left(4\lambda_H^2 - \frac{4}{3}g_E^2\right)\upsilon x + 2\lambda_H A_H \overline{\upsilon} & 0 \\ H_I^0 & & -2\lambda_H A_\lambda \frac{\overline{\upsilon}x}{\upsilon} & 0 & -2\lambda_H A_\lambda x & 0 & -2\lambda_H A_H \overline{\upsilon} \\ \overline{H}_R^0 & & & \left(g_2^2 + \frac{3}{2}g_1^2 + \frac{g_E^2}{15}\right)\overline{\upsilon}^2 - 2\lambda_H A_H \frac{\upsilon x}{\overline{\upsilon}} & 0 & \left(4\lambda_H^2 - \frac{g_E^2}{3}\right)\overline{\upsilon}x + 2\lambda_H A_H \upsilon & 0 \\ \overline{H}_I^0 & SYMMETRIC & & & -2\lambda_H A_H \frac{\upsilon x}{\overline{\upsilon}} & 0 & -2\lambda_H A_H \upsilon \\ N_R & & & & & \frac{5}{3}g_E^2 x^2 - 2\lambda_H A_H \frac{\upsilon\overline{\upsilon}}{x} & 0 \\ N_I & & & & & & -2\lambda_H A_H \frac{\upsilon\overline{\upsilon}}{x} \end{array} \quad (13)$$

respectively. For convenience $H^0, \overline{H}^0, N$ are decomposed into their real and imaginary parts as following:

$$M_R^2 = \begin{array}{c} \\ H_R^0 \\ \\ \overline{H}_R^0 \\ \\ N_R^0 \end{array} \begin{pmatrix} \left(g_2^2+\frac{3}{5}g_1^2+\frac{16}{15}g_E^2\right)\upsilon^2 - 2\lambda_H A_H\,\overline{\upsilon}x/\upsilon & \left(4\lambda_H^2-g_2^2-\frac{3}{5}g_1^2+\frac{4}{15}g_E^2\right)\upsilon\overline{\upsilon} + 2\lambda_H A_H x & \left(4\lambda_H^2-\frac{4}{3}g_E^2\right)\upsilon x + 2\lambda_H A_H\overline{\upsilon} \\ & \left(g_2^2+\frac{3}{2}g_1^2+\frac{1}{15}g_E^2\right)\overline{\upsilon}^2 - 2\lambda_H A_H\,\upsilon x/\overline{\upsilon} & \left(4\lambda_H^2-\frac{1}{3}g_E^2\right)x\overline{\upsilon} + 2\lambda_H A_H\upsilon \\ SYMMETRIC & & \frac{5}{3}g_E^2x^2 - 2\lambda_H A_H\frac{\upsilon\overline{\upsilon}}{x} \end{pmatrix}, \quad \text{columns: } H_R^0,\ \overline{H}_R^0,\ N_R^0 \tag{14}$$

$$M_I^2 = \begin{array}{c} H_I^0 \\ \overline{H}_I^0 \\ N_I \end{array} \begin{pmatrix} -2\lambda_H A_H\frac{\overline{\upsilon}x}{\upsilon} & -2\lambda_H A_H x & -2\lambda_H A_H\upsilon \\ & -2\lambda_H A_H\frac{\upsilon x}{\overline{\upsilon}} & -2\lambda_H A_H\overline{\upsilon} \\ SYMMETRIC & & -2\lambda_H A_H\frac{\overline{\upsilon}\upsilon}{x} \end{pmatrix}. \quad \text{columns: } H_I^0,\ \overline{H}_I^0,\ N_I \tag{15}$$

Discarding the unphysical Higgses (Goldstone bosons), we are left with a charged scalar fields of mass

$$m_{H^\pm}^2 = M_W^2 - \lambda_H A_H x\left(\frac{\overline{\upsilon}}{\upsilon}+\frac{\upsilon}{\overline{\upsilon}}\right) - \lambda_H^2\left(\upsilon^2+\overline{\upsilon}^2\right). \tag{16}$$

The masses of four neutral scalar fields, whose masses can be calculated by diagonalizing the matrix (13). For a viable phenomenology at weak scale, it is required that $x >> \upsilon, \overline{\upsilon}$ and $\frac{\upsilon}{x}, \frac{\overline{\upsilon}}{x} \rightarrow 0$ and $\frac{x}{\upsilon}, \frac{x}{\overline{\upsilon}}$ become very large and yields a heavy $M_{Z'}$. Under these conditions, one finds the eigenvalues of (13) as

$$m_1^2 = -2\lambda_H A_H\frac{\upsilon}{\overline{\upsilon}}x, \tag{17a}$$

$$m_2^2 = -2\lambda_H A_H\frac{\overline{\upsilon}}{\upsilon}x, \tag{17b}$$

$$m_3^2 = \frac{5}{3}g_E^2x^2. \tag{17c}$$

In order to ascertain the mass of the global gauge supersymmetry breaking parameter $m_{1/2}$, we make use of the following relation given in [3, 30, 32, 33]

$$m_{1/2} = -\frac{1}{4} b_0 g^2 \left(\frac{1}{16 S_R T_R^3}\right)^{1/2} \lambda_H \left(\frac{\upsilon}{\overline{\upsilon}} x + \frac{\overline{\upsilon}}{\upsilon} x + \frac{\upsilon\overline{\upsilon}}{x}\right) \tag{18}$$

$$= 0.171 \lambda_H \left(\frac{\upsilon}{\overline{\upsilon}} x + \frac{\overline{\upsilon}}{\upsilon} x + \frac{\upsilon\overline{\upsilon}}{x}\right) \tag{19}$$

with [3, 30] $b_0 = \frac{27}{16\pi^2}$, $S_R = \frac{1}{g^2}$ and $T_R \cong O(g^2)$, where $S_R$ and $T_R$ are established dynamically [30, 32].

After the electroweak and $U(1)'$ breaking, one is left with the nonvanishing VEVs of the Higgs fields. The W- boson mass is given by [30]

$$M_W = g_2 \sqrt{\frac{\upsilon^2 + \overline{\upsilon}^2}{2}} . \tag{20}$$

The VEV's of the scalar Higgs fields, $\upsilon, \overline{\upsilon}, x$, provide the bounds on neutral current phenomenology. Letting the $Z'(Z_\eta)$ be the extra neutral gauge boson associated with U(1)', and the neutral gauge bosons mixing via the mass matrix [30, 33-35] with the analogy of [6, 8, 9, 38]:

$$M^2_{Z-Z'} = \begin{pmatrix} M_Z^2 & \Delta^2 \\ \Delta^2 & M_{Z'}^2 \end{pmatrix}; \tag{21}$$

where

$$\Delta^2 = G g_1 \left(Y_2' \overline{\upsilon}^2 - Y_1' \upsilon^2\right), \tag{22a}$$

$$M_Z^2 = \tfrac{1}{2} G^2 (\upsilon^2 + \overline{\upsilon}^2), \ \sin^2\theta_W = \frac{\frac{3}{5} g_1^2}{g_2^2 + \frac{3}{5} g_1^2}, \tag{22b}$$

$$M_{Z'}^2 = 2 g_1^2 \left(Y_1' \upsilon^2 + Y_2' \overline{\upsilon}^2 + Y_N' x^2\right). \tag{22c}$$

Thus the mass eigen states $Z_1, Z_2$ are given by

$$M^2_{Z_1,Z_2} = \tfrac{1}{2} M_Z^2 + M_{Z'}^2 \mp \sqrt{\left(M_Z^2 - M_{Z'}^2\right)^2 + 4\Delta^4} \tag{23}$$

and the $Z - Z'$ mixing angle by

$$\theta_E = \tan^{-1}\left(\frac{2\Delta^2}{M_{Z'}^2 - M_Z^2}\right). \qquad (24)$$

where $\theta_E$ is expected to be very small by LEP and SLD Z- pole data and by other constraints [39-41].

The masses of top quark and of *D*- quark (at $M_Z$) are given by

$$m_t(M_Z) = \lambda_t(M_Z).\upsilon, \qquad (25)$$

$$m_D(M_Z) = \lambda_D(M_Z).x. \qquad (26)$$

It is worth mentioning that D-quark mass is completely independent of usual quarks and leptons, being proportional not only to different Yukawa couplings but also to different VEVs. However, still we lack in making strong predictions about the D-quark mass because (i) there is still no precise observation of a correct electroweak breaking scale for which the singlet VEV, i.e. $x$ is responsible and (ii) there is no upper bound on the $Z'$ mass, which is also related to the singlet VEV. Since, if an upper bound on $Z'$ mass is found then the same on x can be obtained for the favored values of the $\tan\beta$ of the model i.e. $\tan\beta \leq 1$. This may then lead to a bound on the mass of the D- quark. The latter issue seems more feasible and thus should be explored in future colliders, namely at LHC [42] and LEP [43].

Coming to the supersymmetric fermions, the parameters, gaugino masses, play also an effective role in the charginos and neutralinos sector. The $W^{\pm}$ bosons and charged Higgs bosons $H^{\pm}(\overline{H}^{\pm})$ from the two weak doublets needed in a supersymmetric theory have supersymmetric partners $\tilde{W}^{\pm}$, $\tilde{H}^{+}$, $\tilde{\overline{H}}^{-}$. These are weak eigen states:- $\tilde{W}^{\pm}$ are in an SU(2) triplet, and $\tilde{H}^{+}$, $\tilde{\overline{H}}^{-}$ are in SU(2) doublets. A term $g_2\tilde{W}\tilde{H}^{-}H$ is allowed by $SU(2)\times U(1)$ and when H gets a vacuum expectation value $\upsilon$ and off-diagonal mass term is generated in the $\tilde{W}^{+}\tilde{H}^{-}$ mass matrix.

Remembering $\mu = \lambda_H < N >$, the masses of charginos can be obtained by diagonalizing the mass matrix [30, 31]

$$\begin{array}{cc} & \begin{array}{cc}\widetilde{W}^- & \widetilde{H}^+\end{array} \\ \widetilde{\chi}^{\pm} = \begin{array}{c}\widetilde{W}^- \\ \widetilde{\overline{H}}^-\end{array} & \begin{pmatrix} M_2 & g_2 \upsilon \\ g_2 \overline{\upsilon} & -\mu \end{pmatrix}, \end{array} \tag{27}$$

to obtain the eigenvalues [44, 45]

$$\widetilde{\chi}^{\pm\,2}_{1,2} = \frac{1}{2}\Big[ M_2^2 + \mu^2 + 2M_W^2 \pm \left\{ \left(M_2^2 - \mu^2\right)^2 + 4M_W^4 \cos^2 2\beta + 4M_W^2 \left(M_2^2 + \mu^2 + 2M_2 \mu \sin 2\beta\right) \right\}^{1/2} \Big], \tag{28}$$

where $\cos 2\beta = \frac{\left(\upsilon^2 - \overline{\upsilon}^2\right)}{\left(\upsilon^2 + \overline{\upsilon}^2\right)}$.

The spin ½ partners of gauge bosons and Higgs bosons are perhaps the most promising of the supersymmetry partners for detection and study, because they may give the cleanest experimental signatures. Generally, in the basis $\left(\widetilde{W}^3, \widetilde{B}, \widetilde{B}_E, \widetilde{H}^0, \widetilde{\overline{H}}^0, \widetilde{N}\right)$ the neutralinos (neutral gauginos - higgsinos) sector is given by a complicated $6 \times 6$ matrix,

$$\widetilde{\chi}^0 = \begin{array}{c|cccccc} & \widetilde{W}^3 & \widetilde{B} & \widetilde{B}_E & \widetilde{H}^0 & \widetilde{\overline{H}}^0 & \widetilde{N} \\ \hline \widetilde{W}^3 & M_2 & 0 & 0 & -\frac{g_2}{\sqrt{2}}\upsilon & \frac{g_2}{\sqrt{2}}\overline{\upsilon} & 0 \\ \widetilde{B} & 0 & M_1 & 0 & \frac{\sqrt{3}}{\sqrt{10}}g_1\upsilon & -\frac{\sqrt{3}}{\sqrt{10}}g_1\overline{\upsilon} & 0 \\ \widetilde{B}_E & 0 & 0 & M_E & -\frac{\sqrt{8}}{\sqrt{15}}g_E\upsilon & -\frac{\sqrt{1}}{\sqrt{30}}g_E\overline{\upsilon} & \frac{\sqrt{5}}{\sqrt{6}}g_E x \\ \widetilde{H}^0 & -\frac{g_2}{\sqrt{2}}\upsilon & \frac{\sqrt{3}}{\sqrt{10}}g_1\upsilon & -\frac{\sqrt{8}}{\sqrt{15}}g_E\upsilon & 0 & \mu_H & \lambda_H\overline{\upsilon} \\ \widetilde{\overline{H}}^0 & \frac{g_2}{\sqrt{2}}\overline{\upsilon} & -\frac{\sqrt{3}}{\sqrt{10}}g_1\overline{\upsilon} & -\frac{\sqrt{1}}{\sqrt{30}}g_E\overline{\upsilon} & \mu_H & 0 & \lambda_H\upsilon \\ \widetilde{N} & 0 & 0 & \frac{\sqrt{5}}{\sqrt{6}}g_E x & \lambda_H\overline{\upsilon} & \lambda_H\upsilon & 0 \end{array}. \tag{29a}$$

The minimal set of particles $\left(\widetilde{W}^3, \widetilde{B}, \widetilde{H}^0, \widetilde{\overline{H}}^0\right)$ arises as partners of $W^3, B, H^0, \overline{H}^0$ by SUSY breaking. The partners are all spin ½ , uncolored and electrically neutral particles, differing only in their $SU(2) \times U(1)$ quantum numbers. When $SU(2) \times U(1)$ is spontaneously broken by Higgs mechanism, these states gets off-diagonal contributions to their mass matrix. For example, a term in lagrangian $g_2\tilde{W}^0\tilde{H}^0H$ would give a $\tilde{W}^0\tilde{H}^0$ mass term $g_2\upsilon$ when $H$ gets VEV $\upsilon$. Other term arises when $\overline{H}$ gets VEV $\overline{\upsilon}$. The resulting mass matrix has the form

$$\tilde{\chi}^0 = \begin{matrix} & \tilde{W}^3 & \tilde{B} & \tilde{H}^0 & \tilde{\overline{H}}^0 \\ \begin{matrix}\tilde{W}^3\\ \tilde{B}\\ \tilde{H}^0\\ \tilde{\overline{H}}^0\end{matrix} & \multicolumn{4}{c}{\begin{pmatrix} M_2 & 0 & -\frac{1}{\sqrt{2}}g_2\upsilon & \frac{1}{\sqrt{2}}g_2\overline{\upsilon} \\ 0 & M_1 & \frac{\sqrt{3}}{\sqrt{10}}g_1\upsilon & -\frac{\sqrt{3}}{\sqrt{10}}g_1\overline{\upsilon} \\ -\frac{1}{\sqrt{2}}g_2\upsilon & \frac{\sqrt{3}}{\sqrt{10}}g_1\upsilon vv & 0 & \mu \\ \frac{1}{\sqrt{2}}g_2\overline{\upsilon} & -\frac{\sqrt{3}}{\sqrt{10}}g_1\overline{\upsilon}vv & \mu & 0 \end{pmatrix}} \end{matrix}. \qquad (29b)$$

If there is non-negligible mixing among different fields (e.g. in the case of the stop squarks or of the D-squarks with the coupling $\lambda_D$), the one can have the mass matrix of the following general form [30, 31]

$$\begin{matrix} & \Phi_L^* & \Phi_R^* \\ \begin{matrix}\Phi_L\\ \Phi_R\end{matrix} & \multicolumn{2}{c}{\begin{pmatrix} m_{LL}^2 & m_{LR}^2 \\ m_{LR}^2 & m_{RR}^2 \end{pmatrix}} \end{matrix} \qquad (30)$$

Superstring inspired models usually includes some new coloured particles, "*D-quarks*" and "*D-squarks*". The latter may have a good chance of being relatively light. These have a mass matrix [30, 31]

$$\begin{matrix} & D & \overline{D}^* \\ \begin{matrix}D^*\\ \overline{D}\end{matrix} & \multicolumn{2}{c}{\begin{pmatrix} m_{D_{LL}}^2 & m_{D_{LR}}^2 \\ m_{D_{LR}}^2 & m_{D_{RR}}^2 \end{pmatrix}} \end{matrix}, \qquad (31)$$

with

$$m_{D_{LL}}^2 = \tilde{m}_D^2 + \lambda_D^2 x^2 - \frac{g_1^2}{10}\left(\upsilon^2 - \overline{\upsilon}^2\right) - \frac{g_E^2}{15}\left(5x^2 - 4\upsilon^2 - \overline{\upsilon}^2\right), \qquad (32a)$$

$$m_{D_{RR}}^2 = \tilde{m}_{D^c}^2 + \lambda_D^2 x^2 - \frac{g_1^2}{10}\left(\upsilon^2 - \overline{\upsilon}^2\right) - \frac{g_E^2}{60}\left(5x^2 - 4\upsilon^2 - \overline{\upsilon}^2\right), \qquad (32b)$$

$$m_{D_{LR}}^2 = \lambda_D A_D x + \lambda_H A_D \upsilon\overline{\upsilon}\,. \qquad (32c)$$

The eigenvalues of the D-squarks mass matrix are

$$\tilde{m}^2_{D_1,D_2} = \frac{1}{2}\left[\left(m^2_{D_{LL}} + m^2_{D_{RR}}\right) \mp \sqrt{\left(m^2_{D_{LL}} - m^2_{D_{RR}}\right)^2 + 4m^4_{D_{LR}}}\,\right]. \qquad (33)$$

In the numerical estimation after the RGEs analysis, one often finds that $x \sim \tilde{m}_D, \tilde{m}_{D^c}$ and one of the eigenvalues can be very light, which is usually lighter than the usual D-squarks given by Eq. (26). It has been pointed [31] that the lightest D-squark would be stable if there were no further coupling to other particles. For example, if there were leptoquark coupling $\left(\overline{D}QL\right)$ and $\left(De^c_L u^c_L\right)$ (but no $QQD$ or $u^c_L d^c_L \overline{D}$ to avoid fast proton decay) this problem could be avoided.

Concerning other particles of the third generation, the s-top's may also have substantial "off-diagonal" masses, the relevant mass matrix [15] for them is

$$\begin{array}{c c} & \begin{array}{cc} \tilde{t}_L & \tilde{t}_R \end{array} \\ \begin{array}{c} \tilde{t}^*_L \\ \tilde{t}^*_R \end{array} & \begin{pmatrix} m^2_{t_{LL}} & m^2_{t_{LR}} \\ m^2_{t_{LR}} & m^2_{t_{RR}} \end{pmatrix} \end{array}, \qquad (34)$$

with

$$m^2_{t_{LL}} = \tilde{m}^2_Q + \lambda^2_t \upsilon^2 - \frac{g^2_2}{4}\left(\upsilon^2 - \overline{\upsilon}^2\right) - \frac{g^2_1}{20}\left(\upsilon^2 - \overline{\upsilon}^2\right) + \frac{g^2_E}{15}\left(5x^2 - 4\upsilon^2 - \overline{\upsilon}^2\right), \qquad (35a)$$

$$m^2_{t_{RR}} = \tilde{m}^2_{u^c} + \lambda^2_t \upsilon^2 - \frac{g^2_1}{5}\left(\upsilon^2 - \overline{\upsilon}^2\right) + \frac{g^2_E}{30}\left(5x^2 - 4\upsilon^2 - \overline{\upsilon}^2\right), \qquad (35b)$$

$$m^2_{t_{LR}} = \lambda_t A_t \upsilon + \lambda_t \lambda_H x \overline{\upsilon}\,. \qquad (35c)$$

The eigenvalues of the mass matrix (34) are

$$m^2_{\tilde{t}_1,\tilde{t}_2} = \frac{1}{2}\left[\left(m^2_{t_{LL}} + m^2_{t_{RR}}\right) \mp \sqrt{\left(m^2_{t_{LL}} - m^2_{t_{RR}}\right)^2 + 4m^4_{t_{LR}}}\,\right]. \qquad (36)$$

Contrary to the case of D-squark, in this case $m_t << x \sim M_{Z'}$ so that the s-top's are never very light.

The spectrum of the string models with an extra $U(1)$ is quite decorative. Apart from the usual squarks & sleptons and the sparticles of the third generation ($D$, $D^c$ and physical Higgses $H$, $\overline{H}, N$ + SUSY partners) discussed above so far, the first two generations do also contain extra fermions $\left(H_i^{\pm},\ \tilde{H}_i^0, \tilde{\overline{H}}_i^0, \tilde{N}_i, \tilde{D}_i, \tilde{D}^c{}_i, \tilde{\nu}_R\right)$ (i = 1,2 generations) and their supersymmetric scalar partners. It is worth to make a remark that we have consider in the RGE analysis above that only the $H, \overline{H}, N, \tilde{D}, \tilde{D}^c$ fields of the third generation may have large Yukawa couplings triggering the symmetry breaking.

The eigen values of the mass matrices of charginos, neutralinos, D-squarks and s-top quarks given by Eqs. (27), (29b), (31) and (34) are computed using "Mathematica" and results are shown only for the case of *large x scenario* in Table 4.

The explicit expressions used in computation of the scalar masses, when the mixing is not negligible, as in the case of the top-squarks and of the D-squarks, are as follows [31]:

$$m_{\tilde{d}}^2 = \tilde{m}_Q^2 + \frac{g_2^2}{4}\left(\upsilon^2 - \overline{\upsilon}^2\right) + \frac{g_1^2}{20}\left(\upsilon^2 - \overline{\upsilon}^2\right) + \frac{g_E^2}{30}\left(5x^2 - 4\upsilon^2 - \overline{\upsilon}^2\right), \quad (37)$$

$$m_{\tilde{d}^c}^2 = \tilde{m}_{d^c}^2 + \frac{g_1^2}{10}\left(\upsilon^2 - \overline{\upsilon}^2\right) - \frac{g_E^2}{60}\left(5x^2 - 4\upsilon^2 - \overline{\upsilon}^2\right), \quad (38)$$

$$m_{\tilde{e}}^2 = \tilde{m}_L^2 + \frac{g_2^2}{4}\left(\upsilon^2 - \overline{\upsilon}^2\right) - \frac{3g_1^2}{20}\left(\upsilon^2 - \overline{\upsilon}^2\right) - \frac{g_E^2}{60}\left(5x^2 - 4\upsilon^2 - \overline{\upsilon}^2\right), \quad (39)$$

$$m_{\tilde{e}^c}^2 = \tilde{m}_{e^c}^2 + \frac{3g_1^2}{10}\left(\upsilon^2 - \overline{\upsilon}^2\right) + \frac{g_E^2}{30}\left(5x^2 - 4\upsilon^2 - \overline{\upsilon}^2\right), \quad (40)$$

$$m_{\tilde{\nu}}^2 = \tilde{m}_L^2 - \frac{g_2^2}{4}\left(\upsilon^2 - \overline{\upsilon}^2\right) - \frac{3g_1^2}{20}\left(\upsilon^2 - \overline{\upsilon}^2\right) - \frac{g_E^2}{60}\left(5x^2 - 4\upsilon^2 - \overline{\upsilon}^2\right), \quad (41)$$

$$m_{\tilde{\nu}^c}^2 = \tilde{m}_{\nu^c}^2 + \frac{g_E^2}{12}\left(5x^2 - 4\upsilon^2 - \overline{\upsilon}^2\right). \quad (42)$$

The obtained numerical values are shown in Table 4. The complete set of Renormalization Group Equations (RGE's) of all matter particles pertaining to superpotential (3) used in our analysis, are given in the Appendix of this Chapter.

## 3. RGEs ANALYSIS

Before indulging in the analysis of RGEs in different scenerios, it would be easier for us to distinguish the symmetry breaking scenarios according the position of Singlet VEV i.e. $< N >= x$. Here we will discuss the *Pure Universality* and *Large Tri-linear Coupling Scenerios* under the Universal Boundary Conditions & *Large Singlet VEV Scenario* under the Non-Universal Boundary Conditions of free parameters at Grand Unification scale ( $M_G \approx 2\times10^{16}\,GeV$ ). However, in all respects the primary object is the gauge unification at $M_G$ must be restored.

The Renormalization Group Equations (RGEs) for the gauge couplings, of $SU(3)_C\times SU(2)_L\times U(1)_Y\times U(1)_E$ model, $g_a$ ($\alpha_a \cong g_a^2/4\pi$; $a$ = 3, 2, 1, E) are given by

$$\begin{aligned}\frac{d\alpha_a}{dt} &= -b_a\alpha_a^2 \\ \Rightarrow \frac{dg_a}{dt} &= -\frac{b_a}{8\pi}g_a^3\end{aligned}, \qquad (43)$$

where $$t = \frac{1}{2\pi}\ln\left(\frac{M_G}{\mu_Z}\right). \qquad (44)$$

$M_G$ corresponds to $\cong 2\times10^{16}$ $GeV$ and $\mu_Z$ corresponds to weak scale ( $M_Z$ ).

These equations can be trivially integrated and one finds the value at high scale. The analytic solution for the gauge coupling is

$$\alpha_a(t_Z) = \frac{1}{\alpha_3^{-1} + b_a t_Z}.$$

Where $t_Z$ is the value of 't' at the weak scale ($M_Z$) defined as the mass of Z- boson as

$$t_Z = \frac{1}{2\pi}\ln\left(\frac{M_G}{M_Z}\right) = 5.256.$$

From the deep inelastic scattering of electron from photon, non-abelian gauge theories are asymptotic freedom whereas the abelian theories are not asymptotically free. This is enough to suggest the possible unification of gauge couplings at some high scale, as shown by a classical work of Weinberg, Georgi and Quinn [46]. This gives the possibility where the three different gauge groups at weak scale, merge into a unifying group, when extrapolated to a high scale i.e. at GUT scale. Therefore at the unification scale (*t = 0*) one should expects the equality of the gauge couplings i.e. $g_1(0)=g_2(0)=g_3(0)=g_E(0)$.

The values of the beta function involved in the RGEs of extra U(1) model have been evaluated at the weak scale through the equations (43, 44), are

$$b_3 = 0,\ b_2 = 4.02,\ b_1 = b_E = 9.6. \qquad (45)$$

For $E_6-\eta$ model, at one-loop level the gauge couplings at $M_G$ are $g_1(0)=g_2(0)=g_3(0)=g_E(0)=1.228$. The gauge couplings at weak scale have been evaluated through the running only RGEs of gauge coupling from $M_G$ *to* $M_Z$ scale, as shown in Fig. 1, which are as

$$g_3(M_Z) = 1.228,\ g_2(M_Z)=0.652,\ g_1(M_Z) = g_E(M_Z) = 0.462. \qquad (46)$$

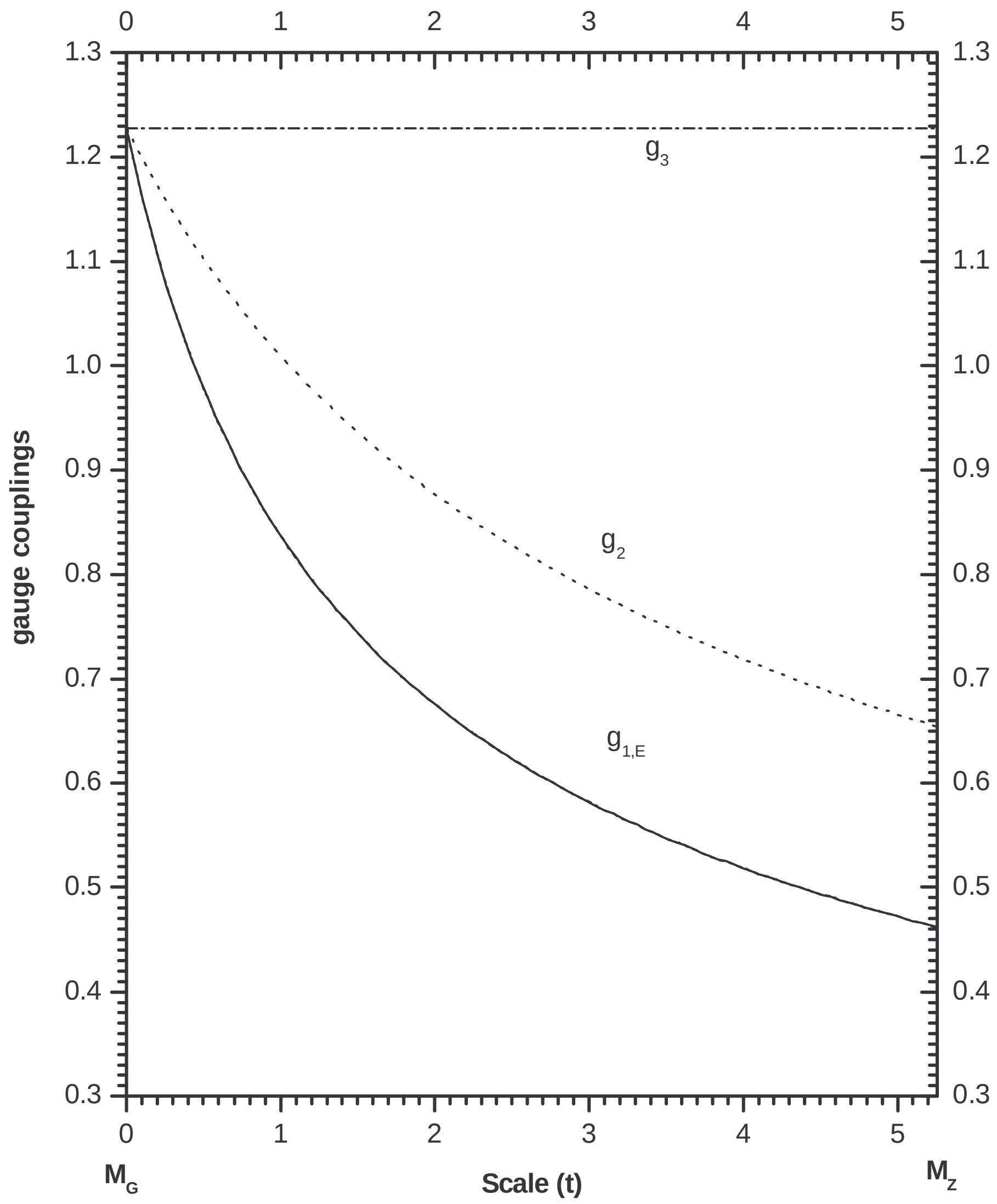

**Fig. 1:** *Unification of gauge couplings at GUT scale and evolution from GUT to weak scale.*

### 3.1. UNIVERSAL BOUNDARY CONDITIONS

For this we consider the universality of the following soft supersymmetry breaking parameters at GUT scale:

| | | |
|---|---|---|
| Gauge couplings | $g_1 = g_2 = g_3 = g_1' = 1.228 = g_0$, | (47a) |
| Soft scalar squared masses | $m_i^2 = m_0^2$, | (47b) |
| Gaugino Masses | $M_1 = M_2 = M_3 = M_1^{'} = C_{1/2} m_{1/2}$, | (47c) |
| Tri-linear couplings | $A_t = A_H = A_D = C_A m_0$, | (47d) |
| Yukawa couplings [6] | $\lambda_t = \lambda_H = \lambda_D = g_0 = 1.228$. | (47e) |

Here $C_A$ and $C_{1/2}$ are the dimensionless parameters, which have been used for twisting or fine-tuning the parameters at GUT scale and the two mass parameters $m_{1/2}$ and $m_0$ of the order of (~1 TeV) are introduced to characterize the soft supersymmetry breaking parameters for the scalars, gauginos and tri-linear couplings. For the consistent minimization of the Higgs potential, the conditions given by Eqs. (8), (9) and (10) must be hold true for the non-vanishing VEVs of the Higgs fields only at weak scale. It is noticed that for the stable vacuum, it is required that soft supersymmetric breaking parameters for the scalar fields are more important than the gaugino masses and therefore $m_{12}/m_0$ should be the order of $\leq 0.5$ [32]. The RGEs have been solved numerically for the running of all soft parameters.

***(a) PURE UNIVERSALITY ($C_A$=1.0, $C_{1/2}$=1.0):***

With this choice, the VEVs of the Higgs doublet and singlet fall in the sequence of $\overline{\upsilon} < \upsilon < x$ at weak scale. On running the RGEs of all scalar soft parameters, the evolution of Yukawa couplings, tri linear couplings (*A*-parameters) and soft scalar mass squared parameters from $M_G$ down to $M_Z$ are shown in Fig. 2, 3 and 4 respectively.

Fig. 2 shows that the Yukawa coupling $\lambda_H$ driven negative faster than $\lambda_t$, $\lambda_D$ since $\lambda_H$ receives the contributions both from $\lambda_t$ and $\lambda_D$ through $H$ and $N$ respectively. In a similar fashion, the parameter $A_H$ goes down negative rapidly than $A_t$, $A_D$ as shown in Fig. 3. Running of all soft scalar masses are shown in Fig. 4, where $m_H^2$ and $m_N^2$ are driven down to negative very fast due to the large Yukawa coupling of top quark ($\lambda_t$) and exotic quark ($\lambda_D$) while other soft mass squares remains positive. This scenario leads to an unsatisfactory, e.g., $M_{Z'}$ is found to be of the order of $M_Z$ and $M_{Z_1} < M_Z(SM)$ with large $Z - Z'$ mixing angle. The effective $\mu$- parameter is $<< M_Z$. The low energy spectrum is shown in Table 2.

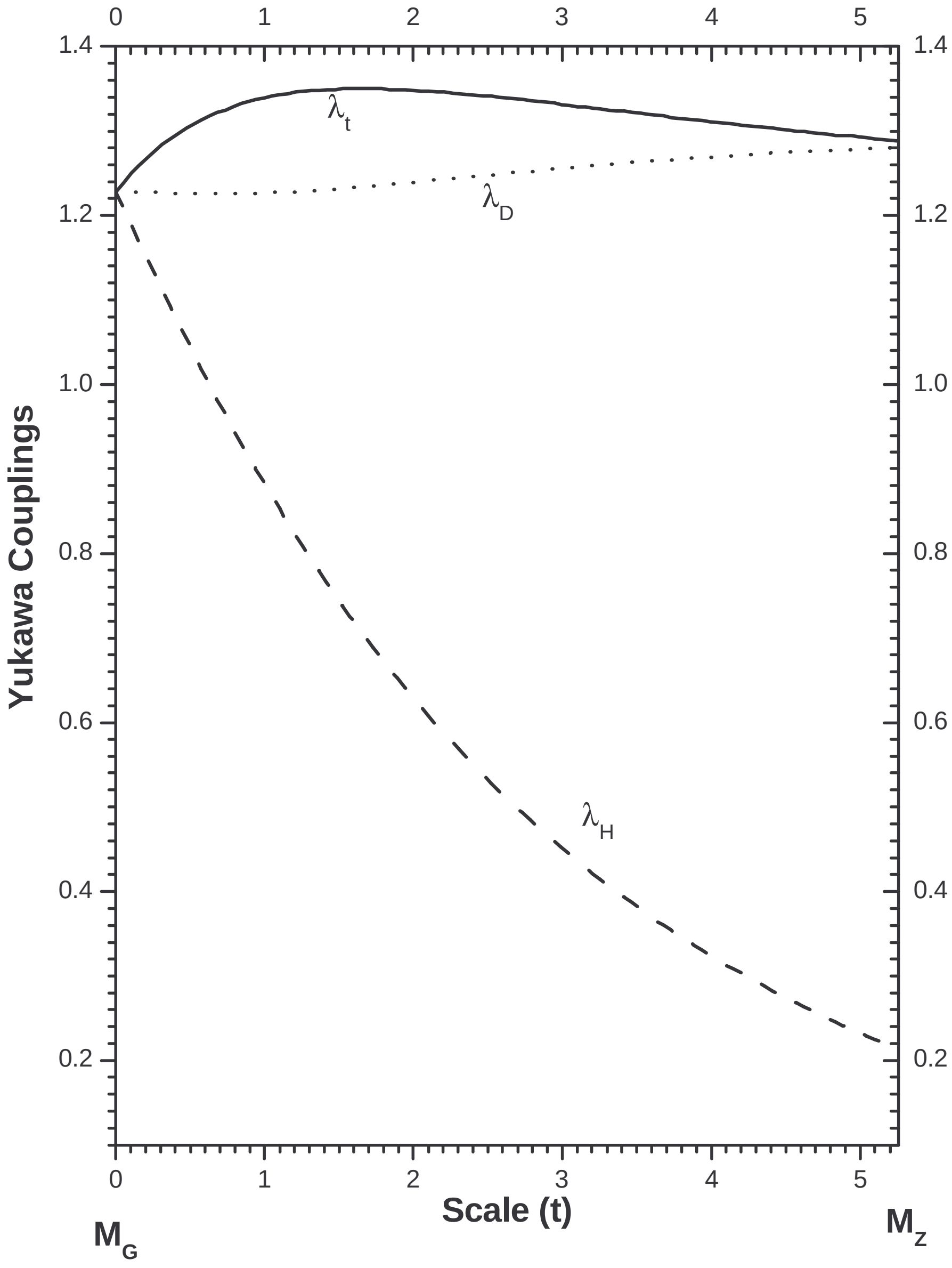

**Fig. 2:** *Evolutions of Yukawa couplings in the RGEs analysis from GUT scale to weak scale, when they are kept universal at GUT scale.*

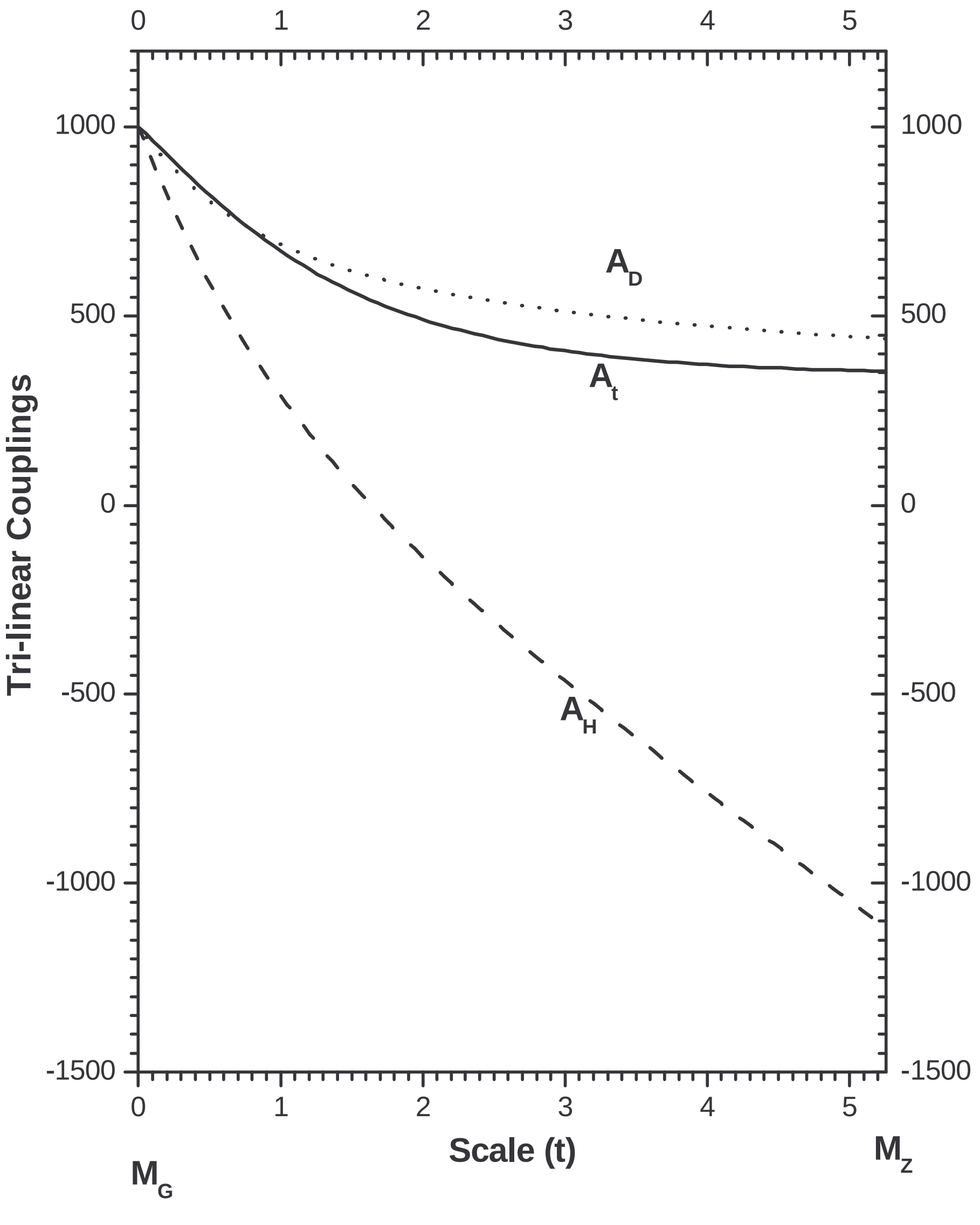

**Fig. 3:** *Behaviour of the tri-linear coupling parameters in the RGEs Analysis from GUT scale to weak scale, when they are universal at GUT scale.*

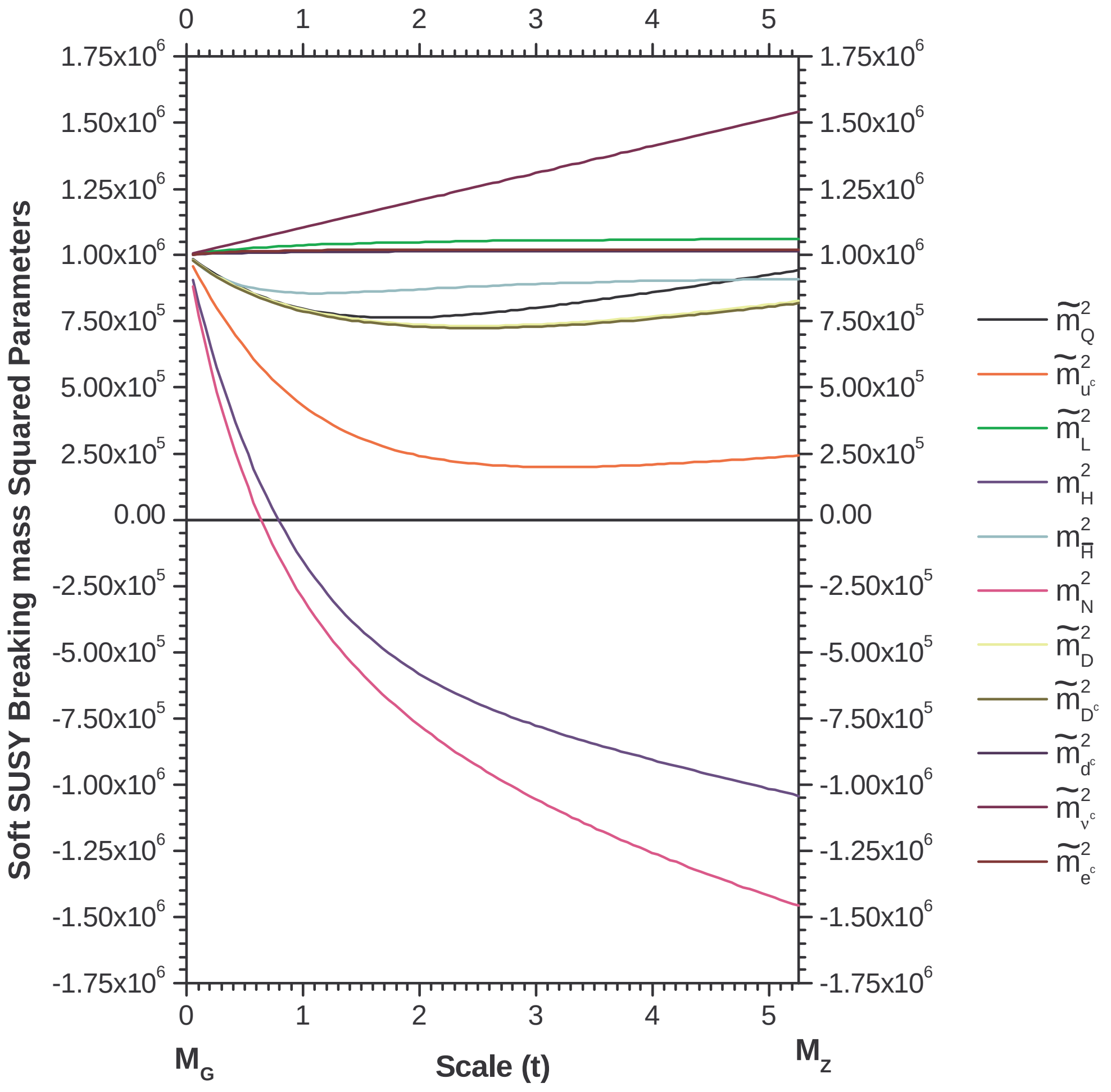

**Fig. 4:** *Running down of soft scalar masses from GUT scale to weak scale, when all the parameters are kept universal at GUT scale.*

| | |
|---|---|
| $g_1,\ g_2,\ g_3,\ g_E$ | 0.462, 0.654, 1.22, 0.462 |
| $M_1,\ M_2,\ M_E$ | 57, 114, 57 |
| $M_3\ (m_{1/2})\ ,\ m_0$ | 18, 42 |
| $\lambda_t\ ,\ \lambda_H\ ,\ \lambda_D$ | 1.228, 0.213, 1.280 |
| $A_t\ ,\ A_H\ ,\ A_D$ | 354, -1118, 440 |
| $\upsilon,\ \overline{\upsilon},\ x$ and $\tan\beta$ | 146, 94, 192 and 0.647 |
| $M_{Z'},\ M_{Z_1},\ M_{Z_2}$ | 95, 73, 110 |
| $Z - Z'$ mix. angle $(\theta_E)$ | -0.738 |
| $m_t\ (= \lambda_t.\upsilon)$ | 188 |
| $m_{\tilde{D}}\ (= \lambda_D.x)$ | 246 |
| $\mu\ (= \lambda_H.x)$ | 41 |

**Table 2:** *Low energy spectrum obtained at weak scale ($M_Z$) for the case of Pure Universality, i.e. when all free parameters are kept universal at GUT Scale.*

***(b) LARGE TRI-LINEAR COUPLING SCENARIO (*** $C_A >> C_{1/2}$***):***

Keeping in view the universality of all parameters at GUT scale as mentioned in Eqs. (47), the analysis of RGEs has been done while varying $C_A$ ( $A_{t,H,D}$) over various values of $C_{1/2}$ (Gaugino masses). In this case the symmetry breaking is governed by large $A_H$ negative values and therefore it is known as large *tri-linear coupling scenario*. Since, if $A_H$ is comparable to the soft masses (specially Higgs boson masses) then the minimization of Higgs potential leads to the unwanted global minima at $\upsilon \approx \upsilon \approx 0$. Therefore, to circumvent this problem $A_H$ is kept relatively large i.e. 5~10 times larger than the soft masses. This is done by varying $C_A$. For the non-vanishing VEVs of the Higgs fields only, the conditions (8) and (9) should be satisfied. $C_A$ should not have the large values otherwise the charge and color breaking minimum may arise at weak scale. Thus the condition (10) must also be true at weak scale, since the charge and color breaking arise due to negative squarks and exotic quarks or by the large values of $A_t$ at weak scale [47], which makes the minima unstable.

When $A_H$ is large, the associated term $\lambda_H A_H N\overline{H}H$ dominates the Higgs potential and pushes the minima to take place $\upsilon \sim \overline{\upsilon}$ (due to the second term in $V_D$) and $\upsilon \sim \overline{\upsilon} \sim x$ (due to the third term in $V_D$) since $Y_1' + Y_2' + Y_N' = 0$. But this large $A_H$ scenario yields a lighter $Z'$ mass as it is controlled by $g_1^2\left(Y_1'\upsilon^2 + Y_2'\overline{\upsilon}^2 + Y_N' x^2\right)$ with large mixing angle ($\theta_E$), proportional to $\left(Y_2'\overline{\upsilon}^2 - Y_1'\upsilon^2\right)$. The effective $\mu$- parameter is found to be very small as compared with $M_Z$. Since $C_A$ and $C_{1/2}$ ($A's$ and $M_i$) controls the difference between the soft mass squared parameters at low energy and thus minimization of the Higgs potential leads the VEVs which decide the bounds on the experimental observables like $M_{Z_1}$, $M_{Z'}$, $\theta_E$, $m_t$ and $\tan\beta$ etc. Some of the observed various sets of $C_A$ and $C_{1/2}$ are listed in Tables 3(a) to 3(d), where the positions of VEVs, $\upsilon \sim \overline{\upsilon}$, $\overline{\upsilon} \sim x$, $\upsilon \sim x$ and $\upsilon \sim \overline{\upsilon} \sim x$ occur at weak scale are shown in the large tri-linear coupling scenario, which may require somewhat more fine tuning between the parameters $C_A$ and $C_{1/2}$ at GUT scale.

| Parameters at Weak Scale | At GUT Scale | | |
|---|---|---|---|
| | $C_A = 4.76$<br>$C_{1/2} = 1.8$ | $C_A = 5.0$<br>$C_{1/2} = 2.1$ | $C_A = 6.08$<br>$C_{1/2} = 1.82$ |
| $g_1 = g_E,\ g_2,\ g_3$ | 0.462, 0.654, 1.228 | 0.462, 0.654, 1.228 | 0.462, 0.654, 1.228. |
| $M_1,\ M_2,\ M_E$ | 57, 114, 57 | 119, 238, 119 | 103, 207, 103 |
| $\lambda_t,\ \lambda_H,\ \lambda_D$ | 1.288, 0.213, 1.280 | 1.288, 0.213, 1.280 | 1.294, 0.190, 1.297 |
| $A_t,\ A_H,\ A_D$ | 422, -1792, 575 | 433, -1811, 584 | 383, -2167, 479 |
| $\upsilon,\ \overline{\upsilon},\ x$ | 123, 123, 144 | 124, 122, 147 | 122, 124, 137 |
| $x/\upsilon,\ \tan\beta$ | 1.178, 1.0 | 1.185, 0.988 | 1.129, 1.02 |
| $m_0,\quad m_{1/2}\ (M_3)$ | 30, 14 | 30, 14 | 30, 13 |
| $M_{Z'},\ M_{Z_1},\ M_{Z_2},$ | 74, 67, 97 | 75, 78, 98 | 72, 65, 97 |
| $\theta_E$ | 0.475 | 0.494 | 0.439 |
| $\mu\ (\cong \lambda_H.x)$ | 31 | 31 | 26 |
| $m_t\ (\cong \lambda_t.\upsilon)$ | 158 | 159 | 158 |
| $m_{\tilde{D}}\ (\cong \lambda_D.x)$ | 185 | 188 | 178 |
| $m_{H^\pm}$ | 340 | 344 | 344 |

**Table 3(a):** *Results are shown here to realize the VEV's position* $\upsilon \cong \overline{\upsilon}$ *occurs in large tri-linear coupling scenario.*

| Parameters at Weak Scale | At GUT Scale | | |
|---|---|---|---|
| | $C_A = 4.92$ $C_{1/2} = 1.04$ | $C_A = 5.30$ $C_{1/2} = 1.4$ | $C_A = 5.90$ $C_{1/2} = 1.9$ |
| $g_1 = g_E,\ g_2,\ g_3$ | 0.462, 0.654, 1.228 | 0.462, 0.654, 1.228 | 0.462, 0.654, 1.228 |
| $M_1,\ M_2,\ M_E$ | 59, 118, 59 | 79, 159, 79 | 108, 216, 108 |
| $\lambda_t,\ \lambda_H,\ \lambda_D$ | 1.288, 0.213, 1.280 | 1.288, 0.213, 1.280 | 1.288, 0.213, 1.280 |
| $A_t,\ A_H,\ A_D$ | 405, -1989, 579 | 419, -1939, 593 | 440, -2010, 616 |
| $\upsilon,\ \overline{\upsilon},\ x$ | 115, 130, 130 | 115, 131, 130 | 115, 131, 131 |
| $x/\upsilon,\ \tan\beta$ | 1.13,1.134 | 1.134, 1.14 | 1.138, 1.137 |
| $m_0,\quad m_{1/2}\,(M_3)$ | 31, 14 | 30, 14 | 29, 14 |
| $M_{Z'},\ M_{Z_1},\ M_{Z_2},$ | 68, 64, 95 | 68, 64, 95 | 68, 64, 95 |
| $\theta_E$ | 0.353 , | 0.352, | 0.353, |
| $\mu\ (\cong \lambda_H .x)$ | 28 | 28 | 28 |
| $m_t\ (\cong \lambda_t .\upsilon)$ | 148 | 148 | 148 |
| $m_{\tilde{D}}\ (\cong \lambda_D .x)$ | 167 | 167 | 167 |
| $m_{H^\pm}$ | 334 | 337 | 344 |

**Table 3(b):** *Results are shown for the illustration of VEV's position* $\overline{\upsilon} \cong x$ *occurs in large tri-linear coupling scenario.*

| Parameters at Weak Scale | At GUT Scale | | |
|---|---|---|---|
| | $C_A = 7.0$<br>$C_{1/2} = 1.0$ | $C_A = 7.5$<br>$C_{1/2} = 1.0$ | $C_A = 7.9$<br>$C_{1/2} = 1.0$ |
| $g_1 = g_E,\ g_2,\ g_3$ | 0.462, 0.654, 1.228 | 0.462, 0.654, 1.228 | 0.462, 0.654, 1.228 |
| $M_1,\ M_2,\ M_E$ | 57, 114, 57 | 57, 114, 57 | 57, 114, 57 |
| $\lambda_t,\ \lambda_H,\ \lambda_D$ | 1.288, 0.213, 1.280 | 1.288, 0.213, 1.280 | 1.288, 0.213, 1.280 |
| $A_t,\ A_H,\ A_D$ | 431, -2317, 652 | 437, -2417, 670 | 442, -2497, 684 |
| $\upsilon,\ \overline{\upsilon},\ x$ | 100, 142, 106 | 97, 144, 101 | 94, 146, 97 |
| $x/\upsilon,\ \tan\beta$ | 1.056, 1.423 | 1.040, 1.492 | 1.029, 1.547 |
| $m_0,\quad m_{1/2}\ (M_3)$ | 31, 13 | 31, 13 | 32, 13 |
| $M_{Z'},\ M_{Z_1},\ M_{Z_2}$ | 58, 56, 93, | 55, 54, 92, | 53, 53, 92 |
| $\theta_E$ | 0.171, | 0.134, | 0.121 |
| $\mu\ (\cong \lambda_H .x)$ | 23 | 21 | 21 |
| $m_t\ (\cong \lambda_t .\upsilon)$ | 130 | 125 | 122 |
| $m_{\tilde{D}}\ (\cong \lambda_D .x)$ | 136 | 129 | 124 |
| $m_{H^{\pm}}$ | 341 | 343 | 344 |

**Table 3(c):** *Results are shown for the illustration of VEV's position $\upsilon \cong x$ occurs in large tri-linear coupling scenario.*

| Parameters at Weak Scale | At GUT Scale | | |
|---|---|---|---|
| | $C_A = 4.9$ $C_{1/2} = 1.5$ | $C_A = 5.3$ $C_{1/2} = 1.7$ | $C_A = 5.7$ $C_{1/2} = 2.0$ |
| $g_1 = g_E,\ g_2,\ g_3$ | 0.462, 0.654, 1.228 | 0.462, 0.654, 1.228 | 0.462, 0.654, 1.228 |
| $M_1,\ M_2,\ M_E$ | 87, 170, 85 | 96, 193, 96 | 113, 227, 113 |
| $\lambda_t,\ \lambda_H,\ \lambda_D$ | 1.288, 0.213, 1.280 | 1.288, 0.213, 1.280 | 1.288, 0.213, 1.280 |
| $A_t,\ A_H,\ A_D$ | 417, -1849, 579 | 427, -1910, 594 | 440, -1961, 609 |
| $\upsilon,\ \overline{\upsilon},\ x$ | 119, 127, 137 | 117, 128, 135 | 117, 128, 135 |
| $x/\upsilon,\ \tan\beta$ | 1.155, 1.069 | 1.150, 1.091 | 1.151, 1.094 |
| $m_0,\quad m_{1/2}\,(M_3)$ | 30, 14 | 30, 14 | 29, 14 |
| $M_{Z'},\ M_{Z_1},\ M_{Z_2}$ | 71, 66, 96 | 70, 65, 96 | 70, 65, 95 |
| $\theta_E$ | 0.412 | 0.393 | 0.394 |
| $\mu\ (\cong \lambda_H .x)$ | 29 | 29 | 29 |
| $m_t\ (\cong \lambda_t .\upsilon)$ | 153 | 151 | 151 |
| $m_{\tilde{D}}\ (\cong \lambda_D .x)$ | 176 | 173 | 173 |
| $m_{H^\pm}$ | 337 | 340 | 344 |

**Table 3(d):** *Results are shown for the illustration of VEV's position* $\upsilon \sim \overline{\upsilon} \sim x$ *occurs in large tri-linear coupling scenario.*

In Fig. 5, $M_{Z'}$ is plotted against $C_{1/2}$ for a various range of $C_A$. Fig. 6 is a plot of $M_{Z_1}$ vs. $C_{1/2}$. From both figures, it is observed that $M_{Z'} << M_Z$ and $M_{Z_1} << M_Z$ as $C_A$ becomes larger which results into the large $Z-Z'$ mixing angle for the large value of $C_A$ as shown in Fig. 7. We have also plotted the effect of $C_A$ on the top quark mass $(m_t)$ as shown in Fig. 8 and found that for the large values of $C_A$ it reduces very sharply. Since the top quark mass is decided by the Eq. (25), where $\lambda_t$ is sufficiently large but the VEV of *H* becomes very small for large values of $C_A$ to provide the top quark mass, i.e., 178 ± 3 *GeV* [48-52]. Fig. 9 reflects the behavior of the effective $\mu$- parameter vs. $C_{1/2}$ for various values of $C_A$ and results into unacceptable scale of $U(1)'$ breaking, theoretically it should at least be the order of few *TeV*. The ratio of the VEVs of the Higgs doublets is plotted for different values of $C_A$ for a wide range of $C_{1/2}$, shown in Fig. 10. It reaches up to 1 or greater than that, which is not favorable for the desired symmetry breaking as such superstring inspired models are favored by low $\tan\beta \le 1.0$ [30-35].

As a result, $M_{Z'} << M_Z$ is observed with large $Z-Z'$ mixing angle and thus excluded for the considered model ($\eta$ model). Therefore, to have heavy $Z'$ to the order of few *TeV* with allowed small $Z-Z'$ mixing angle, the singlet VEV should be very large enough and which provides the effective $\mu$- parameter to the order of *TeV*. In the analysis with the universal boundary conditions at GUT scale, we observe that the large $x$ scenario cannot be obtained and also this scenario do not lead phenomenological acceptable low energy spectrum at weak scale satisfying the experimental observables. This outcome agrees completely with the analysis done in Refs. [6, 19, 38].

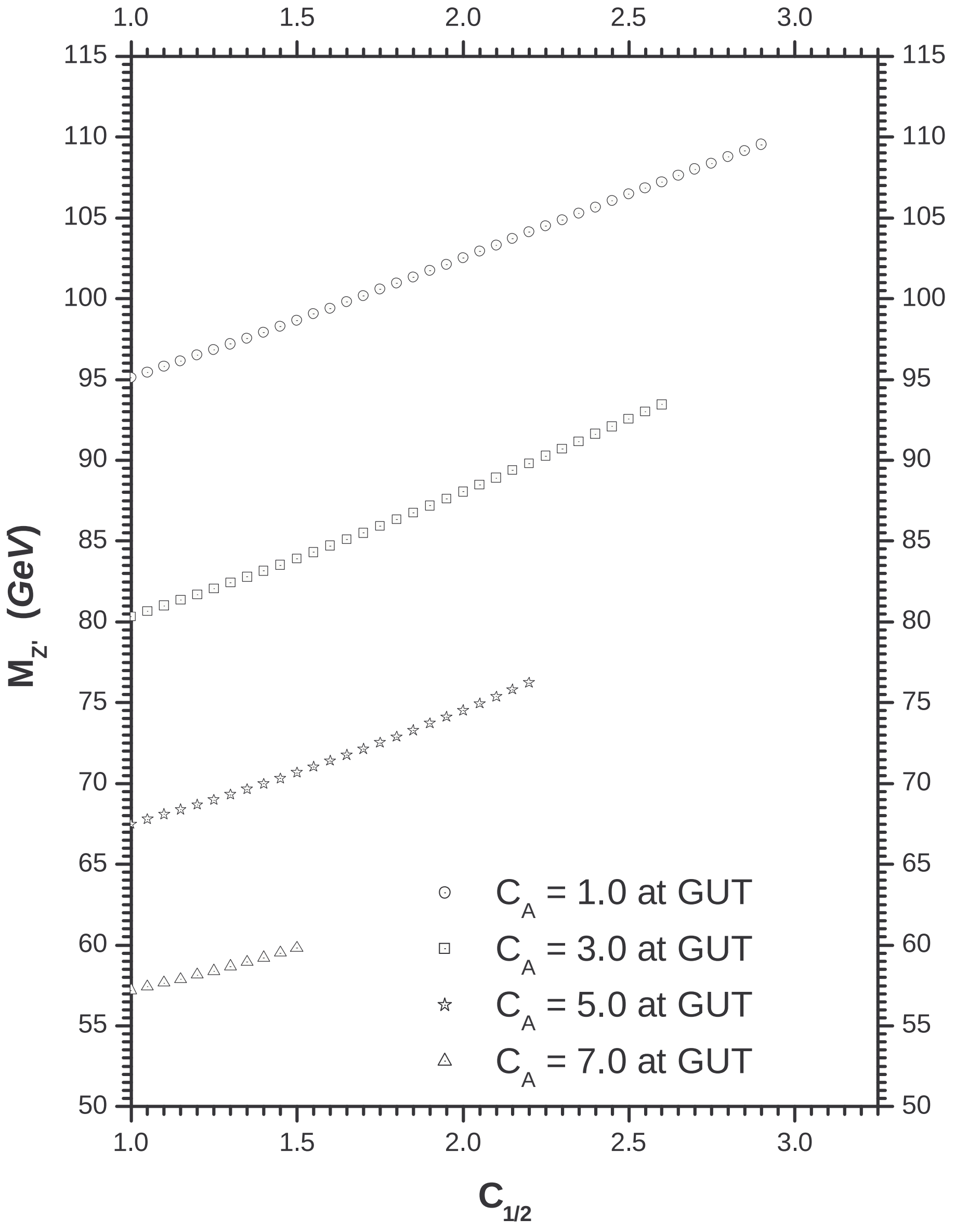

**Fig. 5:** *Possibilities of $Z'$ in large tri-linear coupling scenario by the variations of $Z'$ as a function of $C_{1/2}$ for the various values of $C_A$.*

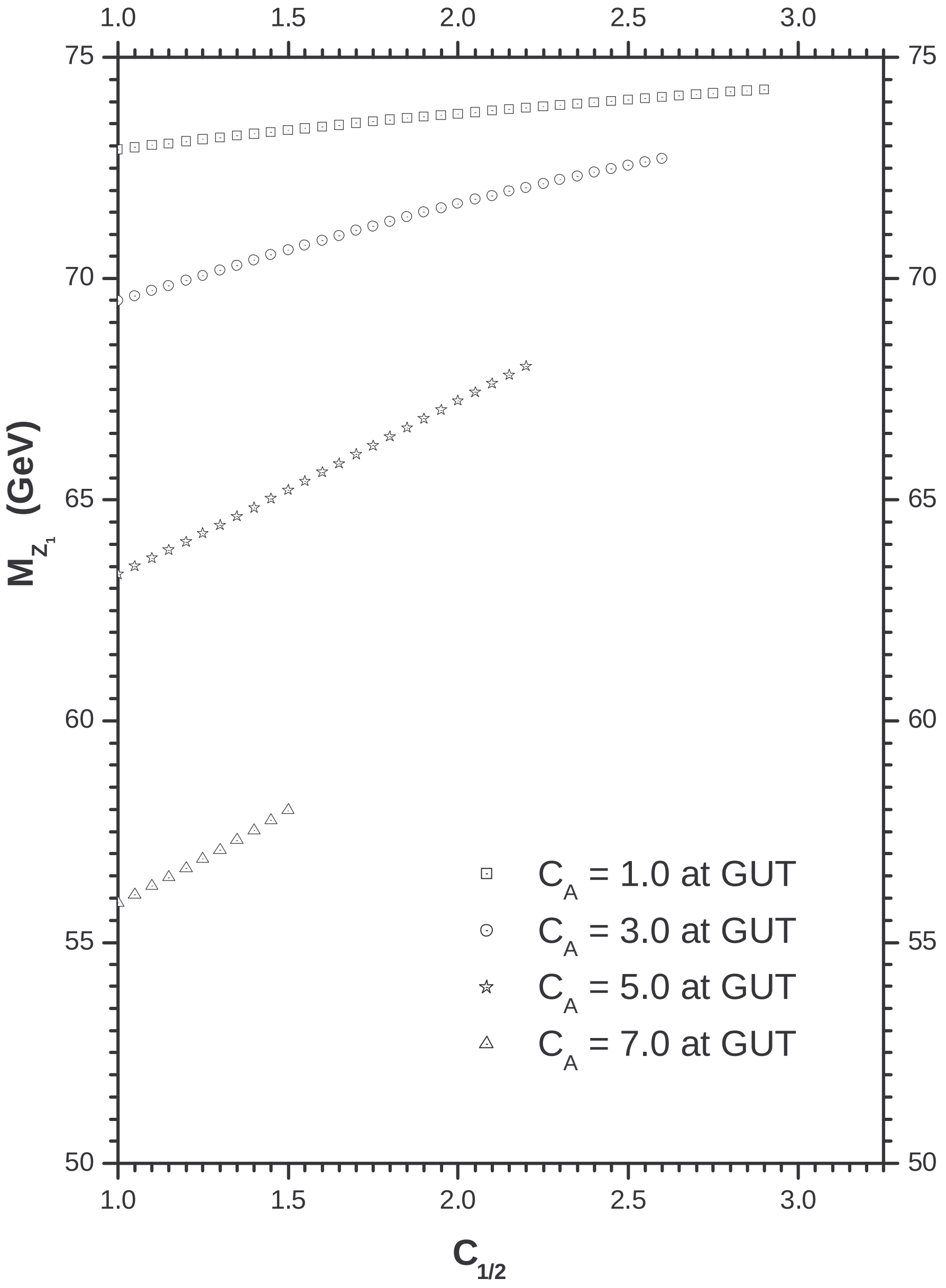

**Fig. 6:** *Effects on $M_{Z_1}$ for the large values of $C_A$ over a wide range of $C_{1/2}$.*

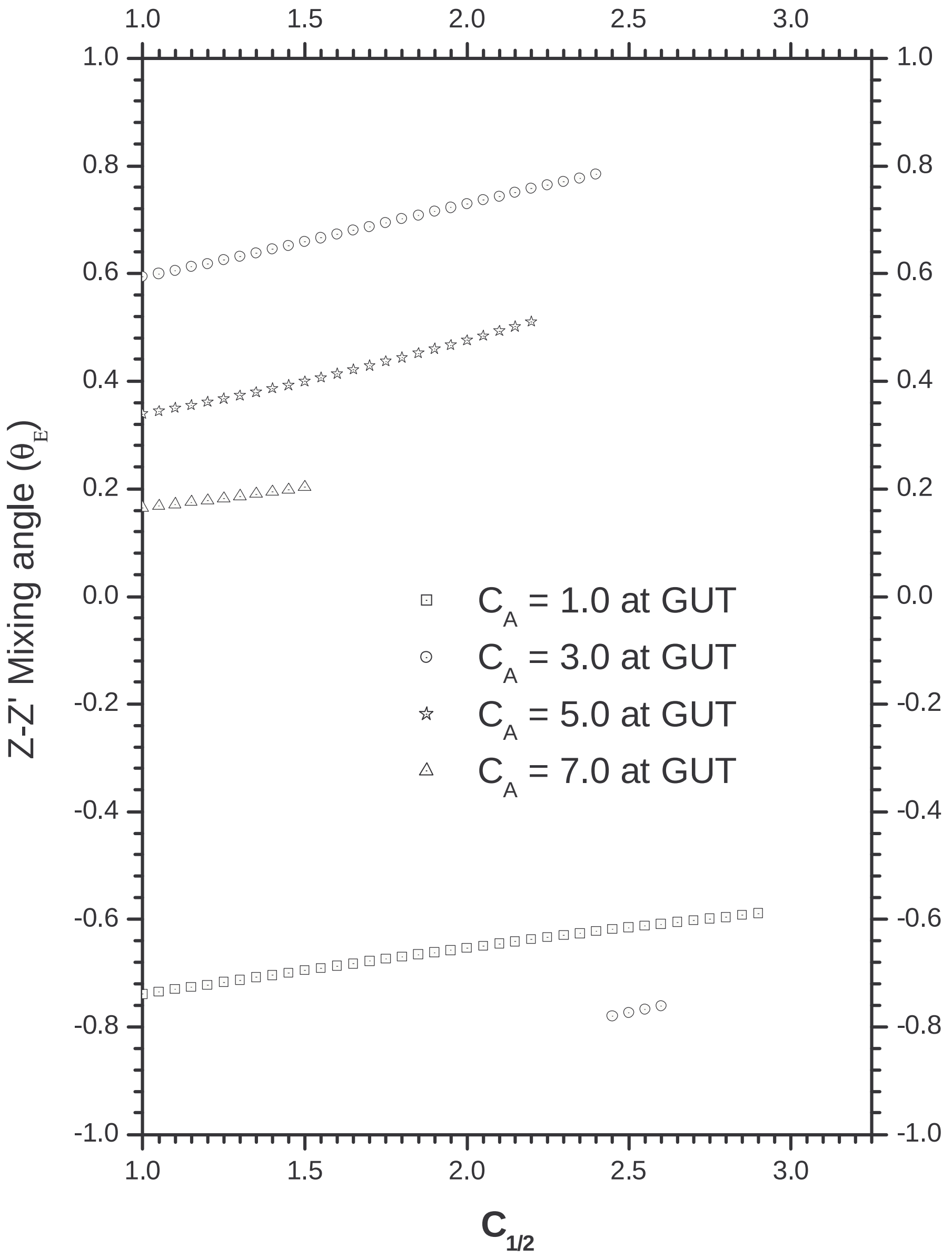

**Fig. 7:** *The* $Z - Z'$ *mixing angle as a function of* $C_{1/2}$ *for the various values of* $C_A$ *in the large tri-linear coupling scenario.*

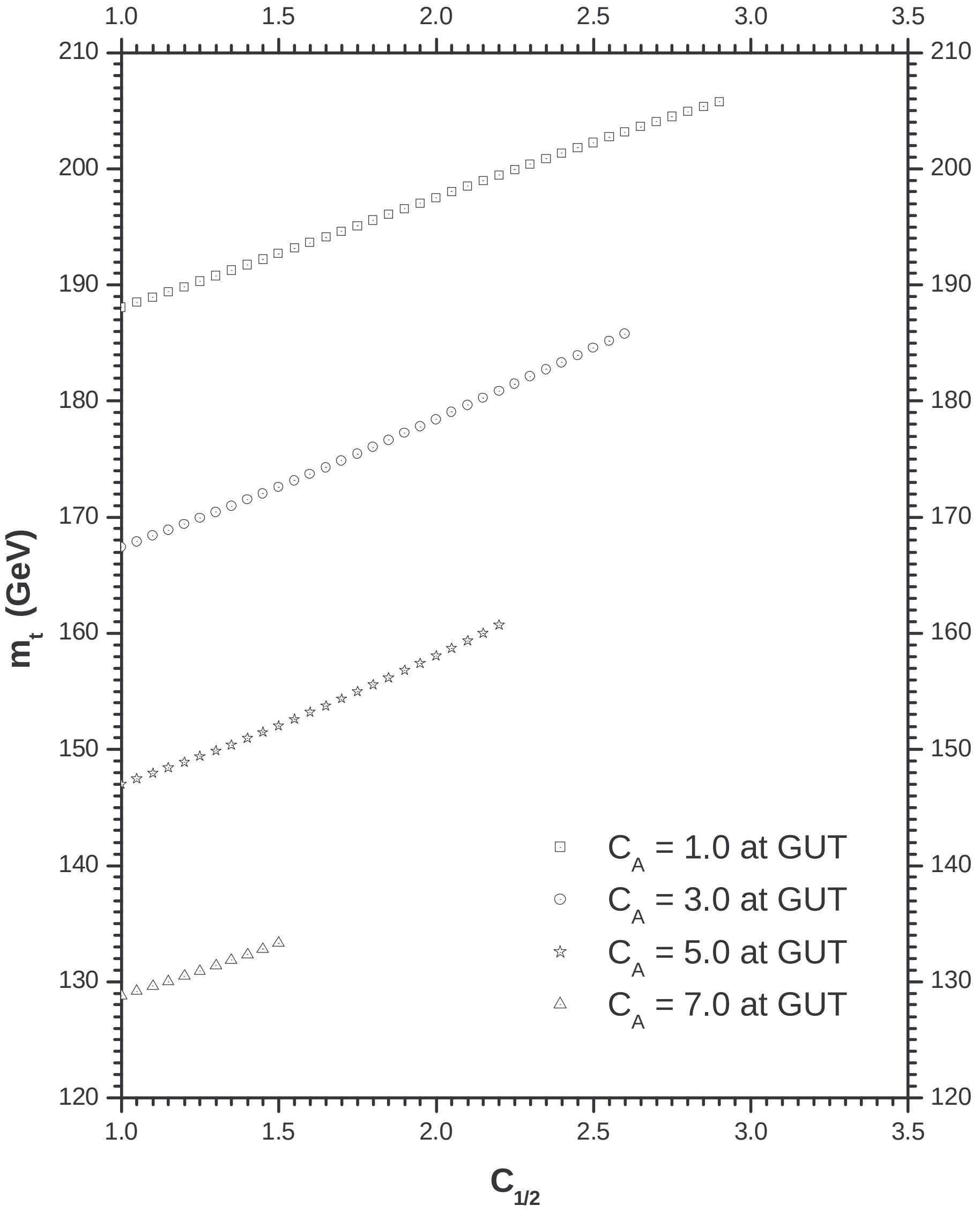

**Fig. 8:** *Theoretical predictions of top quark through the RGE analysis in the large tri-linear coupling scenario.*

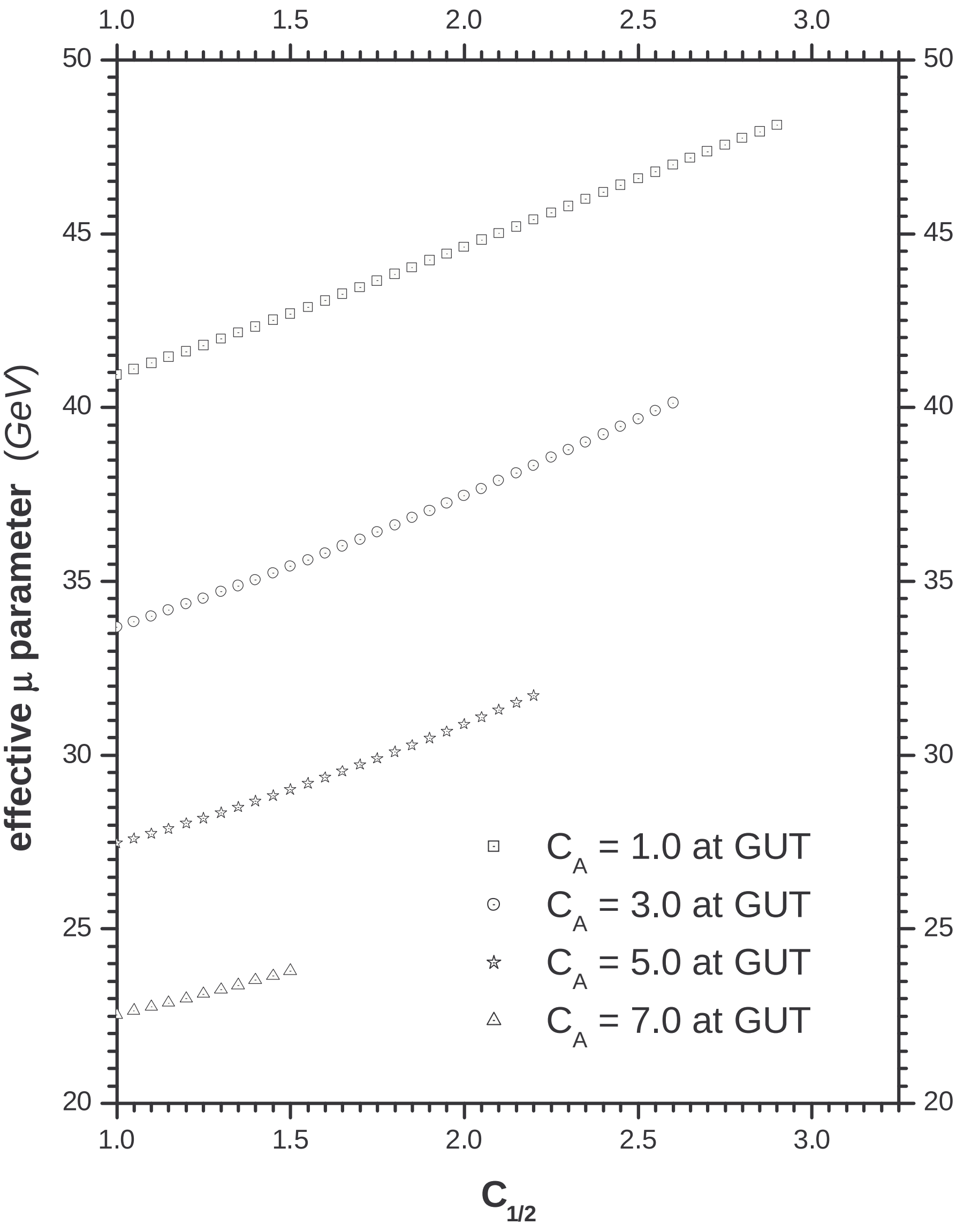

**Fig. 9:** *The tendency of effective $\mu-$ parameter in the large tri-linear coupling scenario.*

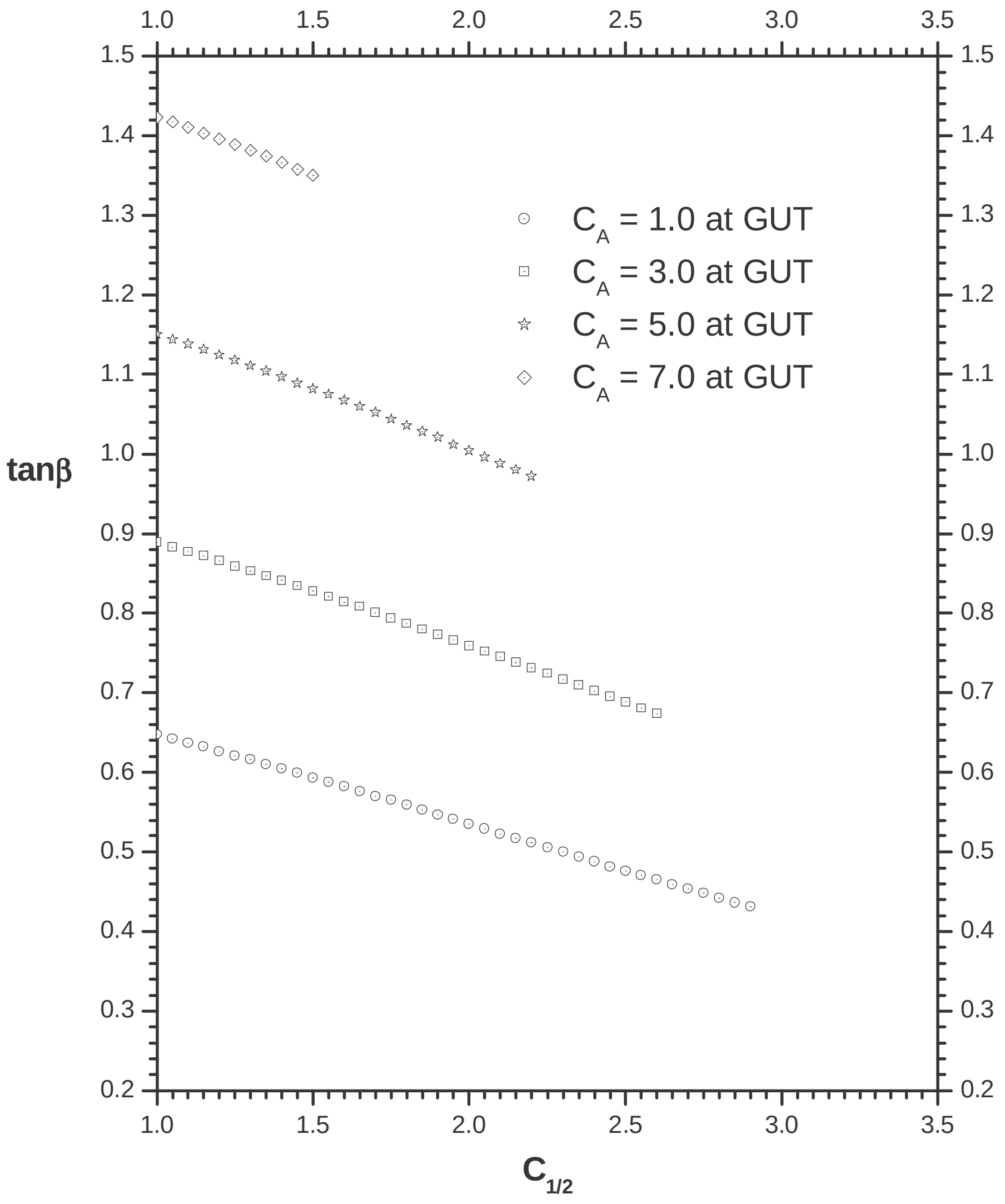

**Fig. 10:** *For the large values of* $C_A$*, the ratio of Higgs doublet VEVs, i.e.* $\tan\beta$ *increases.*

## 3.2. NON-UNIVERSAL BOUNDARY CONDITIONS: LARGE *x* SCENARIO

Relaxing the universal boundary conditions, we now wish to take up the non universal boundary conditions of the free parameters at GUT scale. Out of the five major parameters ($g_i's$: gauge couplings, $M_i's$: Gaugino masses, $\lambda_i's$: Yukawa couplings, $A_i's$: Tri-linear couplings, and $m_i^2's$: soft mass squared parameters), we have only four free parameters at GUT scale since the unification of gauge couplings at GUT scale is the primary and necessary requirement. These four parameters, (i) we choose the universal gaugino masses at GUT scale to be of the order of TeV, characterized by $m_{1/2}$, because the soft supersymmetry breaking terms are expected to be induced through the supergravity by the spontaneous breaking of supersymmetry in the hidden sector [53-55]. These soft supersymmetric terms are subjected to a certain conditions in terms of $m_{1/2}$ and $m_0$ which are of the order of $TeV$ at GUT scale, and are responsible for the gauge and supersymmetry breaking. If the gaugino condensations take place in $E_8'$ before the $E_6$ is broken then the seed of the global supersymmetry breaking in the observable sector will be given by the universal gaugino masses [30], (ii) Keeping the tri-linear couplings universal at GUT scale do not lead the acceptable low energy phenomenological spectrum as this string motivated models desired naturally. Therefore, $A_i's$ are kept universal at GUT scale, (iii) All the soft mass squared masses are considered universal at GUT scale as an assumption. However, with the non-universality of soft scalar masses (assuming universal Yukawa couplings) have been studied for the case of large *x* scenario by Langacker and Wang in Ref. [6], which requires a cumbersome fine tuning among the soft scalar masses at GUT scale, and finally, (iv) the only parameters left are the Yukawa couplings which have been considered non-universal in order to get the large x scenario. There are two basic reasons for choosing the non-universality of Yukawa couplings at GUT scale. First, the $E_6$ relations for the supersymmetric matter couplings will no longer be respected after the compactification with the non-trivial Wilson loop [27,56]. Therefore we regard these couplings in the superpotential as free parameter as long as they do not become extremely large to avoid the perturbative treatment. Second, in our recent work [5], about the investigation of unitarity bounds in the $E_6$ models, we have found that at GUT scale the bounds differ significantly for $\lambda_t$, $\lambda_H$, $\lambda_D$. With these facts in mind, we have invoked the non-universality of Yukawa couplings at GUT scale while others parameters are kept universal at GUT scale as following:

We have now all the tools to analyze low energy gauge and supersymmetry breaking. It is therefore clear how to proceed: given the non-universal Yukawa couplings at scale $M_G$, run the relevant parameters, solving the corresponding RGEs. Then minimizing the effective potential (keeping gaugino masses universal), determine the $\upsilon, \overline{\upsilon}, x$. This further determines the $m_{1/2}$ at weak scale through the relation given by Eq. (19).

In order to run these steps effectively, we examine more closely the choice of initial parameters at $M_G$ by considering the structure of RGEs and the relevant conditions to realize the desired minimum.

To get the large $x$ scenario, i.e. to satisfy the requirement $x >> \upsilon, \overline{\upsilon}$, we evolve the complete renormalization group equations (RGEs) from GUT scale down to Weak scale. It is found that Yukawa couplings at GUT scale follows the constraints $\frac{\lambda_i^2}{4\pi} \leq 1$ [34]. The set of maximum possible values that $\lambda_{t,H,D}$ can take at GUT scale is shown three-dimensionally in Fig. 11, which leads the VEV of the Singlet is large enough compared to the VEVs of the Higgs fields to give the successful low energy phenomenology [57] at weak scale ( $M_Z$ ) along with the exception of $\upsilon >> \overline{\upsilon}$ so that the top quark is maximally heavy for large top Yukawa coupling. Therefore, we obtain the large $x$ scenario with the real and positive VEVs of $H$, $\overline{H}$ and $N$ only satisfying the constraints given by the Eqs. (8), (9) and (10). Thus for $x >> \upsilon, \overline{\upsilon}$, the minimization conditions, given by Eq. (12), reduce to

$$-m_H^2 \cong m_3^2 \tan\beta + \lambda_H^2 x^2 + g_E^2 Y_1' Y_N' x^2 ; \qquad (48a)$$

$$-m_{\overline{H}}^2 \cong m_3^2 \cot\beta + \lambda_H^2 x^2 + g_E^2 Y_2' Y_N' x^2 ; \qquad (48b)$$

and $$-m_N^2 \cong g_E^2 Y_N'^2 x^2 , \qquad (48c)$$

$$\Rightarrow x^2 \cong \frac{m_N^2}{g_E^2 Y_N'^2} \cong O(TeV)^2 . \qquad (48d)$$

This suggests that, in general, absolute magnitude of Singlet mass is larger than that of the $H, \overline{H}$. Also, from Eq. (22c), $Z'$ mass equation reduces to the following form

$$M_{Z'}^2 \cong 2 g_1^2 Y_N'^2 x^2 . \qquad (49)$$

This with the use of Eq. (48d), gives

$$M_{Z'}^2 \cong \left| -2m_N^2 \right| \cong O(TeV)^2 . \qquad (50)$$

i.e. the $Z'$ mass is directly governed by Singlet scalar mass, which requires that $x$ should be the order of $TeV$ so that $Z'$ could be the order of $TeV$. Thus the $Z'$ mass typically lies with in the range of $TeV$ and hence by large $Z'$ mass, the mixing angle ($\theta_E$) is suppressed to the order of $10^{-3}$, which is in the good agreement of the present experimental limits [39-41].

In the whole exercise for large Singlet VEV scenario ($x >> \upsilon, \overline{\upsilon}$) the ratios of VEVs are found such that ratio $\frac{x}{\upsilon}$ is found in the range of 10~30 and $\tan\beta = \frac{\overline{\upsilon}}{\upsilon} < O(1)$. These correspond respectively to the

$$\left(\frac{M_{Z_1}}{M_Z}\right)^2 \approx O(1); \qquad \left(\frac{\upsilon}{\upsilon_{\max.}}\right) \cong O(1).$$

With these above facts and criteria given by Eqs. (8-10), we have done the RGEs analysis pertaining to the top quark mass i.e. for $m_t = 178 \pm 3$ $GeV$ [48-52] as one of the key parameters in experimental observables and found a very tiny region of Yukawa couplings at GUT scale. The set of possible combinations of Yukawa couplings at GUT scale are shown in Fig. 12**.** For the illustrations, the obtained low energy spectrum are listed in Table 4 for few sets of non-universal Yukawa couplings at GUT scale, which leads to the phenomenological acceptable large x scenario.

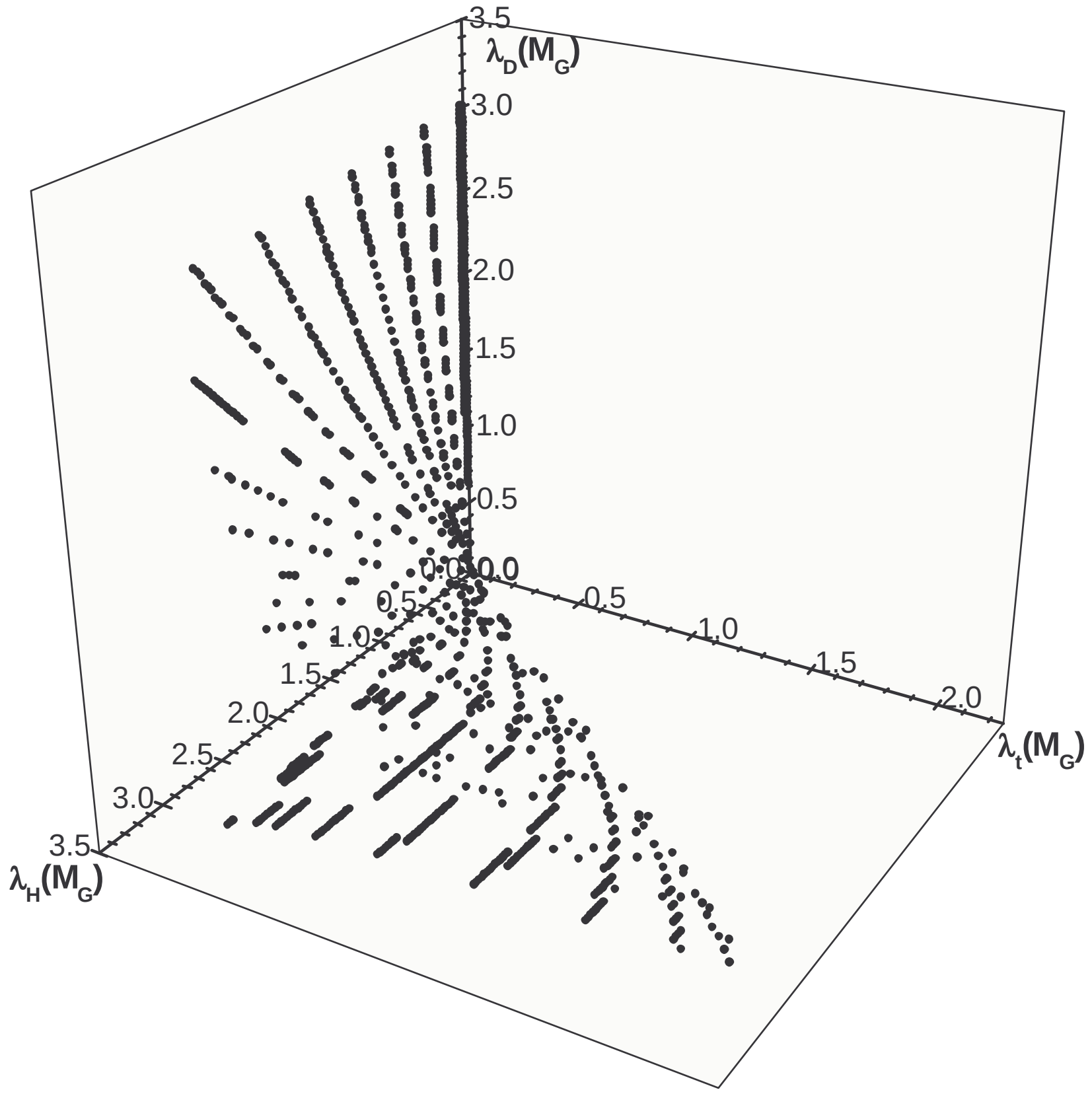

**Fig. 11:** *Three dimensional theoretical scatter-plots of Yukawa couplings* $\lambda_{t,H,D}$ *at GUT scale shows the maximum allowed values of* $\lambda_{t,H,D}$ *which lead the large x scenario with all primary constraints.*

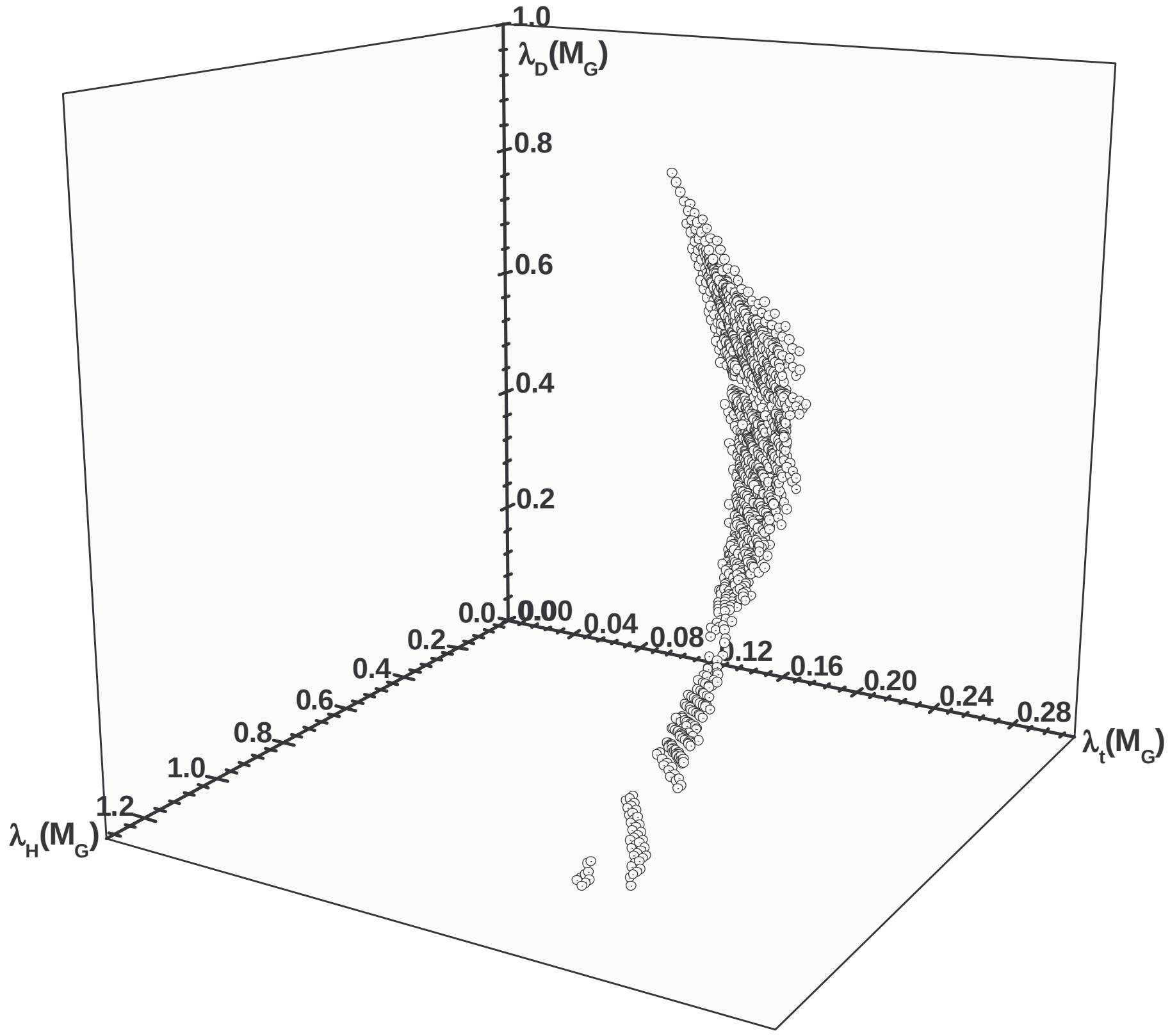

**Fig. 12:** *3-D region of* $\lambda_{t,H,D}$ *at weak scale for the large x scenario corresponding to the top quark* $m_t = 178 \pm 3\ GeV$.

**Table 4:** *For illustration, the few sets of low energy spectrums are shown corresponding to* $m_t \cong 178 \pm 3 GeV$ *in the large x scenario, when the Yukawa couplings are non-universal at* $M_G \cong 2\times 10^{16}\, GeV$ .

| **At Weak Scale ($M_Z$)** | $\lambda_t(M_G) = 0.172$ <br> $\lambda_H(M_G) = 0.384$ <br> $\lambda_D(M_G) = 0.240$ | $\lambda_t(M_G) = 0.184$ <br> $\lambda_H(M_G) = 0.444$ <br> $\lambda_D(M_G) = 0.324$ | $\lambda_t(M_G) = 0.192$ <br> $\lambda_H(M_G) = 0.516$ <br> $\lambda_D(M_G) = 0.408$ |
|---|---|---|---|
| $g_1 = g_E,\ g_2,\ g_3$ | 0.462, 0.654, 1.228 | 0.462, 0.654, 1.228 | 0.462, 0.654, 1.228 |
| $M_1,\ M_2,\ M_E$ | 57, 114, 57 | 57, 114, 57 | 57, 114, 57 |
| $\lambda_t,\ \lambda_H,\ \lambda_D$ | 1.024, 0.390, 0.950 | 1.045, 0.386, 1.061 | 1.056, 0.389, 1.124 |
| $A_t,\ A_H,\ A_D$ | 1149, -289, 1238 | 1088, -481, 978 | 1056, -605, 814 |
| ***VEVs:*** $\upsilon,\ \overline{\upsilon},\ x$ | 172, 27, 1794 | 168, 43, 2812 | 166, 53, 2052 |
| $x/\upsilon,\ \tan\beta$ | 10, 0.158 | 17, 0.255 | 12, 0.317 |
| $m_0,\quad m_{1/2}\,(M_3)$ | 521, 777 | 742, 775 | 521, 475 |
| $M_{Z_1},\ M_{Z_2},\ \theta_E$ | 91, 759, -0.009 | 92, 1188, -0.004 | 92, 868, -0.006 |
| $\mu\ (=\lambda_H .x)$ | 700 | 1087 | 799 |
| $m_t\ (=\lambda_t .\upsilon)$, | 176 | 176 | 175 |
| $m_{\tilde{D}}\ (=\lambda_D .x)$ | 1703 | 2984 | 2307 |
| $m_{H^\pm}$ | 1148 | 1477 | 1297 |
| ***Neutral Real Scalar Higgs:*** | | | |
| $m_{H^0},\ m_{\overline{H}^0},\ m_{N^0}$ | 1073, 124, 1625 | 1679, 124, 2086 | 1226, 129, 1827 |
| ***Soft SUSY Breaking Parameters:*** | | | |
| $\tilde{t}_{LL},\ \tilde{t}_{RR},\ \tilde{t}_{LR}$ | 1066, 699, 471 | 1138, 799, 490 | 1077, 708, 479 |
| $\tilde{t}_1,\ \tilde{t}_2$ | 649, 1098 | 748, 1172 | 655, 1110 |
| $\tilde{D}_{LL},\ \tilde{D}_{RR},\ \tilde{D}_{LR}$ | 1912, 1954, 2086 | 3041, 3108, 2376 | 2429, 2474, 2159 |
| $\tilde{D}_1,\ \tilde{D}_2$ | 1273, 2418 | 2555, 3519 | 2031, 2810 |
| $\tilde{d},\ \tilde{d}^c$ | 1054, 1218 | 1127, 1183 | 1065, 1211 |
| $\tilde{e},\ \tilde{e}^c$ | 1003, 954 | 961, 862 | 994, 935 |
| $\tilde{\nu},\ \tilde{\nu}^c$ | 1000, 1139 | 958, 1304 | 991, 1178 |
| ***Charginos:*** | | | |
| $\tilde{\chi}_1^\pm,\ \tilde{\chi}_2^\pm$ | 104, 710 | 108, 1092 | 105, 808 |
| ***Neutralinos:*** | | | |
| $\tilde{\chi}_1^0,\ \tilde{\chi}_2^0,\tilde{\chi}_3^0,\ \tilde{\chi}_4^0$ | 58, 115, 707, 788 | 1216, 1089, 116, 58 | 58, 118, 802, 897 |

**continue........**

| At Weak Scale ($M_Z$) | $\lambda_t(M_G) = 0.200$ <br> $\lambda_H(M_G) = 0.504$ <br> $\lambda_D(M_G) = 0.372$ | $\lambda_t(M_G) = 0.208$ <br> $\lambda_H(M_G) = 0.588$ <br> $\lambda_D(M_G) = 0.456$ | $\lambda_t(M_G) = 0.216$ <br> $\lambda_H(M_G) = 0.564$ <br> $\lambda_D(M_G) = 0.408$ |
|---|---|---|---|
| $g_1 = g_E,\ g_2,\ g_3$ | 0.462, 0.654, 1.228 | 0.462, 0.654, 1.228 | 0.462, 0.654, 1.228 |
| $M_1,\ M_2,\ M_E$ | 57, 114, 57 | 57, 114, 57 | 57, 114, 57 |
| $\lambda_t,\ \lambda_H,\ \lambda_D$ | 1.071, 0.392, 1.098 | 1.079, 0.398, 1.145 | 1.092, 0.399, 1.118 |
| $A_t,\ A_H,\ A_D$ | 1015, -587, 879 | 988, -686, 755 | 951, -662, 825 |
| ***VEVs:*** $\upsilon,\ \overline{\upsilon},\ x$ | 166, 51, 1905 | 164, 58, 3072 | 165, 56, 3316 |
| $x/\upsilon,\ \tan\beta$ | 11, 0.308 | 19, 0.354 | 20, 0.342 |
| $m_0,\quad m_{1/2}\,(M_3)$ | 490, 454 | 769, 664 | 840, 739 |
| $M_{Z_1},\ M_{Z_2},\ \theta_E$ | 92, 806, -0.007 | 92, 1298, -0.003 | 92, 1401, -0.002 |
| $\mu\ (= \lambda_H . x)$ | 747 | 1222 | 1324 |
| $m_t\ (= \lambda_t . \upsilon)$, | 178 | 177 | 180 |
| $m_{\tilde{D}}\ (= \lambda_D . x)$ | 2092 | 3517 | 3707 |
| $m_{H^\pm}$ | 1249 | 1633 | 1692 |
| ***Neutral Real Scalar Higgs:*** | | | |
| $m_{H^0},\ m_{\overline{H}^0},\ m_{N^0}$ | 1138, 129, 1768 | 1834, 118, 2311 | 1979, 117, 2400 |
| ***Soft SUSY Breaking Parameters:*** | | | |
| $\tilde{t}_{LL},\ \tilde{t}_{RR},\ \tilde{t}_{LR}$ | 1065, 685, 470 | 1158, 821, 502 | 1179, 848, 502 |
| $\tilde{t}_1,\ \tilde{t}_2$ | 634, 1096 | 769, 1194 | 798, 1215 |
| $\tilde{D}_{LL},\ \tilde{D}_{RR},\ \tilde{D}_{LR}$ | 2239, 2280, 2180 | 3547, 3617, 2348 | 3722, 3799, 2470 |
| $\tilde{D}_1,\ \tilde{D}_2$ | 1806, 2637 | 3187, 3937 | 3327, 4150 |
| $\tilde{d},\ \tilde{d}^c$ | 1052, 1215 | 1146, 1172 | 1168, 1160 |
| $\tilde{e},\ \tilde{e}^c$ | 999, 946 | 946, 830 | 931, 796 |
| $\tilde{\nu},\ \tilde{\nu}^c$ | 997, 1155 | 943, 1360 | 929, 1411 |
| ***Charginos:*** | | | |
| $\tilde{\chi}_1^\pm,\ \tilde{\chi}_2^\pm$ | 104, 757 | 108, 1228 | 109, 1328 |
| ***Neutralinos:*** | | | |
| $\tilde{\chi}_1^0,\ \tilde{\chi}_2^0,\ \tilde{\chi}_3^0,\ \tilde{\chi}_4^0$ | 58, 118, 751, 835 | 58, 117, 225, 1327 | 58, 117, 1325, 1430 |

**continue........**

| At Weak Scale ($M_Z$) | $\lambda_t(M_G) = 0.220$ <br> $\lambda_H(M_G) = 0.732$ <br> $\lambda_D(M_G) = 0.588$ | $\lambda_t(M_G) = 0.224$ <br> $\lambda_H(M_G) = 0.720$ <br> $\lambda_D(M_G) = 0.564$ | $\lambda_t(M_G) = 0.228$ <br> $\lambda_H(M_G) = 0.696$ <br> $\lambda_D(M_G) = 0.528$ |
|---|---|---|---|
| $g_1 = g_E,\ g_2,\ g_3$ | 0.462, 0.654, 1.228 | 0.462, 0.654, 1.228 | 0.462, 0.654, 1.228 |
| $M_1,\ M_2,\ M_E$ | 57, 114, 57 | 57, 114, 57 | 57, 114, 57 |
| $\lambda_t,\ \lambda_H,\ \lambda_D$ | 1.089, 0.406, 1.185 | 1.095, 0.406, 1.178 | 1.101, 0.406, 1.167 |
| $A_t,\ A_H,\ A_D$ | 954, -787, 641 | 937, -784, 660 | 919, -774, 690 |
| ***VEVs:*** $\upsilon,\ \overline{\upsilon},\ x$ | 161, 65, 3548 | 162, 64, 2837 | 162, 64, 2445 |
| $x/\upsilon$, $\tan\beta$ | 22, 0.400 | 18, 0.399 | 15, 0.394 |
| $m_0$, $m_{1/2}\,(M_3)$ | 872, 714 | 699, 573 | 605, 498 |
| $M_{Z_1},\ M_{Z_2},\ \theta_E$ | 92, 1498, -0.002 | 92, 1198, -0.003 | 92, 1033, -0.004 |
| $\mu\ (=\lambda_H.x)$ | 1439 | 1152 | 993 |
| $m_t\ (=\lambda_t.\upsilon)$, | 176 | 177 | 178 |
| $m_{\tilde{D}}\ (=\lambda_D.x)$ | 4205 | 3342 | 2853 |
| $m_{H^\pm}$ | 1813 | 1621 | 1501 |
| ***Neutral Real Scalar Higgs:*** | | | |
| $m_{H^0},\ m_{\overline{H}^0},\ m_{N^0}$ | 2118, 111, 2556 | 1694, 118, 2299 | 1460, 122, 2122 |
| ***Soft SUSY Breaking Parameters:*** | | | |
| $\tilde{t}_{LL},\ \tilde{t}_{RR},\ \tilde{t}_{LR}$ | 1205, 887, 519 | 1135, 788, 497 | 1102, 737, 484 |
| $\tilde{t}_1,\ \tilde{t}_2$ | 832, 1244 | 735, 1171 | 685, 1134 |
| $\tilde{D}_{LL},\ \tilde{D}_{RR},\ \tilde{D}_{LR}$ | 4202, 4280, 2328 | 3388, 3449, 2232 | 2932, 2984, 2207 |
| $\tilde{D}_1,\ \tilde{D}_2$ | 3908, 4552 | 3077, 3730 | 2602, 3277 |
| $\tilde{d},\ \tilde{d}^c$ | 1194, 1148 | 1124, 1182 | 1089, 1198 |
| $\tilde{e},\ \tilde{e}^c$ | 916, 759 | 959, 859 | 978, 901 |
| $\tilde{\nu},\ \tilde{\nu}^c$ | 913, 1460 | 957, 1314 | 976, 1242 |
| ***Charginos:*** | | | |
| $\tilde{\chi}_1^\pm,\ \tilde{\chi}_2^\pm$ | 109, 1445 | 108, 1158 | 107, 1000 |
| ***Neutralinos:*** | | | |
| $\tilde{\chi}_1^0,\ \tilde{\chi}_2^0,\tilde{\chi}_3^0,\ \tilde{\chi}_4^0$ | 58, 117, 1441, 1529 | 58, 117, 1153, 1229 | 58, 118, 994, 1064 |

**continue........**

| At Weak Scale ($M_Z$) | $\lambda_t(M_G) = 0.232$ $\lambda_H(M_G) = 0.672$ $\lambda_D(M_G) = 0.492$ | $\lambda_t(M_G) = 0.236$ $\lambda_H(M_G) = 0.912$ $\lambda_D(M_G) = 0.732$ | $\lambda_t(M_G) = 0.236$ $\lambda_H(M_G) = 0.972$ $\lambda_D(M_G) = 0.792$ |
|---|---|---|---|
| $g_1 = g_E,\ g_2,\ g_3$ | 0.462, 0.654, 1.228 | 0.462, 0.654, 1.228 | 0.462, 0.654, 1.228 |
| $M_1,\ M_2,\ M_E$ | 57, 114, 57 | 57, 114, 57 | 57, 114, 57 |
| $\lambda_t,\ \lambda_H,\ \lambda_D$ | 1.107, 0.406, 1.154 | 1.103, 0.413, 1.207 | 1.100, 0.414, 1.214 |
| $A_t,\ A_H,\ A_D$ | 902, -760, 726 | 913, -862, 576 | 919, -875, 556 |
| ***VEVs:*** $\upsilon,\ \overline{\upsilon},\ x$ | 162, 63, 3235 | 159, 69, 1970 | 159, 70, 4083 |
| $x/\upsilon,\ \tan\beta$ | 20, 0.387 | 12, 0.435 | 26, 0.439 |
| $m_0,\quad m_{1/2}\,(M_3)$ | 804, 668 | 481, 381 | 991, 786 |
| $M_{Z_1},\ M_{Z_2},\ \theta_E$ | 92, 1366. -0.002 | 92, 833, -0.006 | 92, 1724, -0.001 |
| $\mu\ (= \lambda_H .x)$ | 1315 | 814 | 1692 |
| $m_t\ (= \lambda_t.\upsilon)$, | 180 | 176 | 175 |
| $m_{\tilde{D}}\ (= \lambda_D.x)$ | 3732 | 2379 | 4958 |
| $m_{H^\pm}$ | 1723 | 1386 | 2006 |
| ***Neutral Real Scalar Higgs:*** | | | |
| $m_{H^0},\ m_{\overline{H}^0},\ m_{N^0}$ | 1931, 113, 2431 | 1177, 127, 1961 | 2437, 103, 2833 |
| ***Soft SUSY Breaking Parameters:*** | | | |
| $\tilde{t}_{LL},\ \tilde{t}_{RR},\ \tilde{t}_{LR}$ | 1170, 835521 | 1069, 690, 471 | 1265, 967, 539 |
| $\tilde{t}_1,\ \tilde{t}_2$ | 783, 1207 | 639, 1100 | 909, 1307 |
| $\tilde{D}_{LL},\ \tilde{D}_{RR},\ \tilde{D}_{LR}$ | 3752, 3824, 2391 | 2302, 2542, 1995 | 4926, 5015, 2306 |
| $\tilde{D}_1,\ \tilde{D}_2$ | 3408, 4132 | 2232, 2783 | 4682, 5246 |
| $\tilde{d},\ \tilde{d}^c$ | 1158, 1164 | 1056, 1213 | 1254, 1115 |
| $\tilde{e},\ \tilde{e}^c$ | 936, 807 | 997, 941 | 875, 657 |
| $\tilde{\nu},\ \tilde{\nu}^c$ | 934, 1394 | 995, 1165 | 873, 1579 |
| ***Charginos:*** | | | |
| $\tilde{\chi}_1^\pm,\ \tilde{\chi}_2^\pm$ | 109, 1319 | 105, 823 | 110, 1694 |
| ***Neutralinos:*** | | | |
| $\tilde{\chi}_1^0,\ \tilde{\chi}_2^0,\tilde{\chi}_3^0,\ \tilde{\chi}_4^0$ | 1397, 1314, 117, 58 | 58, 118, 813, 866 | 58, 117, 1688, 1756 |

**continue........**

| At Weak Scale ($M_Z$) | $\lambda_t(M_G) = 0.240$ $\lambda_H(M_G) = 1.044$ $\lambda_D(M_G) = 0.852$ | $\lambda_t(M_G) = 0.248$ $\lambda_H(M_G) = 0.768$ $\lambda_D(M_G) = 0.552$ | $\lambda_t(M_G) = 0.252$ $\lambda_H(M_G) = 1.128$ $\lambda_D(M_G) = 0.888$ |
|---|---|---|---|
| $g_1 = g_E, g_2, g_3$ | 0.462, 0.654, 1.228 | 0.462, 0.654, 1.228 | 0.462, 0.654, 1.228 |
| $M_1, M_2, M_E$ | 57, 114, 57 | 57, 114, 57 | 57, 114, 57 |
| $\lambda_t, \lambda_H, \lambda_D$ | 1.103, 0.416, 1.219 | 1.121, 0.412, 1.169 | 1.113, 0.420, 1.219 |
| $A_t, A_H, A_D$ | 911, -892, 541 | 858, -820, 680 | 880, -916, 539 |
| ***VEVs:*** $\upsilon, \overline{\upsilon}, x$ | 159, 71, 2882 | 161, 67, 3006 | 158, 72, 5086 |
| $x/\upsilon$, $\tan\beta$ | 18, 0.447 | 19, 0.414 | 32, 0.456 |
| $m_0$, $m_{1/2}(M_3)$ | 699, 551 | 741, 600 | 1229, 967 |
| $M_{Z_1}, M_{Z_2}, \theta_E$ | 92, 1218, -0.003 | 92, 1270, -0.003 | 92, 2147, -0.001 |
| $\mu\ (= \lambda_H.x)$ | 1199 | 1239 | 2135 |
| $m_t\ (= \lambda_t.\upsilon)$, | 175 | 180 | 176 |
| $m_{\tilde{D}}\ (= \lambda_D.x)$ | 3514 | 3515 | 6201 |
| $m_{H^\pm}$ | 1695 | 1696 | 2276 |
| ***Neutral Real Scalar Higgs:*** | | | |
| $m_{H^0}, m_{\overline{H}^0}, m_{N^0}$ | 1721, 115, 2398 | 1795, 114, 2394 | 3035, 92, 3221 |
| ***Soft SUSY Breaking Parameters:*** | | | |
| $\tilde{t}_{LL}, \tilde{t}_{RR}, \tilde{t}_{LR}$ | 1141, 798, 504 | 1149, 803, 498 | , 1122, 571 |
| $\tilde{t}_1, \tilde{t}_2$ | 743, 1178 | 751, 1183 | 1062, 1435 |
| $\tilde{D}_{LL}, \tilde{D}_{RR}, \tilde{D}_{LR}$ | 3555, 3616, 2107 | 35489, 3615, 2326 | 6123, 6234, 2432 |
| $\tilde{D}_1, \tilde{D}_2$ | 3307, 3845 | 3229, 3904 | 5897, 6451 |
| $\tilde{d}, \tilde{d}^c$ | 1129, 1180 | 1136, 1175 | 1378, 1040 |
| $\tilde{e}, \tilde{e}^c$ | 957, 854 | 950, 838 | 776, 322 |
| $\tilde{\nu}, \tilde{\nu}^c$ | 954, 1323 | 947, 1347 | 774, 1820 |
| ***Charginos:*** | | | |
| $\tilde{\chi}_1^\pm, \tilde{\chi}_2^\pm$ | 108, 1204 | 109, 1244 | 111, 2139 |
| ***Neutralinos:*** | | | |
| $\tilde{\chi}_1^0, \tilde{\chi}_2^0, \tilde{\chi}_3^0, \tilde{\chi}_4^0$ | 58, 118, 1196, 1251 | 58, 118, 1238, 1302 | 58, 116, 2130, 2181 |

**continue........**

| At Weak Scale ($M_Z$) | $\lambda_t(M_G) = 0.256$ $\lambda_H(M_G) = 1.080$ $\lambda_D(M_G) = 0.828$ | $\lambda_t(M_G) = 0.260$ $\lambda_H(M_G) = 0.900$ $\lambda_D(M_G) = 0.648$ | $\lambda_t(M_G) = 0.260$ $\lambda_H(M_G) = 1.224$ $\lambda_D(M_G) = 0.948$ |
|---|---|---|---|
| $g_1 = g_E,\ g_2,\ g_3$ | 0.462, 0.654, 1.228 | 0.462, 0.654, 1.228 | 0.462, 0.654, 1.228 |
| $M_1,\ M_2,\ M_E$ | 57, 114, 57 | 57, 114, 57 | 57, 114, 57 |
| $\lambda_t,\ \lambda_H,\ \lambda_D$ | 1.119, 0.420, 1.213 | 1.128, 0.417, 1.188 | 1.118, 0.422, 1.222 |
| $A_t,\ A_H,\ A_D$ | 863, -912, 556 | 835, -876, 625 | 863, -935, 530 |
| ***VEVs:*** $\upsilon,\ \overline{\upsilon},\ x$ | 158, 72, 2761 | 159, 70, 2255 | 158, 73, 5607 |
| $x/\upsilon,\ \tan\beta$ | 17, 0.455 | 14, 0.440 | 36, 0.464 |
| $m_0,\quad m_{1/2}(M_3)$ | 669, 526 | 551, 437 | 1352, 1060 |
| $M_{Z_1},\ M_{Z_2},\ \theta_E$ | 92, 1167, -0.003 | 92, 935, -0.005 | 92, 2367, -0.001 |
| $\mu\ (=\lambda_H.x)$ | 1159 | 941 | 2367 |
| $m_t\ (=\lambda_t.\upsilon)$, | 177 | 180 | 176 |
| $m_{\tilde{D}}\ (=\lambda_D.x)$ | 3350 | 2681 | 6852 |
| $m_{H^\pm}$ | 1675 | 1497 | 2407 |
| ***Neutral Real Scalar Higgs:*** | | | |
| $m_{H^0},\ m_{\overline{H}^0},\ m_{N^0}$ | 1649, 115, 2368 | 1347, 123, 2114 | 3346, 87, 3409 |
| ***Soft SUSY Breaking Parameters:*** | | | |
| $\tilde{t}_{LL},\ \tilde{t}_{RR},\ \tilde{t}_{LR}$ | 1129, 779, 496 | 1086, 712, 473 | 1457, 1207, 588 |
| $\tilde{t}_1,\ \tilde{t}_2$ | 725, 1164 | 663, 1117 | 1145, 1507 |
| $\tilde{D}_{LL},\ \tilde{D}_{RR},\ \tilde{D}_{LR}$ | 3401, 3459, 2126 | 2776, 2823, 2139 | 6753, 6876, 2492 |
| $\tilde{D}_1,\ \tilde{D}_2$ | 3145, 3695 | 2481, 3087 | 6534, 7083 |
| $\tilde{d},\ \tilde{d}^c$ | 1167, 1185 | 1073, 1204 | 1448, 991 |
| $\tilde{e},\ \tilde{e}^c$ | 963, 868 | 986, 918 | 710, 307 |
| $\tilde{\nu},\ \tilde{\nu}^c$ | 961, 1300 | 984, 1210 | 707, 1951 |
| ***Charginos:*** | | | |
| $\tilde{\chi}_1^\pm,\ \tilde{\chi}_2^\pm$ | 108, 1166 | 948, 107 | 111, 2369, |
| ***Neutralinos:*** | | | |
| $\tilde{\chi}_1^0,\ \tilde{\chi}_2^0,\tilde{\chi}_3^0,\ \tilde{\chi}_4^0$ | 58, 117, 1155, 1202 | 58, 118, 938,987 | 58, 116, 2358, 2403 |

**continue........**

| At Weak Scale ($M_Z$) | $\lambda_t(M_G)=0.260$<br>$\lambda_H(M_G)=1.260$<br>$\lambda_D(M_G)=0.984$ | $\lambda_t(M_G)=0.264$<br>$\lambda_H(M_G)=1.356$<br>$\lambda_D(M_G)=1.056$ | $\lambda_t(M_G)=0.268$<br>$\lambda_H(M_G)=1.164$<br>$\lambda_D(M_G)=0.864$ |
|---|---|---|---|
| $g_1=g_E,\ g_2,\ g_3$ | 0.462, 0.654, 1.228 | 0.462, 0.654, 1.228 | 0.462, 0.654, 1.228 |
| $M_1,\ M_2,\ M_E$ | 57, 114, 57 | 57, 114, 57 | 57, 114, 57 |
| $\lambda_t,\ \lambda_H,\ \lambda_D$ | 1.117, 0.422, 1.224 | 1.118, 0.423, 1.227 | 1.127, 0.423, 1.214 |
| $A_t,\ A_H,\ A_D$ | 866, -938, 523 | 862, -949, 514 | 836, -934, 552 |
| ***VEVs:*** $\upsilon,\ \overline{\upsilon},\ x$ | 158, 74, 2272 | 157, 74, 2792 | 158, 73, 2328 |
| $x/\upsilon$, $\tan\beta$ | 14, 0.467 | 18, 0.471 | 15, 0.465 |
| $m_0$, $m_{1/2}\,(M_3)$ | 549, 428 | 673, 525 | 564, 441 |
| $M_{Z_1},\ M_{Z_2},\ \theta_E$ | 92, 960, -0.005 | 92, 1179, -0.003 | 92, 984, -0.004 |
| $\mu\ (=\lambda_H.x)$ | 959 | 1182 | 984 |
| $m_t\ (=\lambda_t.\upsilon)$, | 176 | 176 | 178 |
| $m_{\tilde{D}}\ (=\lambda_D.x)$ | 2782 | 3427 | 2827 |
| $m_{H^\pm}$ | 1532 | 1707 | 1552 |
| ***Neutral Real Scalar Higgs:*** | | | |
| $m_{H^0},\ m_{\overline{H}^0},\ m_{N^0}$ | 1357, 123, 2165 | 1668, 114, 2411 | 1391, 122, 2200 |
| ***Soft SUSY Breaking Parameters:*** | | | |
| $\tilde{t}_{LL},\ \tilde{t}_{RR},\ \tilde{t}_{LR}$ | 1090, 723, 482 | 1131, 784, 485 | 1094, 726, 480 |
| $\tilde{t}_1,\ \tilde{t}_2$ | 672, 1123 | 732, 1169 | 675, 1126 |
| $\tilde{D}_{LL},\ \tilde{D}_{RR},\ \tilde{D}_{LR}$ | 2874, 2921, 2009 | 3475, 3534, 2070 | 2915, 2963, 2062 |
| $\tilde{D}_1,\ \tilde{D}_2$ | 2633, 3142 | 3241, 732 | 2658, 3196 |
| $\tilde{d},\ \tilde{d}^c$ | 1078, 1204 | 1121, 1184 | 1081, 1202 |
| $\tilde{e},\ \tilde{e}^c$ | 985, 916 | 961, 864 | 983, 911 |
| $\tilde{\nu},\ \tilde{\nu}^c$ | 983, 1212 | 959, 1306 | 981, 1222 |
| ***Charginos:*** | | | |
| $\tilde{\chi}_1^\pm,\ \tilde{\chi}_2^\pm$ | 107, 966 | 108, 1187 | 107, 992 |
| ***Neutralinos:*** | | | |
| $\tilde{\chi}_1^0,\ \tilde{\chi}_2^0,\ \tilde{\chi}_3^0,\ \tilde{\chi}_4^0$ | 58, 118, 953, 997 | 58, 118, 1174, 1217 | 58, 118, 979, 1021 |

**continue.......**

| At Weak Scale ($M_Z$) | $\lambda_t(M_G) = 0.268$ <br> $\lambda_H(M_G) = 1.440$ <br> $\lambda_D(M_G) = 1.116$ | $\lambda_t(M_G) = 0.272$ <br> $\lambda_H(M_G) = 1.044$ <br> $\lambda_D(M_G) = 0.744$ | $\lambda_t(M_G) = 0.272$ <br> $\lambda_H(M_G) = 1.260$ <br> $\lambda_D(M_G) = 0.936$ |
|---|---|---|---|
| $g_1 = g_E,\ g_2,\ g_3$ | 0.462, 0.654, 1.228 | 0.462, 0.654, 1.228 | 0.462, 0.654, 1.228 |
| $M_1,\ M_2,\ M_E$ | 57, 114, 57 | 57, 114, 57 | 57, 114, 57 |
| $\lambda_t,\ \lambda_H,\ \lambda_D$ | 1.120, 0.424, 1.229 | 1.135, 0.422, 1.200 | 1.128, 0.424, 1.219 |
| $A_t,\ A_H,\ A_D$ | 857, -958, 508 | 815, -918, 590 | 832, -948, 538 |
| ***VEVs:*** $\upsilon,\ \overline{\upsilon},\ x$ | 157, 75, 2399 | 158, 72, 2644 | 157, 74, 2047 |
| $x/\upsilon$, $\tan\beta$ | 15, 0.476 | 17, 0.457 | 13, 0.471 |
| $m_0$, $m_{1/2}\ (M_3)$ | 578, 449 | 642, 505 | 495, 385 |
| $M_{Z_1},\ M_{Z_2},\ \theta_E$ | 92, 1014, -0.004 | 92, 1117, -0.003 | 92, 866, -0.006 |
| $\mu\ (= \lambda_H . x)$ | 1018 | 1115 | 868 |
| $m_t\ (= \lambda_t . \upsilon)$, | 176 | 179 | 177 |
| $m_{\tilde{D}}\ (= \lambda_D . x)$ | 2949 | 3174 | 2494 |
| $m_{H^\pm}$ | 1587 | 1646 | 1462 |
| ***Neutral Real Scalar Higgs:*** | | | |
| $m_{H^0},\ m_{\overline{H}^0},\ m_{N^0}$ | 1433, 121, 2239 | 1579, 116, 2331 | 1223, 126, 2067 |
| ***Soft SUSY Breaking Parameters:*** | | | |
| $\tilde{t}_{LL},\ \tilde{t}_{RR},\ \tilde{t}_{LR}$ | 1101, 739, 486 | 1117, 760, 487 | 1073, 696, 469 |
| $\tilde{t}_1,\ \tilde{t}_2$ | 687, 1134 | 708, 1151 | 647, 1104 |
| $\tilde{D}_{LL},\ \tilde{D}_{RR},\ \tilde{D}_{LR}$ | 3030, 3079, 2008 | 3235, 3291, 2169 | 2613, 2654, 1998 |
| $\tilde{D}_1,\ \tilde{D}_2$ | 2797, 3294 | 2961, 3541 | 2361, 2878 |
| $\tilde{d},\ \tilde{d}^c$ | 1088, 1199 | 1105, 1190 | 1061, 1210 |
| $\tilde{e},\ \tilde{e}^c$ | 980, 905 | 969, 881 | 994, 935 |
| $\tilde{\nu},\ \tilde{\nu}^c$ | 978, 1234 | 967, 1278 | 992, 1177 |
| ***Charginos:*** | | | |
| $\tilde{\chi}_1^\pm,\ \tilde{\chi}_2^\pm$ | 107, 1024 | 108, 1122 | 106, 876 |
| ***Neutralinos:*** | | | |
| $\tilde{\chi}_1^0,\ \tilde{\chi}_2^0,\ \tilde{\chi}_3^0,\ \tilde{\chi}_4^0$ | 58, 118, 1010, 1052 | 58, 118, 1110, 1154 | 58, 119, 862, 904 |

**continue........**

| At Weak Scale ($M_Z$) | $\lambda_t(M_G) = 0.272$<br>$\lambda_H(M_G) = 1.500$<br>$\lambda_D(M_G) = 1.152$ | $\lambda_t(M_G) = 0.276$<br>$\lambda_H(M_G) = 1.140$<br>$\lambda_D(M_G) = 0.816$ | $\lambda_t(M_G) = 0.280$<br>$\lambda_H(M_G) = 1.056$<br>$\lambda_D(M_G) = 0.732$ |
|---|---|---|---|
| $g_1 = g_E,\ g_2,\ g_3$ | 0.462, 0.654, 1.228 | 0.462, 0.654, 1.228 | 0.462, 0.654, 1.228 |
| $M_1,\ M_2,\ M_E$ | 57, 114, 57 | 57, 114, 57 | 57, 114, 57 |
| $\lambda_t,\ \lambda_H,\ \lambda_D$ | 1.122, 0.425, 1.230 | 1.135, 0.424, 1.208 | 1.141, 0.423, 1.197 |
| $A_t,\ A_H,\ A_D$ | 850, -965, 506 | 812, -936, 568 | 795, -924, 598 |
| ***VEVs:*** $\upsilon,\ \overline{\upsilon},\ x$ | 157, 75, 2186 | 158, 73, 2342 | 158, 73, 2425 |
| $x/\upsilon$, $\tan\beta$ | 14, 0.479 | 15, 0.465 | 15, 0.460 |
| $m_0$, $m_{1/2}\,(M_3)$ | 526, 409 | 567, 444 | 589, 462 |
| $M_{Z_1},\ M_{Z_2},\ \theta_E$ | 92, 924, -0.005 | 92, 990, -0.004 | 92, 1025, -0.004 |
| $\mu\ (=\lambda_H.x)$ | 930 | 992 | 1026 |
| $m_t\ (=\lambda_t.\upsilon)$, | 176 | 179 | 180 |
| $m_{\tilde{D}}\ (=\lambda_D.x)$ | 2690 | 2829 | 2903 |
| $m_{H^\pm}$ | 1519 | 1559 | 1581 |
| ***Neutral Real Scalar Higgs:*** | | | |
| $m_{H^0},\ m_{\overline{H}^0},\ m_{N^0}$ | 1306, 124, 2148 | 1399, 121, 2210 | 1449, 120, 2232 |
| ***Soft SUSY Breaking Parameters:*** | | | |
| $\tilde{t}_{LL},\ \tilde{t}_{RR},\ \tilde{t}_{LR}$ | 1085, 717, 484 | 1094, 725, 477 | 1100, 732, 478 |
| $\tilde{t}_1,\ \tilde{t}_2$ | 664, 1117 | 675, 1125 | 682, 1131 |
| $\tilde{D}_{LL},\ \tilde{D}_{RR},\ \tilde{D}_{LR}$ | 2791, 2835, 1973 | 2917, 2966, 2094 | 2984, 3035, 2157 |
| $\tilde{D}_1,\ \tilde{D}_2$ | 2558, 3048 | 2651, 3205 | 2703, 3287 |
| $\tilde{d},\ \tilde{d}^c$ | 1072, 1207 | 1081, 1201 | 1087, 1198 |
| $\tilde{e},\ \tilde{e}^c$ | 989, 924 | 982, 910 | 979, 903 |
| $\tilde{\nu},\ \tilde{\nu}^c$ | 987, 1199 | 980, 1224 | 977, 1239 |
| ***Charginos:*** | | | |
| $\tilde{\chi}_1^\pm,\ \tilde{\chi}_2^\pm$ | 107, 937 | 107, 1000 | 107, 1033 |
| ***Neutralinos:*** | | | |
| $\tilde{\chi}_1^0,\ \tilde{\chi}_2^0,\ \tilde{\chi}_3^0,\ \tilde{\chi}_4^0$ | 58, 119, 922, 963 | 58, 118, 987, 1028 | 58, 118, 1020, 1062 |

**continue........**

| At Weak Scale ($M_Z$) | $\lambda_t(M_G)=0.284$ <br> $\lambda_H(M_G)=1.212$ <br> $\lambda_D(M_G)=0.852$ | $\lambda_t(M_G)=0.284$ <br> $\lambda_H(M_G)=1.440$ <br> $\lambda_D(M_G)=1.044$ | $\lambda_t(M_G)=0.292$ <br> $\lambda_H(M_G)=1.464$ <br> $\lambda_D(M_G)=1.032$ |
|---|---|---|---|
| $g_1=g_E,\ g_2,\ g_3$ | 0.462, 0.654, 1.228 | 0.462, 0.654, 1.228 | 0.462, 0.654, 1.228 |
| $M_1,\ M_2,\ M_E$ | 57, 114, 57 | 57, 114, 57 | 57, 114, 57 |
| $\lambda_t,\ \lambda_H,\ \lambda_D$ | 1.140, 0.425, 1.209 | 1.134, 0.427, 1.223 | 1.140, 0.428, 1.220 |
| $A_t,\ A_H,\ A_D$ | 798, -951, 563 | 815, -972, 525 | 797, -980, 592 |
| ***VEVs:*** $\upsilon,\ \overline{\upsilon},\ x$ | 157, 74, 2203 | 157, 75, 2578 | 157, 76, 2305 |
| $x/\upsilon,\ \tan\beta$ | 14, 0.472 | 16, 0.481 | 15, 0.484 |
| $m_0,\quad m_{1/2}\,(M_3)$ | 533, 416 | 621, 483 | 555, 431 |
| $M_{Z_1},\ M_{Z_2},\ \theta_E$ | 92, 931, -0.005 | 92, 1089, -0.003 | 92, 974, -0.004 |
| $\mu\ (=\lambda_H.x)$ | 937 | 1101 | 987 |
| $m_t\ (=\lambda_t.\upsilon)$, | 179 | 178 | 178 |
| $m_{\tilde{D}}\ (=\lambda_D.x)$ | 2665 | 3151 | 2813 |
| $m_{H^\pm}$ | 1521 | 1656 | 1571 |
| ***Neutral Real Scalar Higgs:*** | | | |
| $m_{H^0},\ m_{\overline{H}^0},\ m_{N^0}$ | 1316, 123, 2150 | 1540, 117, 2346 | 1377, 122, 2221 |
| ***Soft SUSY Breaking Parameters:*** | | | |
| $\tilde{t}_{LL},\ \tilde{t}_{RR},\ \tilde{t}_{LR}$ | 1083, 710, 471 | 1113, 756, 489 | 1092, 723, 478 |
| $\tilde{t}_1,\ \tilde{t}_2$ | 661, 1114 | 705, 1148 | 674, 1124 |
| $\tilde{D}_{LL},\ \tilde{D}_{RR},\ \tilde{D}_{LR}$ | 2766, 2811, 2069 | 3220, 3274, 2072 | 2904, 2951, 2051 |
| $\tilde{D}_1,\ \tilde{D}_2$ | 2504, 3048 | 2977, 3494 | 2658, 3175 |
| $\tilde{d},\ \tilde{d}^c$ | 1071, 1206 | 1101, 1193 | 1079, 1203 |
| $\tilde{e},\ \tilde{e}^c$ | 988, 923 | 972, 888 | 984, 914 |
| $\tilde{\nu},\ \tilde{\nu}^c$ | 986, 1202 | 970, 1266 | 982, 1218 |
| ***Charginos:*** | | | |
| $\tilde{\chi}_1^\pm,\ \tilde{\chi}_2^\pm$ | 107, 944 | 108, 1107 | 107, 994 |
| ***Neutralinos:*** | | | |
| $\tilde{\chi}_1^0,\ \tilde{\chi}_2^0,\tilde{\chi}_3^0,\ \tilde{\chi}_4^0$ | 58, 119, 929, 970 | 58, 118, 1091, 1130 | 58, 119, 977, 1016 |

The typical scale dependence of $m_H^2, m_{\bar{H}}^2$ and $m_N^2$ are shown in Fig 13-15. Keeping constant $\lambda_H$ and $\lambda_D$ at GUT (e.g. $\lambda_H = 0.650; \lambda_D = 0.450$), the flows of $m_H^2$, $m_{\bar{H}}^2$ and $m_N^2$ are shown in Fig. 13 for the different $\lambda_t(M_G)$ values. It is clearly shown that the evolution of $m_H^2$ is mainly governed by top Yukawa couplings whereas $m_{\bar{H}}^2$ and $m_N^2$ have almost unique pattern. From Fig. 14, for the different values of $\lambda_H(M_G)$ (keeping others constant) we observe that $m_{\bar{H}}^2$ has only a little dependence while the evolutions of $m_H^2$, $m_N^2$ also effected due to the contribution of $\lambda_H$ in the RGE of $m_H^2$, $m_N^2$ significantly. Fig. 15 shows the effective variations in the evolutions of $m_N^2$ for different $\lambda_D(M_G)$ while the flows of $m_H^2$, $m_{\bar{H}}^2$ remains almost same. From Figs. 13-15, one may put a remark on the choices of Yukawa couplings at $M_G$ that the large values of $\lambda_t, \lambda_H$ not only compel $m_N^2$ goes down to negative values faster as it should be but also force $m_H^2$ to move down negative rapidly, which may cause the electroweak breaking larger than the *TeV*. Therefore, $\lambda_t, \lambda_H$ should not be very large at GUT scale. On the other hand, keeping constant $\lambda_t, \lambda_H$ at reasonable value, the $m_N^2$, $m_H^2$ show the trends opposite. The large values of $\lambda_D$ allow $m_N^2$ to go down negative rapidly while for $m_H^2$ it occurs with delay. Therefore, a particular reasonable choices of Yukawa couplings at GUT scale can lead only the breaking of $U(1)'$ and electroweak breaking in the TeV region by going down to negative values of $m_N^2$, $m_H^2$, where $m_N^2$ becomes first negative and which allows the solution with $x >> \upsilon$ in accordance with the phenomenological requirements i.e. $M_{Z_2}$ to be the order of (*TeV*) or may be larger than TeV and $M_{Z_1}$ should approach *Z*(SM). Note also that due to the presence of the top Yukawa coupling $\lambda_t$, in general one has $m_H^2 > m_{\overline{H}}^2$ as a consequence of this fact that one always ends up with $\upsilon > \overline{\upsilon}$. Therefore neglecting Yukawa couplings $\lambda_b$ and $\lambda_\tau$ with respect to $\lambda_t$ is justified posteriori [34].

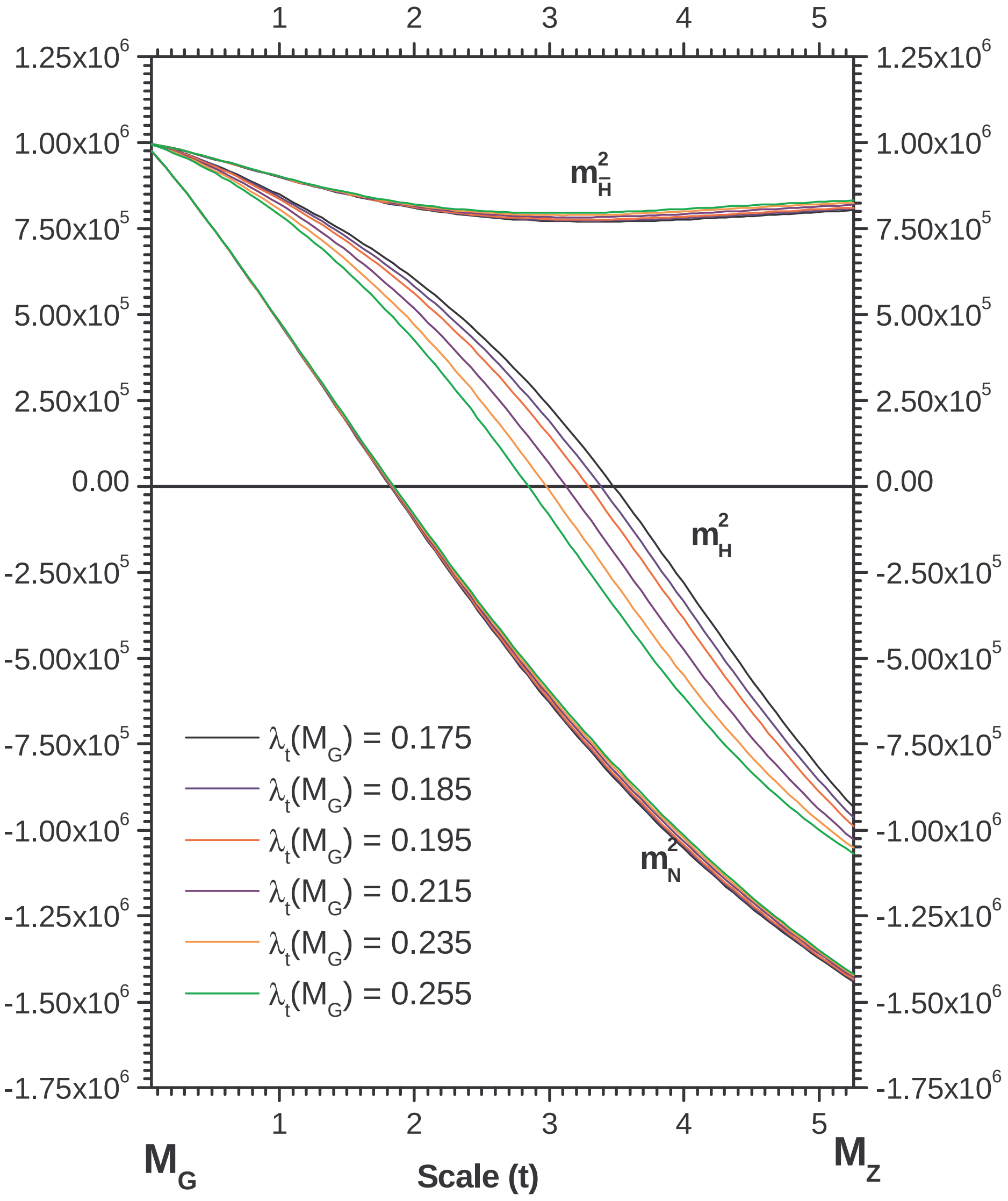

**Fig. 13:** *Typical scale dependence of the soft scalar masses* $m^2_H$, $m^2_{\bar{H}}$ *and* $m^2_N$ *according to the renormalization group equations. Figure shows the effect in the evolution of* $m^2_H$ *for different values of* $\lambda_t(M_G)$ *when* $\lambda_H(M_G)=0.650$, $\lambda_D(M_G)=0.450$.

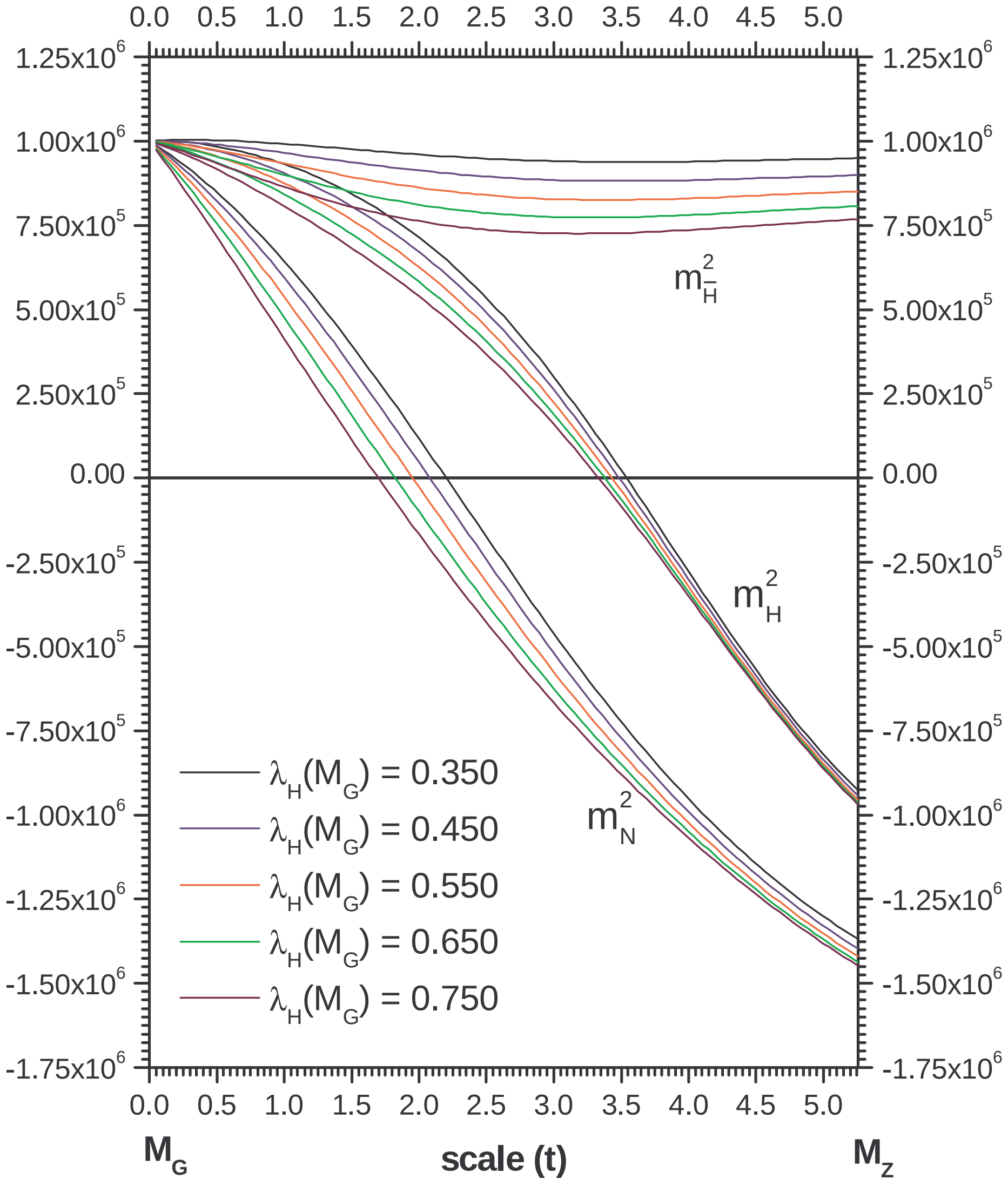

**Fig. 14:** *Figure shows the trends in the evolution of* $m_H^2$, $m_{\bar{H}}^2$ *and* $m_N^2$ *for different values of* $\lambda_H(M_G)$ *from GUT scale to weak scale, when* $\lambda_t(M_G)=0.175$, $\lambda_D(M_G)=0.450$.

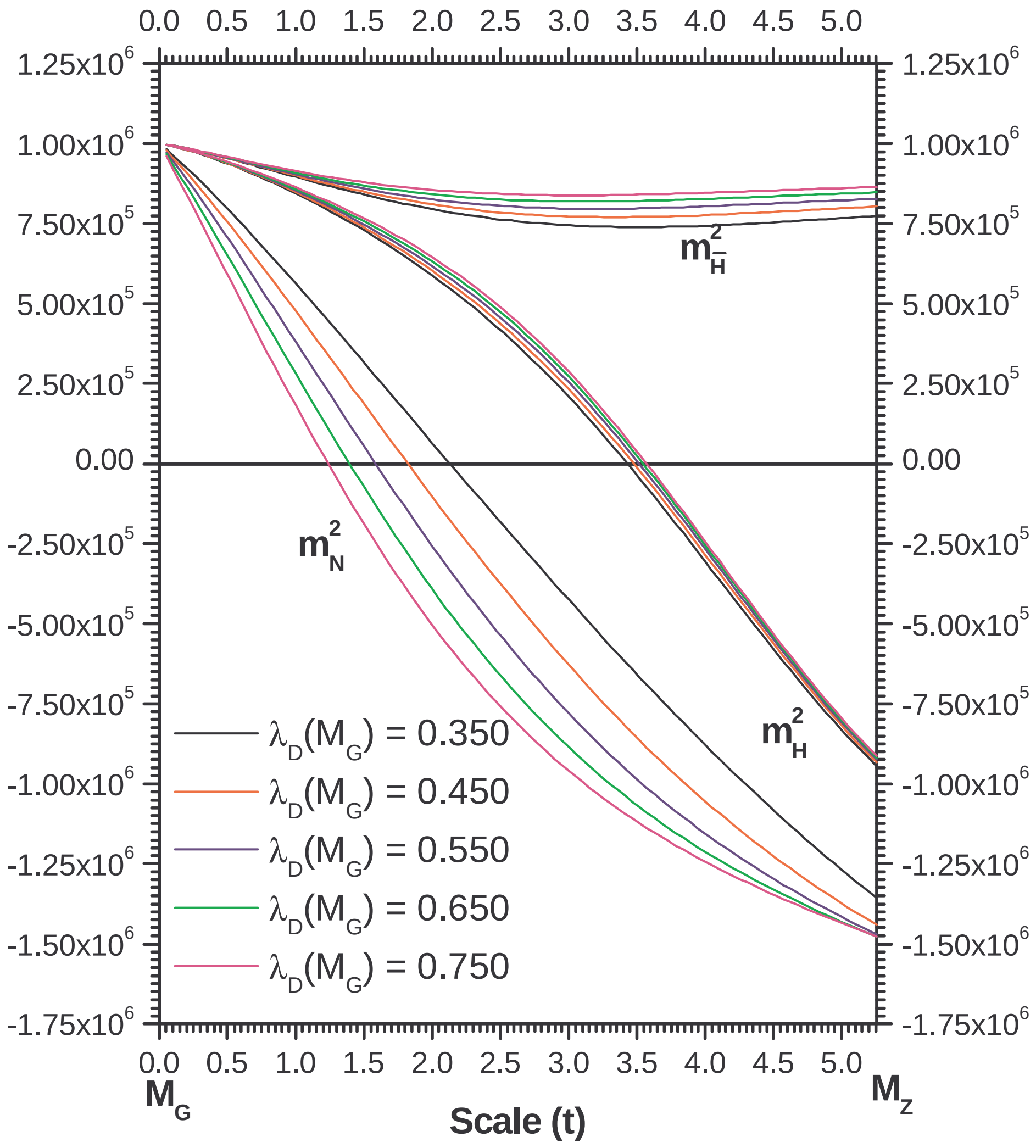

**Fig. 15:** *Figure shows the significant dependence of* $m_N^2$ *on* $\lambda_D(M_G)$ *when* $\lambda_t(M_G)=0.175$, $\lambda_H(M_G)=0.650$.

Further in order to discuss the general features of the analysis, the two parameters of greatest phenomenological interest are the ratio $\frac{x}{\upsilon}$ and gaugino mass $m_{1/2}$. Because $\frac{x}{\upsilon}$ combines with $\frac{\overline{\upsilon}}{\upsilon} \leq 1$ determines all the FCNC phenomenology as associated with $Z$, $Z'$ and $m_{1/2}$ lead the masses of supersymmetric particles. In the large *x* scenario, for the heavy $Z'$ up to the order of 2 TeV, we notice that $\frac{x}{\upsilon}$ take the values from 10~30 as shown in Fig. 16. This result is well consistent with the limit quoted in Ref. [13] and in our recent work [35].

Fig. 17 shows the typical correlations between $Z'$ and $\theta_E$ ($Z$-$Z'$ mixing angle), which provides more stringent lower bound for heavy $Z'$ ($E_6 - \eta$ model) for the small allowed mixing angle. The behavior reflects that $Z'$ should at least be the order of 800 GeV or greater than that which is higher than the prescribed limit by the experiments.

The scattered plot of $\frac{x}{\upsilon}$ vs. $m_{1/2}$ are shown in Fig. 18 corresponding to the points in the plane of $\lambda_t, \lambda_H$, $\lambda_D$ as shown in Fig.12. We obtain $m_{1/2}$ values from 300 GeV to 2 TeV under the upper limit of 3 TeV as quoted in Ref. [30]. As $\frac{x}{\upsilon}$ increases $m_{1/2}$ also increases and vice versa. The values of $\frac{x}{\upsilon}$ and $m_{1/2}$ are unrelated with each other because (i) for a particular allowed range of $\lambda_D$, the magnitude of *x* is fixed (ii) on the other hand, the magnitude of $\upsilon$ and $m_{1/2}$ should be in appropriate proportion to provide the experimentally bound on the $M_W$. The distributions of $\tan\beta$ with respect to $\frac{x}{\upsilon}$ are shown in Fig. 19.

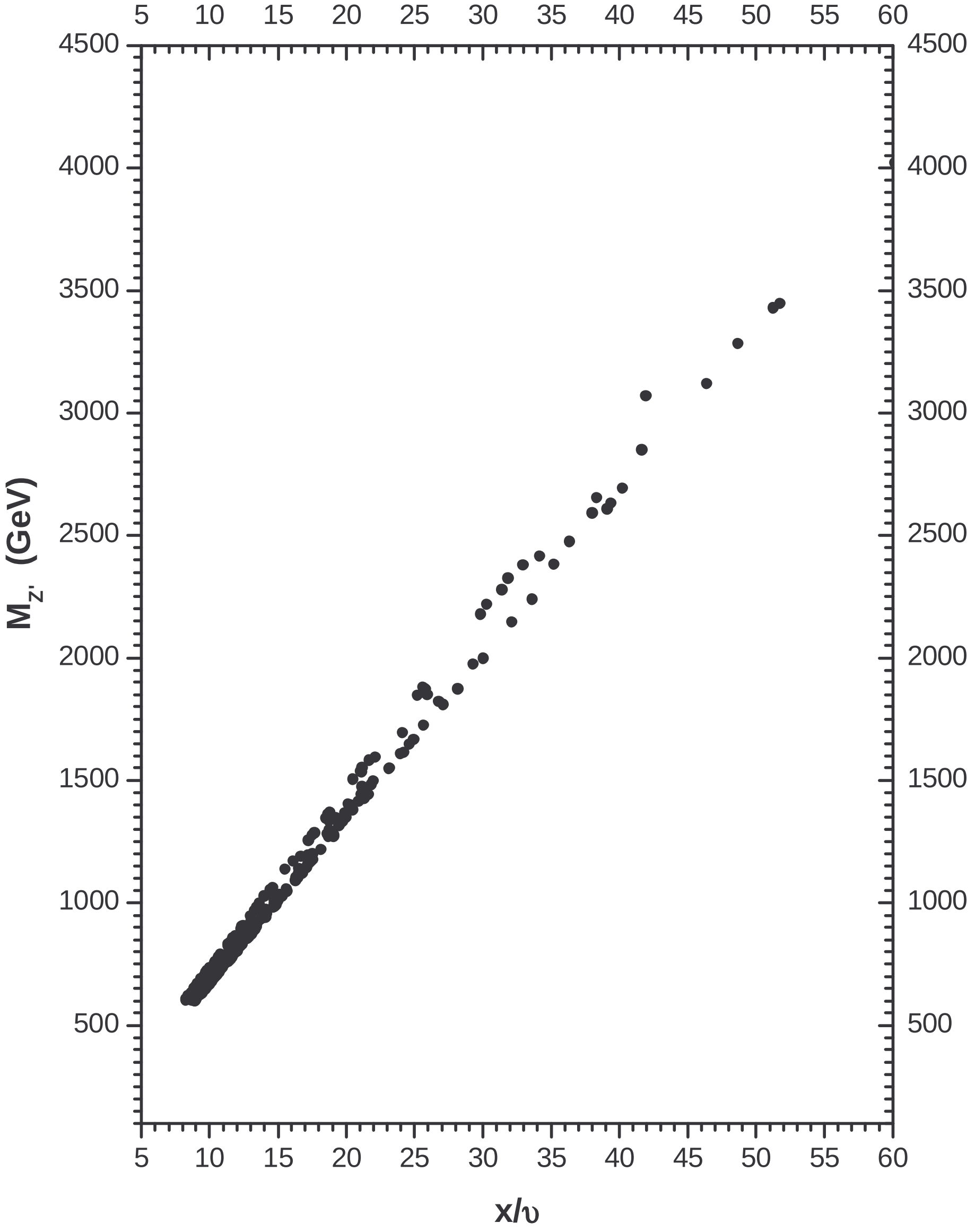

**Fig. 16:** *Theoretical scatter-plot of the ratio of the VEVs* $x/\upsilon$ *and* $M_{Z'}$ *for the large x scenario obtained with the non-universal Yukawa couplings at GUT scale.*

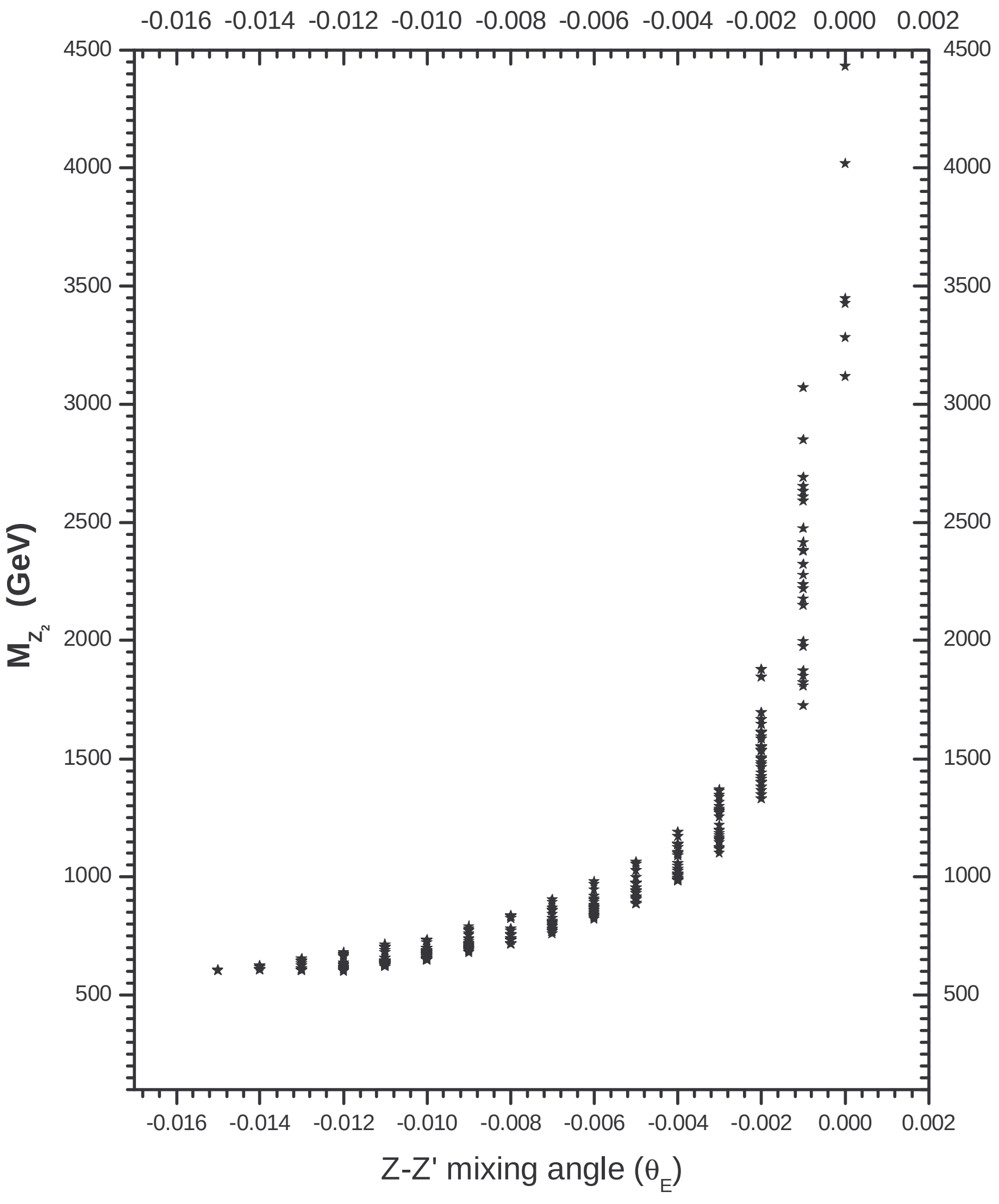

**Fig. 17:** *Figure shows the direct relation of $M_{Z'}$ and $Z - Z'$ mixing angle and reflects that $M_{Z'} \geq 800\ GeV$ theoretically.*

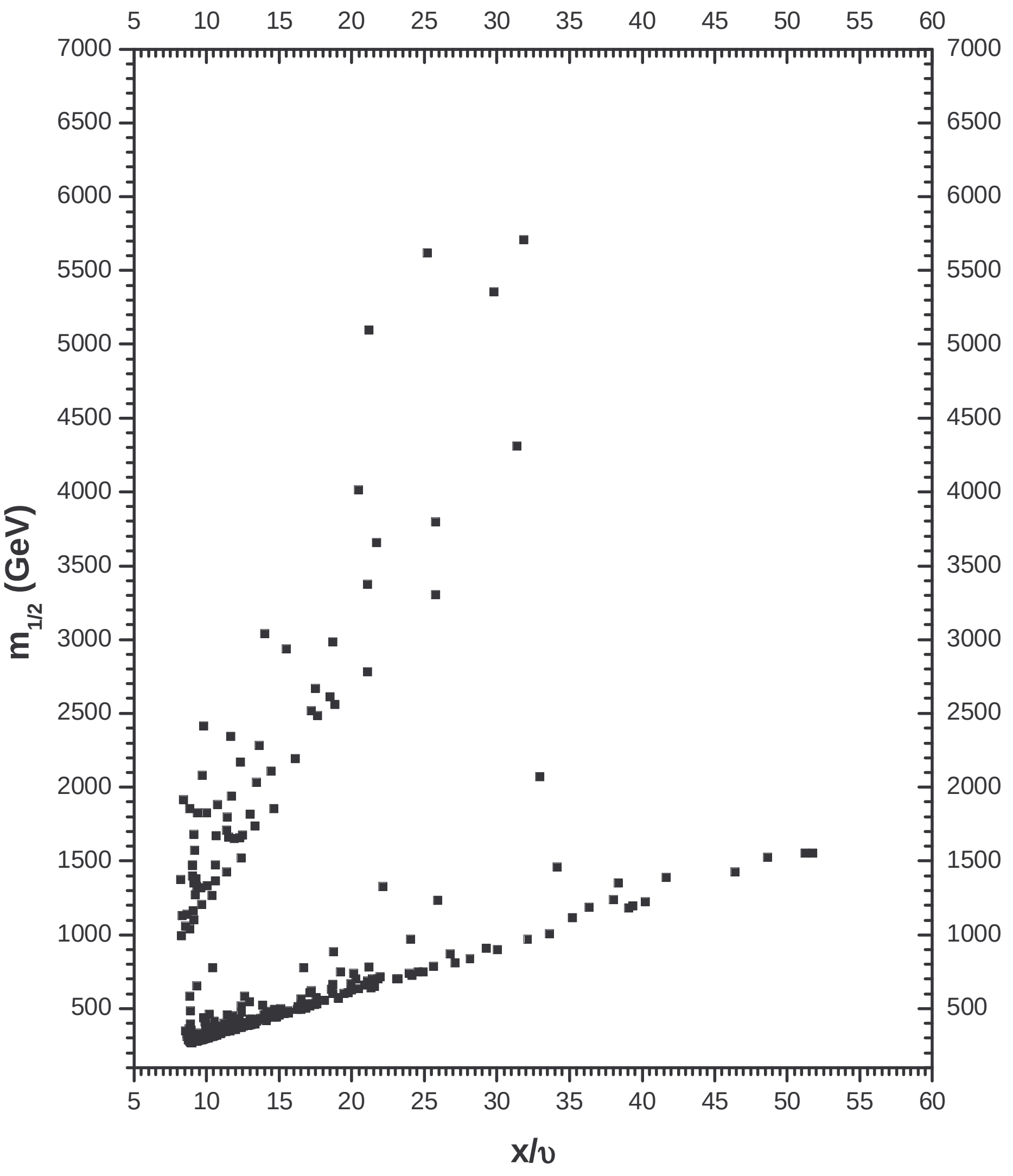

**Fig. 18:** *Theoretical scatter-plot showing the correlation between the ratio of VEVs* $x/\upsilon$ *and the gaugino mass* $m_{1/2}$ *for the allowed points in the palne of Fig. 12 corresponding to* $m_t = 178 \pm 2\ GeV$ .

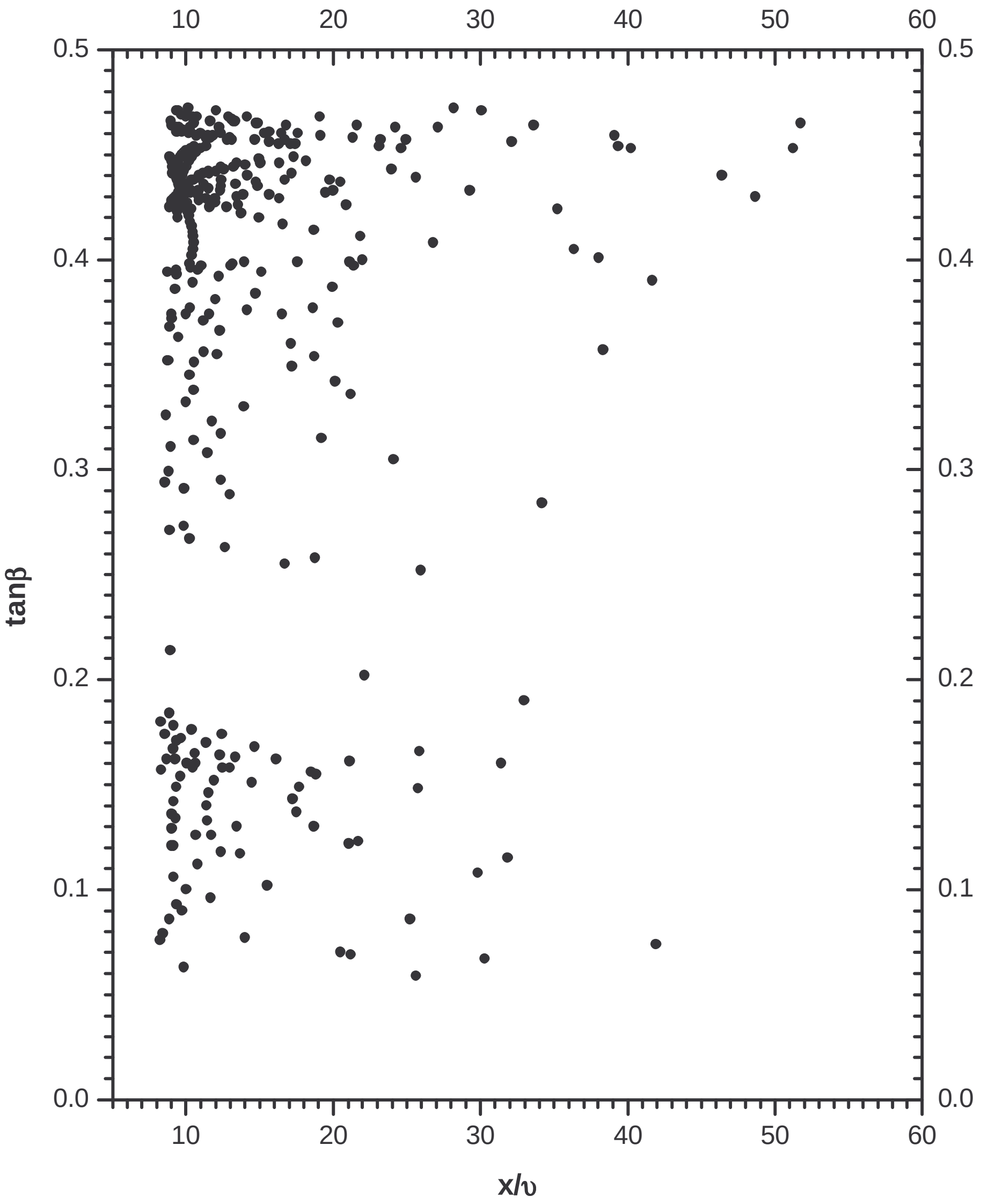

**Fig. 19:** *Theoretical scatter-plot showing the region of the* $\overline{\upsilon}/\upsilon,\ x/\upsilon$ *plane favoured by our model.*

## 4. CONCLUSION

A systematic numerical analysis of RGEs of extra U(1) superstring inspired model is done, with the non-universal Yukawa coupling at GUT scale ($M_G \sim 2\times10^{16}\ GeV$), where the gauge couplings are unified at GUT scale and gaugino masses, tri-linear coupling parameters, and soft SUSY breaking mass parameters are kept universal, i.e., $M_a = m_{1/2}$, $A_i = m_0$ and $m_i^2 = m_0^2$ at GUT scale. For completeness, we have discussed the nature of low energy spectrum obtained through the RGEs analysis in different scenarios, for example, *pure universality, large tri-linear coupling scenario* under the universal boundary conditions and the *large x scenario* under the non-universal boundary conditions. The correct electroweak symmetry breaking provides an effective $\mu$ parameter at weak scale, $\mu = \lambda_H x$, which should be at least of the order of few *TeV* and the presence of the extra neutral gauge boson ($Z'$) also fall in the *TeV* region with the small $Z - Z'$ mixing angle( $\sim 10^{-3}$) [39-41]. Our conclusions are:

(i) The universality of soft breaking parameters either in *Pure universality* or *Large tri-linear scenario* at GUT scale does not provide phenomenologically acceptable scenario. Since both scenarios provides $M_{Z'} < M_Z(SM)$ with large mixing angle which is excluded for the $E_6 - \eta$ model. Besides this, the analysis provide the mass of the top quark less than the experimental mass limits and the effective $\mu$ parameter found is less than $M_Z$. Thus, these scenarios are not able to provide large singlet VEV, which leads to the heavy $Z'$ and effective $\mu$ parameter of the order of *TeV* and therefore discarded.

(ii) Large singlet VEV is obtained by invoking the non-universality of non-zero Yukawa couplings at GUT scale for the desired low energy phenomenology, i.e., large *x* scenario.

(iii) With the fact that non-zero Yukawa couplings (should not be greater than the order of unity) are essential for triggering the electroweak symmetry breaking, it is noticed that a tiny parametric space of non-universal Yukawa couplings at GUT scale provides $Z'$ of the order of *TeV* with allowed mixing angle $\alpha_{Z-Z'}$ $\left(\approx 10^{-3}\right)$ where we have ignored the kinetic mixing of $U(1)_\chi$ and $U(1)_\psi$. The effective $\mu$ parameter is $\sim O(TeV)$. In the scalar Higgs sector, the lightest neutral scalar Higgs satisfies the unitarity constrained mass limits [5].

(iv) In *large x scenario*, the stable vacuum is obtained with the necessary condition $<N>^2 >> <H>^2 >> <\overline{H}>^2$ which is generated when the model parameters have the specific values at $M_G$: (a) the top quark Yukawa coupling is large enough, i.e., $\lambda_t^2/4\pi(t_G) \geq \alpha_3$, (b) the Yukawa coupling with extra *D* quark is not so large $\lambda_D^2/4\pi(t_G) \leq \alpha_3$. This vacuum condition is specified more precisely by putting more constraints at $M_Z$ i.e.: (a) $(\upsilon/x)^2 \leq 0.015$, (b) $(\overline{\upsilon}/\upsilon)^2 \leq 0.0015$ to provide $M_{Z_1} \cong M_Z(SM)$ and $M_{Z_2} >> M_{Z_1}$, i.e., *O*(2 *TeV*) pertaining to the top quark mass $m_t \cong 178 \pm 3\ GeV$ [48-52].

(v) Further, in brief, we would like to put some remarks on the charginos and neutralinos masses obtained in the *large x scenario*. The phenomenology of charginos and neutralinos depends on their field content: (a) they tend to be "gaugino-like" (for $M_2 \leq \mu$) or (b) "higgsino-like" ($\mu \leq M_2$), with a mixed field content available only for a relatively small region of parameter space. The large x scenario leads the charginos and neutralinos to be "gaugino-like" region where the chargino mass is driven by $M_2$ and neutralino mass by $M_1$ such that $M_1$ and $M_2$ are unified at GUT scale with $M_1 = M_2/2$ at weak scale. In this analysis, the lightest chargino and the neutralino satisfy the experimental lower bounds [58] of $\tilde{\chi}_1^{\pm}$ and $\tilde{\chi}_1^{0}$ that is 94 $GeV$ and 45 $GeV$ respectively. The other soft supersymmetric breaking mass parameters, like, $\tilde{\nu}$, $\tilde{e}$, $\tilde{t}$, $\tilde{d}(\tilde{b})$, $\tilde{l}$ are found in order to satisfy the experimental lower bounds [59].

Thus we have realized that out of two scenarios, namely, large $A_H$ and large singlet VEV $(<0|N|0> \equiv x)$, only the latter case induces an acceptable low energy spectrum for doing the phenomenology at weak scale and also may explain the new physics beyond the standard model.

.

# APPENDIX

The relevant one loop renormalization group equations (RGE) of extra U(1) superstring inspired model are given below, where the $U(1)'$ :quantum numbers are specified in Table 1.

***Gauge Couplings***

$$\frac{dg_1}{dt} = -\frac{b_1}{8\pi} g_1^3, \ \frac{dg_2}{dt} = -\frac{b_2}{8\pi} g_2^3, \ \frac{dg_3}{dt} = -\frac{b_3}{8\pi} g_3^3, \ \frac{dg_E}{dt} = -\frac{b_E}{8\pi} g_E^3$$

***Gaugino Masses***

$$\frac{dM_1}{dt} = -\frac{b_1}{4\pi} g_1^2 M_1, \ \frac{dM_2}{dt} = -\frac{b_2}{4\pi} g_2^2 M_2, \ \frac{dM_E}{dt} = -\frac{b_E}{4\pi} g_E^2 M_E$$

***Yukawa Couplings***

$$\frac{d\lambda_t}{dt} = \frac{\lambda_t}{8\pi}\left[\frac{16}{3} g_3^2 + 3g_2^2 + \frac{13}{15} g_1^2 + 2g_E^2\left(Y_1'^2 + Y_Q'^2 + Y_u'^2\right) - 6\lambda_t^2 - \lambda_H^2\right]$$

$$\frac{d\lambda_H}{dt} = \frac{\lambda_H}{8\pi}\left[3g_2^2 + \frac{3}{5} g_1^2 + 2g_E^2\left(Y_1'^2 + Y_2'^2 + Y_N'^2\right) - 3\lambda_t^2 - 4\lambda_H^2 - 3\lambda_D^2\right]$$

$$\frac{d\lambda_D}{dt} = \frac{\lambda_D}{8\pi}\left[\frac{16}{3} g_3^2 + \frac{4}{15} g_1^2 + 2g_E^2\left(Y_N'^2 + Y_D'^2 + Y_{D^C}'^2\right) - 2\lambda_H^2 - 5\lambda_D^2\right]$$

***Tri-linear Scalar Couplings***

$$\frac{dA_t}{dt} = \frac{1}{4\pi}\left[\frac{16}{3} g_3^2 m_{1/2} + 3g_2^2 M_2 + \frac{13}{15} g_1^2 M_1 + 2g_E^2\left(Y_1'^2 + Y_Q'^2 + Y_u'^2\right)M_E - 6\lambda_t^2 A_t - \lambda_H^2 A_H\right]$$

$$\frac{dA_H}{dt} = \frac{1}{4\pi}\left[3g_2^2 M_2 + \frac{3}{5} g_1^2 M_1 + 2g_E^2\left(Y_1'^2 + Y_2'^2 + Y_N'^2\right)M_E - 3\lambda_t^2 A_t - 4\lambda_H^2 A_H - 3\lambda_D^2 A_D\right]$$

$$\frac{dA_D}{dt} = \frac{1}{4\pi}\left[\frac{16}{3} g_3^2 m_{1/2} + \frac{4}{15} g_1^2 M_1 + 2g_E^2\left(Y_N'^2 + Y_D'^2 + Y_{D^C}'^2\right)M_E - 2\lambda_H^2 A_H - 5\lambda_D^2 A_D\right]$$

***Soft Supersymmetry-Breaking Scalar Masses***

$$\frac{d\tilde{m}_Q^2}{dt} = \frac{1}{4\pi}\left[\frac{16}{3}g_3^2 m_{1/2}^2 + 3g_2^2 M_2^2 + \frac{1}{15}g_1^2 M_1^2 + 4Y_Q'^2 g_E^2 M_E^2 - \lambda_t^2\left(m_{t^c}^2 + m_q^2 + m_H^2 + A_t^2\right)\right]$$

$$\frac{d\tilde{m}_{t^c}^2}{dt} = \frac{1}{4\pi}\left[\frac{16}{3}g_3^2 m_{1/2}^2 + \frac{16}{15}g_1^2 M_1^2 + 4Y_u'^2 g_E^2 M_E^2 - 2\lambda_t^2\left(m_{t^c}^2 + m_q^2 + m_H^2 + A_t^2\right)\right]$$

$$\frac{d\tilde{m}_L^2}{dt} = \frac{1}{4\pi}\left[3g_2^2 M_2^2 + \frac{3}{5}g_1^2 M_1^2 + 4Y_L'^2 g_E^2 M_E^2\right]$$

$$\frac{dm_H^2}{dt} = \frac{1}{4\pi}\left[3g_2^2 M_2^2 + \frac{3}{5}g_1^2 M_1^2 + 4Y_1'^2 g_E^2 M_E^2 - 3\lambda_t^2\left(m_{t^c}^2 + m_q^2 + m_H^2 + A_t^2\right) - \right.$$
$$\left. - \lambda_h^2\left(m_H^2 + m_{\overline{H}}^2 + m_N^2 + A_H^2\right)\right]$$

$$\frac{dm_{\overline{H}}^2}{dt} = \frac{1}{4\pi}\left[3g_2^2 M_2^2 + \frac{3}{5}g_1^2 M_1^2 + 4Y_2'^2 g_E^2 M_E^2 - \lambda_h^2\left(m_H^2 + m_{\overline{H}}^2 + m_N^2 + A_H^2\right)\right]$$

$$\frac{dm_N^2}{dt} = \frac{1}{4\pi}\left[4Y_N'^2 g_E^2 M_E^2 - 2\lambda_h^2\left(m_H^2 + m_{\overline{H}}^2 + m_N^2 + A_H^2\right) - 3\lambda_D^2\left(m_N^2 + m_D^2 + m_{D^c}^2 + A_D^2\right)\right]$$

$$\frac{d\tilde{m}_D^2}{dt} = \frac{1}{4\pi}\left[\frac{16}{3}g_3^2 m_{1/2}^2 + \frac{4}{15}g_1^2 M_1^2 + 4Y_D'^2 g_E^2 M_E^2 - \lambda_D^2\left(m_N^2 + m_D^2 + m_{D^c}^2 + A_D^2\right)\right]$$

$$\frac{d\tilde{m}_{D^c}^2}{dt} = \frac{1}{4\pi}\left[\frac{16}{3}g_3^2 m_{1/2}^2 + \frac{4}{15}g_1^2 M_1^2 + 4Y_{D^C}'^2 g_E^2 M_E^2 - \lambda_D^2\left(m_N^2 + m_D^2 + m_{D^c}^2 + A_D^2\right)\right]$$

$$\frac{d\tilde{m}_{d^c}^2}{dt} = \frac{1}{4\pi}\left[\frac{16}{3}g_3^2 m_{1/2}^2 + \frac{4}{15}g_1^2 M_1^2 + 4Y_d'^2 g_E^2 M_E^2\right]$$

$$\frac{d\tilde{m}_{e^c}^2}{dt} = \frac{1}{4\pi}\left[\frac{12}{15}g_1^2 M_1^2 + 4Y_e'^2 g_E^2 M_E^2\right]$$

$$\frac{d\tilde{m}_{\nu^c}^2}{dt} = \frac{1}{4\pi}\left[4Y_{\nu^c}'^2 g_E^2 M_E^2\right].$$

## REFERENCES

1. Usha Bhatia, N.K. Sharma, Phys. Rev. **D39**, 2502 (1989).
2. Usha Bhatia, N.K. Sharma, Int. J. Mod. Phys. **A5**, 501 (1990).
3. N. K. Sharma, P. Saxena, Sardar Singh, and Ashok K. Nagawat and R.S. Sahu, Phys. Rev. **D56**, 4152 (1997).
4. R.S. Sahu, N.K. Sharma, Sardar Singh, Ashok K. Nagawat, and P. Saxena, PRAMANA – Journal of physics, vol. **52**, (1999).
5. N.K. Sharma, Pranav Saxena, Prachi Parashar, Ashok K. Nagawat, and Sardar Singh, Phys. Rev. **D72**, 095016 (2005).
6. P. Langacker and J. Wang, Phys. Rev. **D58**, 115010 (1998); P. Langacker, N. Polonsky, J. Wang, Phys. Rev. **D60**, 115005 (1999).
7. M. Cvetic, P. Langacker, Phys. Rev. **D54**, 3570 (1996); Mod. Phys. Lett. **A11**, 1247 (1996).
8. G. Cleaver, M. Cvetic, J.R. Espinosa, L. Everett and P. Langacker, Phys. Rev. **D57**, 2701 (1998).
9. M. Cvetic, D. A. Demir, J. R. Espinosa, L. Everett and P. Langacker, Phys. Rev. **D56**, 2861 (1997)..
10. J.A. Casas, C. Munoz, Phys. Lett. **B306**, 288 (1993); H.P. Nilles, N. Polonsky, Nucl. Phys. **B484**, 33 (1997).
11. G.F. Giudice, A. Masiero, Phys. Lett. **B206**, 480 (1988); E.J. Chun, J. E. Kim, H. P. Nilles, Nucl. Phys. **B370**, 105 (1992); J. E. Kim, H. P. Nilles, Mod. Phys. Lett. **A9**, 3575 (1994); ibid, Phys. Lett. **B138**, 150 (1984).
12. D. Suematsu, Phys. Rev. **D57**, 1738 (1998); D. Suematsu and Y. Yamagishi, Int. J. Mod. Phys. **A10**, 4521 (1995)., T. Matsuda, Phys. Rev. **D58**, 035004 (1998).

13. Y. Nir, Phys. Lett. B**354**, 107 (1995); V. Jain & R. Shrock, Phys. Lett. **B352**, 83 (1995).

14. P. Chiappetta etal., Phys. Rev. **D54**, 789 (1996); G. Altarelli etal., Phys. Lett. **B375**, 292 (1996); K. Agashe etal., Physlett. **B385**, 218 (1996); V. Barger, K. Cheung and P. Langacker, Phys. Lett. **B381**, 226 (1996); J. Rosner, Phys. Lett. **B387**, 113 (1996); P.H. Frampton, M. Wise, and B. Wright, Phys. Lett. B54, 5820 (1996).

15. K. S. Babu, Chris Kolda, J.M. Russel, Phys. Rev. **D54**, 4635 (1996).

16. LEP Collaboration, LEP Electroweak Working Group, and SLD Heavy Flavor Group, Report No. CERN-PPE/96-183.

17. For reviews, see M. Cvetic, and S. Godfrey, hep-ph/9504216; A. Lieke, Phys. Rep. **317**, 149 (1999).

18. CDF Collaboration, F. Abe et al., Phys. Rev. Lett. **79**, 2192 (1997); D∅ Collaboration, S. Abachi et al., Phys. Rev. **B3856**, 471 (1996).

19. For a review, see M. Cvetic, and P. Langacker, **hep-ph/9707451**.

20. NuTeV Collaborations, G.P. Zeller et al., Phys. Rev. Lett. **88**, 091802 (2002).

21. C.S. Wood et al., Science **275**, 1759 (1997); S.C. Benett and C.E. Wieman, Phys. Rev. Lett. **82**, 2484 (1999); V.A. Dzuba, V.V. Flambaum, and J.S.M. Ginges, **hep-ph/0204134**.

22. R. Casalbuoni, S. De Curtis, D. Dominici, and R. Gato, Phys. Lett. **B460**, 135 (1999); J.L. Rosner, Phys. Rev. **D61**, 016006 (2000); J. Erler, P. Langacker, Phys. Rev. Lett. **84**, 212 (2000).

23. For a review, e.g., M.Cvetic, S. Godfrey, in *Proceedings of electroweak symmetry breaking and beyond standard model*, edited by T. Barklow, S.

Dawson, H. Haber and J. Seigrist (World Scientific, Singapore 1995), and references there in.

24. For reviews, see, H.P. Nilles, Phys. Rep. **110C**, 1 (1984).
25. K. Inoue, A. Kakuto, H. Homatsu and S. Takeshita, Prog. Theor. Phys. **68**, 927 (1982).
26. Y. Hosotani, Phys. Lett. **B126**, 309 (1983).
27. E. Witten, Nucl. Phys. **B258**, 76 (1985).
28. M. Dine, V. Kaplunovsky, M. Mangano, C. Nappi and N. Seiberg, Nucl. Phys. **B259**, 549 (1985).
29. S. Cecotti, J.P. Derendinger, S. Ferrara, L. Girardello and M. Roncadelli, Phys. Lett. **B156**, 318 (1985).
30. J. Ellis, K. Enqvist, D.V. Nanopoulos and F. Zwirner, Nucl. Phys. **B276**, 14 (1986); Mod. Phys. Lett. **A1**, 57 (1986).
31. L. E. Ibanez and J. Mas, Nucl. Phys. **B286**, 107 (1987).
32. J. Ellis, D.V. Nanopoulos, S.T. Petcov and F. Zwirner, Nucl. Phys. **B283**, 93 (1987); J. Ellis, A. B. Lahanas, D. V. Nanopoulos, K. Tamvakis,Phys. Lett. **B134**, 429 (1984).
33. E. Cohen, J. Ellis, K. Enqvist, D.V. Nanopoulos and F. Zwirner, Phys. Lett. **B165**, 76 (1985).
34. N. Nakamura, Y. Umemura and K. Yamamoto, Prog. Theor. Phys. **79**, 502 (1988).
35. J.L. Hewett, T.G. Rizzo, Phys. Rep. **183**, 193 (1989).
36. J.P. Derendinger and C.A. Savoy, Nucl. Phys. **B237**, 307 (1984).

37. L. Alvarez-Gaume, J.Polchinski and M.B. Wise, Nucl. Phys. **B221**, 495 (1983); C. Kounnas, A.B. Lahanas, D.V. Nanopoulos, M. Quiros, Nucl. Phys. **B236**, 438 (1984).

38. J. Erler,P. Langacker, T. Li, Phys. Rev. **D66**, 015002 (2002).

39. P. Langacker, M. Luo, Phys. Rev. **D45**, 278 91992).

40. R. Baratte et al., (*ALEPHCollaboration*), Eur. Phys. J. **C12**, 183 92000); M. Aciarri et al., (*L3 Collaboration*), Phys. Lett. **B479**, 101 (2000); P. Abreu et al., (*DELPHI Collaboration*), Phys. Lett. **B485**, 45 (2000); G. Abbiendi et al., (*OPAL Collaboration*), Eur. Phys. J. **C33**, 174 (2004); D. Abbaneo et al., (*LEP Collaboration*), **hep-ex/0312023**.

41. Gi-Chol Cho, Mod. Phys. Lett. **A15**, 311 (2000); K. Cheung, Phys. Lett. **B517**, 167 (2001).

42. W. Armstrong et al. (*ATLAS Collaboration*) ATLAS Report No. CERN/LHCC92-3 (unpublished); Report No. CERN/LHCC94-93/LHCC/P2, 1994 (unpublished); G.L. Bayatian et al.(CMS Collaboration), Report No. CERN/LHCC94-38/P1, 1994 (unpublished); M. Dittmar, A Djoudai, and A.S. Nicollart, Phys. lett. **B583**, 111 (2004).

43. (LEP Electroweak Working Group, SLD Heavy Flavor Group), hep-ph/0212036; http://lepwwg.web.cern.ch/LEPEWWG.

44. H. Haber, G. Kane, Phys. Rep. **C117**, 75 (1980).

45. V. Barger, M.S. Berger, P. Ohmann, Phys. Rev. **D49**, 4908 (1994).

46. H. Georgi, H. Quinn & S. Weinberg, Phys. Rev. Lett. **39**, 461 (1974).

47. A. Kusenko, P. Langacker and G. Segre, Phys. Rev. **D54**, 5824 (1996); A. Kusenko and P. Langacker, Phys. Lett. **B391**, 29 (1997).

48. T. Affolder et al., (CDF Collaboration), Phys. Rev. D63, 032003 (2001);

A. Abe et al., (CDF Collaboration), Phys. Rev. Lett. **82**, 271 (1999).

49. B. Abott et al., (D∅ Collaboration), Phys. Rev. **D60**, 052001 (1999).

50. V. M. Abazov et al., Nature (London) **429**, 638 (2004); Maria Florecia Canelli, in Intersections of Particle and Nuclear Physics, AIP Conf. Proc. No. **698** (AIP, New York, 2004), p.469; V. M. Abazov et al., Phys. Lett. **B600**, 25 (2005); L. Demortier et al., (*The CDF Collaborations, The D∅ Collaboration, and The Tevatron Electroweak Working Group*) hep-ex/0404010.

51. T. Aaltonen et al., (*CDF Collaboration: Measurement of the top quark mass in all hadronic decays in* $p\bar{p}$ *collision at CDF II*), **hep-ex/0612026**.

52. V.M. Abazov et al., (*D∅ Collaboration: Measurement of the top quark mass in di-lepton channel*), **hep-ex/0609056**.

53. M. Dine, R. Rohm, N. Seiberg, E. Wittne, Phys. Lett. **B155**, 55 (1985); E. Cohen, J. Ellis, C. Gomez, D.V. Nanopoulos, Phys. Lett. **B161**, 85 (1981).

54. P. Binetruy, M.K. Gaillard, Phys. Lett. **B165**, 347 (1986); E. Cohen, J. Ellis, K. Enqvist, D.V. Nanopoulos, Phys. Lett. **B161**, 85 (1985).

55. S. Ferrara, A. Font, F. Quevedo, M. Quiros, and M. Villasante, Nucl. Phys. **B288**, 233 (1987).

56. A. Strominger and E. Witten, Comm. Math. Phys. **101**,, 341 (1985).

57. S.M. Barr, Phys. Rev. Lett. **55**, 2778 (1985); V. Barger, N. G. Deshpande, Phys. Lett. **56**, 30 (1986); L. S. Durkin and P. Langacker, Phys. Lett. **B166**, 136 (1986); T. Tasuaka, H. Miro, D. Seumatsu and S. Watanbe, Prog. Theor. Phys. **68**, 927 (1987); F. Del. Augila, G. A. Blair, M. Daniel and G.G. Ross, Nucl. Phys. **B283**, 50 (1987).

58. *LEPSUSYWG, ALEPH, DELPHI, L3, OPAL Collaborations*: LEPSUSYWG/ 02-01.1, 02-02.1, 02-04.1, 02-06.2, 02-07.1, 02-08.1, 02-09.2, 02-10.1, 01-

03.1, 01-07.1; and see also http://wwww.cern.ch/lepsusy/; J. Abdallah et al., (*DELPHI Collaborations*), Eur. J. Phys. **C31**, 421 (2003).

59. A. Heister, Phys. Lett. **B544**, 73 (2002); *L3 Collaboration*, hep-ex/031007; *OPAL Collaboration*, Phys. Lett. **B545**, 272 (2002), Err. Ibid. **B548**, 258 (2002); *CDF Collaboration*, Phys. Rev. Lett. **88**, 041801 (2002); *D∅ Collaboration*, Phys. Rev. Lett. **83**, 4937 (1999) in *SUPERSYYMETRY*, Part-II by M. Schmitt, pg. 1120-1142, Review of Particle Physics, Journal of Physics **G33**, (2006).